\documentclass[english,aps,prx,notitlepage,twocolumn,longbibliography]{revtex4-1}
\usepackage[T1]{fontenc}
\usepackage[latin9]{inputenc}
\usepackage{babel}
\usepackage{float}
\usepackage{multirow}
\usepackage{amsmath}
\usepackage{graphicx}
\usepackage{xcolor}
\usepackage[unicode=true,
 bookmarks=true,bookmarksnumbered=false,bookmarksopen=false,
 breaklinks=false,pdfborder={0 0 1},backref=false,colorlinks=false]
 {hyperref}

\makeatletter

\providecommand{\tabularnewline}{\\}

\usepackage{babel}

\newcommand{\sm}[1]{{\scriptscriptstyle#1}}

\makeatother

\begin{document}
\title{Efficient slave-boson approach for multiorbital two-particle response functions
and superconductivity}
\author{Tsung-Han Lee$^{1}$, Nicola Lanat\`a$^{2,3}$, Minjae Kim$^{1,4}$, Gabriel
Kotliar$^{1,5}$}
\affiliation{$^{1}$Physics and Astronomy Department, Rutgers University, Piscataway,
New Jersey 08854, USA}
\affiliation{$^{2}$Department of Physics and Astronomy, Aarhus University, 8000
Aarhus C, Denmark}
\affiliation{$^{3}$Nordita, KTH Royal Institute of Technology and Stockholm University,
Hannes Alfv\'ens v\"ag 12, SE-106 91 Stockholm, Sweden}
\affiliation{$^{4}$Department of Chemistry, Pohang University of Science and Technology (POSTECH), Pohang 37673, Korea}
\affiliation{$^{5}$Condensed Matter Physics and Materials Science Department,
Brookhaven National Laboratory, Upton, New York 11973, USA}
\begin{abstract}
We develop an efficient approach for computing two-particle response
functions and interaction vertices for multiorbital strongly correlated
systems based on the rotationally-invariant slave-boson framework.
%, known as one of the quantum embedding theories.
The method is applied to the degenerate three-orbital
Hubbard-Kanamori model for investigating the origin of the $s$-wave
orbital antisymmetric spin-triplet superconductivity in the Hund's
metal regime, previously found in the dynamical mean-field theory
studies. By computing the pairing interaction considering the particle-particle
and the particle-hole scattering channels, we identify the mechanism
leading to the pairing instability around Hund's metal crossover arises
from the particle-particle channel, containing the local electron
pair fluctuation between different particle-number sectors of the atomic Hilbert
space. On the other hand, the particle-hole spin fluctuations induce the $s$-wave pairing instability before entering the Hund's regime. Our approach paves the way for investigating the pairing mechanism
in realistic correlated materials. 
%The possible applications of our formalism to density matrix embedding theory are also discussed
\end{abstract}
\maketitle

\section{Introduction}

Slave-boson approaches are among the most widely used theories
for describing strongly correlated systems~ \citep{Barnes_1976,Piers_1984,Kotliar1986,Li_1989_PRB_SRISB,Fresard_CRISB_1992,Florens_2004,deMedici_2005}.
In particular, the saddle-point approximation of the slave-boson method provides a reliable description of the local correlation effects, while requiring a relatively low computational cost compared to dynamical mean-field theory (DMFT)~\citep{DMFT_RMP_1996}.
The development of the rotationally-invariant slave-boson (RISB) saddle-point
approximation~\citep{Lecherman_2007}, equivalent to the Gutzwiller approximation
(GA)~\citep{Gutzwiller1963,Bunemann_GA_RISB}, has also been extended to realistic multiorbital
systems, in combination with density functional theory~\citep{Christoph_2011,Lechermann_2018}, uncovering
many intriguing phenomena, including selective-Mott transition~\citep{deMedici_2005,Lanata_2017_PRL,Lanata2019_NPJ},
Hund's metal behavior~\citep{Medici_Janus_PRL,Medici_2014,Facio_2018_PRB,Lechermann_2019_SRO},
valence fluctuations, and correlation induced topological materials~\citep{Lanata_2013_Ce,Lanata_2015_PRX,Dai_Topological}. 

Recently, RISB has been reformulated as a quantum embedding theory, where the interacting lattice problem is mapped to an impurity problem coupled to a self-consistently determined environment~\citep{Lanata_2015_PRX}, similar to DMFT and density matrix  embedding theory (DMET)~\citep{DMFT_RMP_1996,DMFT_RMP_2006,DMET_2012}. In particular, the RISB saddle-point equations are equivalent to the {\color{black}``non-interacting bath'' DMET (NIB-DMET)} self-consistent equations when setting the quasiparticle renormalization matrix to unity and enforcing an additional constraint on the structure of the physical density matrix \citep{Ayral_2017,RISB_DMET_Lee_2019}. 
%Due to this equivalency, the new development in one technique often inspires the new applications in the other, e.g., the finite-temperature, the non-equilibrium, the excited states, and the dynamics extensions, and different embedding schemes in both approaches
%Both RISB and DMET were originally proposed to describe the low temperature ground state properties. However, the recent development in both approaches allow further investigation of the finite-temperature effects, non-equilibrium dynamics, excited states,  and single-particle spectral properties.
In addition, the two methods, originally proposed for describing the ground state or low-temperature properties, have been extended to study the finite-temperature effects, the non-equilibrium dynamics, the excited states, and the single-particle spectral functions in correlated systems~\citep{FTDMET_2020,Lanata_finite_T_2015,Schiro_2010,Kretchmer_2018,Ye2021,EwDMET_2021,gGA_2017}.

%While RISB captures
%reliable single-particle observables, many experimental techniques
%rely on the two-particle response functions. Moreover, the two-particle
%interaction vertices are also important for disentangling the pairing
%mechanism of unconventional superconductors, and for the diagrammatic
%extension of the mean-field theory \citep{Rohringer2018}. 
%It is natural to ask if the slave-boson saddle-point approaches can produce these quantities. 
So far, RISB is mostly used for investigating the single-particle spectral functions and the static local observables. However, the two-particle response functions and the corresponding interaction vertices are also important for explaining the emergent phenomena in correlated materials, e.g., the spin-fluctuation mediated pairing in unconventional superconductors~\citep{Scalapino2012}. It is, therefore, important to extend RISB to study these quantities.
Indeed, it is possible to compute the two-particle response functions with the Gaussian fluctuation approach around the slave-boson saddle-point~\citep{Read_1983_Gaussian,Kotliar1986,Li_1989_PRB_SRISB,Lilly1990,Jolicoeur1991,Lavagna1990,Li1991,Arrigoni1995,Raimondi1993,Zimmermann1997,Dao2017,Riegler2020}.
However, the technique is so far restricted to the single-orbital Hubbard model. 
On the other hand, the development of the time-dependent Gutzwiller
approximation has been extended to multiorbital systems and applied
to the two-orbital Hubbard model for spin susceptibilities \citep{Seibold2001_charge,Seibold2004_magnetic,Fabrizio2017,Seibold_2008_PRL,Seibold_2008_PRB,von_Oelsen_2011_NJP,von_Oelsen_2011_PRL}.
To the best of our knowledge, the theories have not been generalized to compute
arbitrary two-particle response functions and quasiparticle interaction
vertices for multiorbital systems.

%Resolving the pairing mechanisms for unconventional superconductors is one of the central topics of condensed matter physics. The two-particle response functions, e.g., spin, charge, orbital, and pairing susceptibility, are especially important for disentangling the pairing mechanism. Dynamical mean-field theory (DMFT) is particularly successful for investigating these two-particle response functions in strongly correlated materials , e.g., Hund's physics in iron-based and ruthenate superconductors  and the valence fluctuation in heavy fermion systems .

%While the methodologies for evaluating the two-particle response function in DMFT are well established , the evaluation of them in multiorbital systems is typically heavy. Consequently, it is desirable to seek for an alternative approach that is efficient and reliable for strongly correlated system. 

In this work, we develop an efficient approach to compute general
susceptibilities and quasiparticle interaction vertices based on fluctuation
around the RISB saddle-point, allowing a diagrammatic analysis for
the pairing mechanism. We apply our method to the degenerate three-orbital
Hubbard-Kanamori model to investigate the origin of the $s$-wave
orbital-antisymmetric spin-triplet pairing instability in the Hund's
metal regime, previously found in the DMFT and GA studies \citep{Werner_power_law,Hoshino_Werner_2015_PRL,Zegrodnik_2013,Zegrodnik_2014}.
We show that, in agreement with DMFT \citep{Hoshino_Werner_2015_PRL},
our approach captures the $s$-wave spin-triplet pairing instability
around the Hund's metal crossover. By investigating the pairing interaction
considering the particle-particle and the particle-hole scattering
channel, we identify that the mechanism leading to the local $s$-wave
orbital-antisymmetric spin-triplet pairing arises from the particle-particle
channel, containing the local electron pair fluctuation between different
particle-number sectors of the local Hilbert space. Interestingly, the
particle-hole spin-fluctuation mechanism for the $s$-wave pairing, considered also in previous works \citep{Capone_science,Capone_PRL,Hoshino_Werner_2015_PRL}, induces the $s$-wave pairing instability slightly before entering the Hund's regime.
%Our new development can be combined with density functional theory to investigate the two-particle response functions and superconductivity in realistic materials.
Possible applications of our formalism to {\color{black} NIB-DMET} are also discussed.

\section{Model}

{\color{black}We consider the following generic multi-orbital Hubbard-Kanamori model: }
\begin{equation}
\hat{H}=\sum_{\mathbf{k}\alpha\beta\sigma}\epsilon_{\mathbf{k}\alpha\beta}d_{\mathbf{k}\alpha\sigma}^{\dagger}d_{\mathbf{k}\beta\sigma}+\sum_{i}\hat{H}_{\text{loc}}\big[\{d_{i\alpha\sigma}^{\dagger},d_{i\alpha\sigma}\}\big],\label{hamiltonian}
\end{equation}
where $\alpha$ is the orbital index, $\sigma$
is the spin index, $i$ is the unit-cell label, and $\mathbf{k}$ is the momentum conjugate to $i$. {\color{black} As a proof of principle and for pedagogical reason, we will assume a three-orbital degenerate model with the energy
dispersion of a two-dimensional square lattice with the nearest neighbor hopping:}
\begin{equation}
\epsilon_{\mathbf{k}\alpha\beta}=-2t(\text{cos}(\mathbf{k}_{x})+\text{cos}(\mathbf{k}_{y}))\delta_{\alpha,\beta}\,,
\end{equation}
{\color{black}where $\alpha\in\{1,2,3\}$, and we will set $t=1$ as the energy unit.  However, we note that our formalism applies to  multiorbital Hubbard models with general hopping matrix and arbitrary number of orbitals.} The term $H_{\text{loc}}$
represents the following operator: 
\begin{align}
\hat{H}_{\text{loc}} & \big[\{d_{i,\alpha}^{\dagger},d_{i\alpha}\}\big]=U\sum_{\alpha}n_{i\alpha\uparrow}n_{i\alpha\downarrow}+U'\sum_{\alpha<\alpha',\sigma}n_{i\alpha\sigma}n_{i\alpha'\bar{\sigma}}\nonumber \\
+ & (U'-J)\sum_{\alpha<\alpha',\sigma}n_{i\alpha\sigma}n_{i\alpha'\sigma}-J\sum_{\alpha<\alpha'}\big(d_{i\alpha\uparrow}^{\dagger}d_{i\alpha\downarrow}d_{i\alpha'\downarrow}^{\dagger}d_{i\alpha'\uparrow}\nonumber \\
+ & d_{i\alpha\uparrow}^{\dagger}d_{i\alpha\downarrow}^{\dagger}d_{i\alpha'\uparrow}d_{i\alpha'\downarrow}+\text{H.c.}\big)-\mu_0\sum_{\alpha\sigma}n_{\alpha\sigma},\label{eq:Kanamorai_int}
\end{align}
which contains the Kanamori interaction~\citep{Kanamori_1963} in
the cubic-harmonic basis. The first term is the intra-orbital Coulomb
interaction, the second term and the third term is the inter-orbital
Coulomb interaction, and the last term contains the spin-flip and
the pairing hopping interaction. Throughout our paper, we assume the
rotationally invariant condition $U'=U-2J$ and set $J=U/4$. Note
that, with this choice of parameters, the bare orbital-antisymmetric
spin-triplet pairing interaction is repulsive, i.e., $U'-J>0$. %%
The electron occupancy is controlled by the chemical potential $\mu_0$.

Due to the $O(3)\otimes SU(2)$ symmetry in the degenerate three-orbital
model, the orbital-antisymmetric spin-triplet pairing channels ~\citep{Cheung_2019,Agterberg_2020_PRR,Kaba_group_2019}
are related to each other by a rotation in the orbital and the spin
space. Consequently, we focus on the pairing fluctuation in one of
the orbital-antisymmetric spin-triplet pairing channels: 
\begin{equation}
\hat{\mathcal{O}}_{\text{P}}=\sum_{\alpha\beta}\sum_{\sigma\sigma'}\big[\lambda_{6}\big]_{\alpha\beta}[-i\sigma_{y}\sigma_{z}]_{\sigma\sigma'}d_{i,\alpha\sigma}^{\dagger}d_{i,\beta\sigma'}^{\dagger}.\label{eq:3orb_triplet}
\end{equation}
Similarly, we have the following independent {\color{black}operators} for the charge, spin, orbital, and spin-orbital fluctuation channels: 
\begin{equation}
\mathcal{\hat{O}}_{s}=\begin{cases}
\,\underset{\alpha\beta}{\sum}\big[\lambda_{0}\big]_{\alpha\beta}\big[\sigma_{0}\big]_{\sigma\sigma'}d_{i,\alpha\sigma}^{\dagger}d_{i,\beta\sigma'} & s=\text{ch}\\
\,\underset{\alpha\beta}{\sum}\big[\lambda_{0}\big]_{\alpha\beta}\big[\sigma_{z}\big]_{\sigma\sigma'}d_{i,\alpha\sigma}^{\dagger}d_{i,\beta\sigma'} & s=\text{sp}\\
\,\underset{\alpha\beta}{\sum}\big[\lambda_{4}\big]_{\alpha\beta}\big[\sigma_{0}\big]_{\sigma\sigma'}d_{i,\alpha\sigma}^{\dagger}d_{i,\beta\sigma'} & s=\text{orb}\\
\,\underset{\alpha\beta}{\sum}\big[\lambda_{4}\big]_{\alpha\beta}\big[\sigma_{z}\big]_{\sigma\sigma'}d_{i,\alpha\sigma}^{\dagger}d_{i,\beta\sigma'} & s=\text{so}\\
\,\underset{\alpha\beta}{\sum}\big[\lambda_{1}\big]_{\alpha\beta}\big[\sigma_{0}\big]_{\sigma\sigma'}d_{i,\alpha\sigma}^{\dagger}d_{i,\beta\sigma'} & s=\text{orb}^{*}\\
\,\underset{\alpha\beta}{\sum}\big[\lambda_{1}\big]_{\alpha\beta}\big[\sigma_{z}\big]_{\sigma\sigma'}d_{i,\alpha\sigma}^{\dagger}d_{i,\beta\sigma'} & s=\text{so}^{*},
\end{cases}\label{eq:3orb_flucs}
\end{equation}
where we label the fluctuation channels by $s\in\{\text{ch},\ \text{sp},\ \text{orb},\ \text{so},\ \text{orb*},\ \text{so*},\ \text{P}\}$
throughout the paper. {\color{black}Here, $\lambda_0$ is the $3\times 3$ identity matrix and $\lambda_{i}$ are the Gell-Mann matrices (see Appx. \ref{sec:Gell-mann-matrices}), while $\sigma_0$ is the $2\times2$ identity matrix and $\sigma_{i}$ ($i=x,y,z$) are the Pauli matrices.}

\section{Method}

{\color{black}
Our fluctuation approach around the RISB normal-state saddle-point is entirely encoded in the following Lagrange function~\citep{Isidori_RISB_SC_2009} (see Appx.~\eqref{sec:RISB_Nambu}):
%{\color{black}(see Appx.~\ref{sec:RISB_Nambu} and Ref.~\citep{Isidori_RISB_SC_2009} for the derivations)}
\begin{align}
&\mathcal{L}[|\Phi\rangle,\mathbf{R},\boldsymbol{\Lambda};\mathbf{D},
\boldsymbol{\Lambda}^{c} ,E^{c},\boldsymbol{\Delta}] = \nonumber\\
&\mathcal{L}_{\text{qp}}[\mathbf{R},\boldsymbol{\Lambda}] + \mathcal{L}_{\text{emb}}[\mathbf{D},
\boldsymbol{\Lambda}^c,|\Phi\rangle,E^c] + \mathcal{L}_{\text{mix}}[\mathbf{D},\mathbf{R},\boldsymbol{\Lambda}^c],\label{eq:Lemb}
\end{align}
where:
\begin{align}
\mathcal{L}_{\text{qp}}[\mathbf{R},\boldsymbol{\Lambda}]=\frac{-T}{N}\frac{1}{2}\sum_{\mathbf{k}_{1}\mathbf{k}_{2}\omega_{n}}\text{Tr log}\Big[-i\omega_{n}+H_{\mathbf{k}_{1}\mathbf{k}_{2}}^{\text{qp}}\Big]e^{i\omega_{n}0^{+}}\label{eq:Lemb1}
\end{align}
\begin{align}
\mathcal{L}_{\text{emb}}[\mathbf{D},\boldsymbol{\Lambda}^c,|\Phi\rangle,E^c]  = \sum_i\langle\Phi_i|\hat{H}_{i,\text{emb}}[\mathbf{D}_i,\boldsymbol{\Lambda}^{c}_i]|\Phi_i\rangle\nonumber\\
+E^{c}_i\big(\langle\Phi_i|\Phi_i\rangle-1\big)\label{eq:Lemb2}
\end{align}
\begin{align}
\mathcal{L}_{\text{mix}}&[\mathbf{D},\mathbf{R},\boldsymbol{\Lambda}^c] = -\sum_i\Big[\frac{1}{2}\sum_{ab}\big(\big[\boldsymbol{\mathbf{\mathbf{\Lambda}}}_i\big]_{ab}+\big[\boldsymbol{\mathbf{\mathbf{\Lambda}}}^{c}_i\big]_{ab}\big)\big[\boldsymbol{\Delta}_i\big]_{ab}\nonumber\\
&+\sum_{ac\alpha}\big(\big[\mathcal{\mathbf{D}}_i\big]_{a\alpha}\big[\mathbf{R}_i\big]_{c\alpha}\big[\boldsymbol{\Delta}_i(1-\boldsymbol{\Delta}_i)\big]_{ca}^{1/2}+\text{c.c.}\big)\Big].\label{eq:Lemb3}
\end{align}
Equation~\eqref{eq:Lemb1} encodes the contribution of the so-called ``quasiparticle fermionic'' degrees of freedom. Specifically, the matrix:
%\begin{align}
%\big[H_{\mathbf{k}_{1}\mathbf{k}_{2}}^{\text{qp}}]_{ab}=\frac{1}{N}\sum_{\mathbf{k}}\sum_{\alpha\beta}\big[\mathbf{R}_{\mathbf{k}_{1}-\mathbf{k}}\big]_{a\alpha}\big[\tilde{\epsilon}_{\mathbf{k}}\big]_{\alpha\beta}\big[\mathbf{R}_{\mathbf{k}-\mathbf{k}_{2}}^{\dagger}\big]_{\beta b}+\big[\boldsymbol{\mathbf{\mathbf{\Lambda}}}_{\mathbf{k}_{1}-\mathbf{k}_{2}}]_{ab},\label{eq:Hqp}
%\end{align}
\begin{align}
\big[H_{\mathbf{k}_{1}\mathbf{k}_{2}}^{\text{qp}}]_{ab}=\frac{1}{N}\sum_{\mathbf{k}}\big[\mathbf{R}_{\mathbf{k}_{1}-\mathbf{k}}\tilde{\epsilon}_{\mathbf{k}}\mathbf{R}_{\mathbf{k}_2-\mathbf{k}}^{\dagger}\big]_{a b}+\big[\boldsymbol{\mathbf{\mathbf{\Lambda}}}_{\mathbf{k}_{1}-\mathbf{k}_{2}}]_{ab},\label{eq:Hqp}
\end{align}
with {\color{black} the hopping term in the Nambu basis}
\begin{equation}
\tilde{\mathbf{\epsilon}}_{\mathbf{k}}=\begin{pmatrix}\epsilon_{\mathbf{k}} & 0\\
0 & -\epsilon_{\mathbf{-k}}^{*}
\end{pmatrix},
\end{equation}
characterizes the ``quasiparticle Hamiltonian'':
\begin{equation}
\hat{H}^\text{qp} = \sum_{\mathbf{k}_1,\mathbf{k}_2}\big[H_{\mathbf{k}_{1}\mathbf{k}_{2}}^{\text{qp}}]_{ab}\Psi^\dagger_{\mathbf{k}_1 a} \Psi_{\mathbf{k}_2 b},\label{eq:Hqp_op}
\end{equation}
where $\Psi_{\mathbf{k}}^{\dagger}=(f_{\mathbf{k}1\uparrow}^{\dagger}f_{\mathbf{k}1\downarrow}^{\dagger}...f_{\mathbf{k}M\uparrow}^{\dagger}f_{\mathbf{k}M\downarrow}^{\dagger}f_{\mathbf{k}1\uparrow}f_{\mathbf{k}1\downarrow}...f_{\mathbf{k}M\uparrow}f_{\mathbf{k}M\downarrow})$ is a Nambu spinor, $f_{\mathbf{k}a\sigma}$ are the fermionic quasiparticle modes, and $M$ is the total number of orbitals.
The matrix $\mathbf{R}$ is the so-called ``quasiparticle renormalization matrix'' and $\boldsymbol{\Lambda}$ is a matrix of Lagrange multipliers enforcing the RISB constraints~\citep{Lecherman_2007,Isidori_RISB_SC_2009}:
\begin{align}
\big[\boldsymbol{\Delta}_i\big]_{ab}\equiv \langle \Psi_{ia}^{\dagger} \Psi_{ib} \rangle_{T},\label{eq:const1}
\end{align}
where $\boldsymbol{\Delta}_i$ corresponds to the local quasiparticle density matrices~\citep{Lanata_2017_PRL}, and the symbol $\langle...\rangle_{T}$ denotes the thermal average of the non-interacting quasiparticle Hamiltonian $\hat{H}^\text{qp}$ at temperature $T$.

The second term $\mathcal{L}_{\text{emb}}$ (Eq.~\eqref{eq:Lemb2}) encodes the contribution of the slave-boson amplitudes, that here we expressed directly in terms of the corresponding ``quantum embedding'' states $|\Phi_i\rangle$ and \emph{interacting} embedding Hamiltonians~\citep{Lanata_2015_PRX} (see Appx.~\ref{subsec:Embedding-mapping}):
\begin{align}
\hat{H}_{i,\text{emb}} & =H_{i,\text{loc}}\big[\{\hat{d}_{i\alpha}^{\dagger},\hat{d}_{i\alpha}\}\big]+\big(\sum_{a\alpha b}\mathbf{D}_{ia\alpha}\hat{\Xi}_{i\alpha}^{\dagger}\hat{\Psi}_{ib}\bar{I}_{ba}+\text{H.c.}\big)\nonumber \\
 & +\sum_{abcd}\frac{1}{2}\boldsymbol{\Lambda}_{iab}^{c}\bar{I}_{bc}\hat{\Psi}_{ic}^{\dagger}\hat{\Psi}_{id}\bar{I}_{da}\Big],\label{eq:Hemb}
\end{align}
where $\hat{\Xi}_{i}^{\dagger}=(\hat{d}_{i1\uparrow}^{\dagger}\hat{d}_{i1\downarrow}^{\dagger}...\hat{d}_{iM\uparrow}^{\dagger}\hat{d}_{iM\downarrow}^{\dagger}\hat{d}_{i1\uparrow}\hat{d}_{i1\downarrow}...\hat{d}_{iM\uparrow}\hat{d}_{iM\downarrow})$
is the impurity Nambu spinor and $\hat{\Psi}_{i}^{\dagger}=(\hat{f}_{i1\uparrow}^{\dagger}\hat{f}_{i1\downarrow}^{\dagger}...\hat{f}_{iM\uparrow}^{\dagger}\hat{f}_{iM\downarrow}^{\dagger}\hat{f}_{i1\uparrow}\hat{f}_{i1\downarrow}...\hat{f}_{iM\uparrow}\hat{f}_{iM\downarrow})$ is the Nambu spinor for the bath orbitals. {\color{black} The matrix
\begin{equation}
\bar{I}=\begin{pmatrix}\mathbf{1} & 0\\
0 & \mathbf{-1}
\end{pmatrix}\label{eq:-13}
\end{equation}
is the sign exchange matrix generated from the embedding
mapping (see Appx. \ref{subsec:Embedding-mapping}), where $\mathbf{1}$ is the $2M\times 2M$ identity matrix.} The variable $E^c_i$ is a Lagrange multiplier enforcing the normalization of $|\Phi_i\rangle$:
\begin{equation}
\langle \Phi_i|\Phi_i\rangle \equiv 1.\label{eq:const2}
\end{equation}
The matrix $\boldsymbol{\Lambda}^c_i$, describing the embedding Hamiltonian bath potential, is a matrix of Lagrange multipliers enforcing the RISB constraints:
\begin{equation}
\big[\boldsymbol{\Delta}_i\big]_{ab} \equiv \langle\Phi_i|\hat{\Psi}_{ib}\hat{\Psi}_{ia}^{\dagger}|\Phi_i\rangle.\label{eq:const3}
\end{equation}
The matrix $\mathbf{D}_i$, describing the hybridization between the impurity and the bath orbitals, is a matrix of Lagrange multipliers, enforcing the definition of the renormalization matrix~\citep{Lecherman_2007,Isidori_RISB_SC_2009,Lanata_2017_PRL}
\begin{equation}
\mathbf{R}_{ia\alpha}=\sum_{b}\langle\Phi_i|\hat{\Xi}_{i\alpha}^\dagger\hat{\Psi}_{ib}|\Phi_i\rangle[\boldsymbol{\Delta}_i(1-\boldsymbol{\Delta}_i)\big]_{ba}^{-\frac{1}{2}}.\label{eq:R_mat}
\end{equation}
The third term $\mathcal{L}_\text{mix}$ (Eq.~\eqref{eq:Lemb3}) contains the Lagrange multipliers from both $\mathcal{L}_\text{qp}$ and $\mathcal{L}_\text{emb}$.
}
%Note that we have grouped the spin and orbital indices into a single index $\alpha$ and $a$ for the physical and the quasiparticle degrees of freedom, respectively.

{\color{black} %The derivation of the RISB Lagrange function are outlined in Appx.~\ref{sec:RISB_Nambu}.
All physical observables can be obtained from the above variational variables at the saddle-point solution of Eq.~\eqref{eq:Lemb}. The total energy is equal to the Lagrange function (Eq.~\eqref{eq:Lemb}) evaluated at the saddle-point. The expectation value of generic local operators {\color{black}$\hat{\mathcal{O}_i}[\{d_{i\alpha},d^\dagger_{i\alpha}\}]$} is determined from:
\begin{equation}
{\color{black}\langle\hat{\mathcal{O}_i}[\{d_{i\alpha},d^\dagger_{i\alpha}\}]\rangle}\equiv\langle\Phi_i\big|\hat{\mathcal{O}_i}[\{\hat{d}_{i\alpha},\hat{d}^\dagger_{i\alpha}\}]\big|\Phi_i\rangle\label{eq:Op}.
\end{equation}
In particular, the local (physical) single-particle density matrix is obtained from:
\begin{equation}
\rho_{i,\alpha\beta} \equiv \langle\Phi_i\big|\hat{\Xi}^\dagger_{i\alpha}\hat{\Xi}_{i\beta}\big|\Phi_i\rangle,\label{eq:dm_phys}
\end{equation}
%The self-energy is determined from the $\mathbf{R}$ and $\boldsymbol{\Lambda}$ matrices through:
%\begin{equation}
%\boldsymbol{\Sigma}_i(\omega)=-\omega(\mathbf{1}-[\mathbf{R}_i^\dagger\mathbf{R}_i]^{-1})+[\mathbf{R}_i^\dagger]^{-1}\boldsymbol{\Lambda_i}[\mathbf{R}_i]^{-1},
%\end{equation}
The quasiparticle weight is determined from the $\mathbf{R}$ matrix through $Z_i=\mathbf{R}_i^\dagger\mathbf{R}_i$.
}

Note that within the context of {\color{black} NIB-DMET}, Eq.~\eqref{eq:Hqp_op} corresponds to the so-called ``low-level mean-field'' Hamiltonian when setting $\mathbf{R}=I$, and $\boldsymbol{\Lambda}$ is termed ``correlation potential''. Equation~\eqref{eq:Hemb} corresponds to the so-called ``high-level many-body Hamiltonian'' in {\color{black}NIB-DMET}, where the two-particle interaction on the bath orbitals is set to zero~\citep{DMET_2012}.

%\begin{align}
%\mathcal{L}[\mathbf{x}_{\mu},|\Phi_{i}\rangle,E_{i}]= & \mathcal{L}_{\text{qp}}[\mathbf{x}_{\mu}]+\mathcal{L}_{\text{mix}}[\mathbf{x}_{\mu}]+\mathcal{L}_{\text{emb}}[\mathbf{x}_{\mu},|\Phi_{i}\rangle,E_{i}],\label{eq:Ltot}
%\end{align}
%where
%\begin{align}
%\mathcal{L}_{\text{qp}}[\mathbf{x}_{\mu}]= & \frac{-T}{N}\frac{1}{2}\sum_{\mathbf{k}\omega_{n}}\text{Tr log}\Big[-i\omega_{n}+H_{\mathbf{k}ab}^{\text{qp}}\Big]e^{i\omega_{n}0^{+}},\\
%\mathcal{L}_{\text{emb}}[\mathbf{x}_{\mu},|\Phi_{i} & \rangle,E_{i}]=\text{\ensuremath{\langle}\ensuremath{\Phi_{i}}\ensuremath{\big|}}\hat{H}_{\text{emb}}[\mathbf{x}_{\mu}]\big|\Phi_{i}\rangle+E_{i}^{c}\big(1-\langle\Phi_{i}\big|\Phi_{i}\rangle\big),
%\label{lagemb}
%\\
%\mathcal{L}_{\text{mix}}[\mathbf{x}_{\mu}]= & -\Big[\sum_{ab}\frac{1}{2}(\boldsymbol{\Lambda}_{i,ab}+\boldsymbol{\Lambda}_{i,ab}^{c})\boldsymbol{\Delta}_{i,ab}\nonumber \\
% & +\sum_{a\alpha c}\big(\mathcal{\mathbf{D}}_{i,a\alpha}\mathbf{R}_{i,c\alpha}[\boldsymbol{\Delta}(1-\boldsymbol{\Delta})]_{i,ca}^{\frac{1}{2}}+\text{c.c.}\big)\Big].
%\end{align}

\subsection{Parameterization of the single-particle matrices}

{\color{black} To enforce the symmetry conditions of the Lagrange function,} we introduce the following parameterization of the renormalization matrix, $\mathbf{R}_i$, and the Lagrange
multipliers, $\boldsymbol{\Lambda}_i$, $\boldsymbol{\Delta}_i$, $\mathbf{D}_i$, and $\boldsymbol{\Lambda}^{c}_i$
\citep{Lanata_2017_PRL}: 
\begin{align}
\mathbf{R}_i & =\sum_{s}r_{i,s}\tilde{\mathbf{h}}_{s},\label{eq:R}\\
\boldsymbol{\Lambda}_i & =\sum_{s}l_{i,s}\mathbf{h}_{s},\label{eq:Lam}\\
\boldsymbol{\Delta}_i & =\frac{1}{2}\mathbf{1}+\sum_{s}d_{i,s}\mathbf{h}_{s}^{t},\label{eq:Delta}\\
\mathcal{\mathbf{D}}_i & =\sum_{s}D{}_{i,s}\tilde{\mathbf{h}}_{s},\label{eq:D}\\
\boldsymbol{\Lambda}^{c}_i & =\sum_{s}l_{i,s}^{c}\mathbf{h}_{s},\label{eq:Lamc}
\end{align}
where {\color{black} $\mathbf{1}$ is the $4M\times 4M$ identity matrix, and $\mathbf{h}_s$ and $\tilde{\mathbf{h}}_s$ are the symmetry-adapted matrix basis of the above single-particle matrices. The structure of the matrix basis $\mathbf{h}_s$ and $\tilde{\mathbf{h}}_s$ is determined from the group symmetry analysis of the model in the presence of the {\color{black}fluctuating operators} (e.g., Eqs.~\eqref{eq:3orb_triplet}-\eqref{eq:3orb_flucs})~\citep{Lanata_2017_PRL}. This parameterization allows us to classify the fluctuations of the variational parameters ($r_s$, $l_s$, etc.) to a specific symmetry channel $s$, associated to $\mathbf{h}_s$ and $\tilde{\mathbf{h}}_s$. For example, in the degenerate three-orbital Hubbard-Kanamori model,  the $\mathbf{h}_s$ and $\tilde{\mathbf{h}}_s$ (see Appx.~\ref{subsec:Single-particle-basis}) are associated to the fluctuation channels $s\in\{\text{ch},\ \text{sp},\ \text{orb},\ \text{so},\ \text{orb*},\ \text{so*},\ \text{P}\}$ in Eqs.~\eqref{eq:3orb_triplet}-\eqref{eq:3orb_flucs}.
%in the presence of the fluctuating order parameters
%whose structure is determined from the group theory analysis (discussed in the supplement of Ref.~\citep{Lanata_2017_PRL}). 
%${${\mathbf{h}_{s}}$ and $\tilde{\mathbf{h}}_{s}}$ span the most general variational parameters consistent with the symmetry of the model in the presence of the symmetry broken fluctuations in Eqs.~\eqref{eq:3orb_triplet} and \eqref{eq:3orb_flucs}.
%The explicit form of $\mathbf{h}_s$ and $\tilde{\mathbf{h}}_s$ for the three-orbital degenerate Hubbard-Kanamori model are given in Appendix \ref{subsec:Single-particle-basis}.
In addition, for computing the susceptibility of a given channel $s$,
the embedding wavefunction $|\Phi_i\rangle$ has to break the corresponding
symmetry, e.g., the particle number conservation of $|\Phi_i\rangle$ has to be broken
for the pairing susceptibility calculations. %This procedure is described in Appx. \ref{sec:Variational-basis}. 

%It is convenient to group the variational variables into a vector for the conciseness of the following derivations. Taking the three-orbital Hubbard-Kanamori model as example, we have the following vector of variational variables:{\color{black} 
For later convenience, we introduce the following vector of parameters:
\begin{align}
\mathbf{x}_{i}=( & r_{i,\text{ch}},l_{i,\text{ch}},d_{i,\text{ch}},D_{i,\text{ch}},l_{i,\text{ch}}^{c},...,r_{i,s},l_{i,s},\nonumber \\
 & d_{i,s},D_{i,s},l_{i,s}^{c},...,r_{i,\text{P}},l_{i,\text{P}},d_{i,\text{P}},D_{i,\text{P}},l_{i,\text{P}}^{c}),\label{eq:X}
\end{align}and assume that all of its entries are real, which is sufficient for static quantities (e.g.,
static susceptibilities and Landau parameters \citep{Li1991,Zimmermann1997}).}
%The introduction of the complex variables will be important for the dynamic Hubbard excitations \cite{Jolicoeur1991,Dao2017}, which will be discussed in the future work.
Note that our assumption of real
variables is applicable for our model without spin-orbit coupling. 
%When considering the spin-orbital coupling, one has to include
%the fluctuation of the imaginary part of $\mathbf{R}$ and $\mathbf{D}$. The derivation, in this
%case, is straightforward, following the same procedure presented below.
The generalization to spin-orbit coupled systems can be straightforwardly obtained using the same procedure proposed in this work, by including in the Lagrangian also the imaginary part of $\mathbf{R}$ and $\mathbf{D}$.

\subsection{Saddle-point approximation}

The first step of our fluctuation approach is to determine the normal-state
saddle-point solution without any ordering. {\color{black} We assume a spatially homogeneous saddle-point solution, where $\mathbf{x}_i$ does not depend on $i$.}
% Hence, the Lagrangian can be written as:
%\begin{equation}
%\mathcal{L}[\mathbf{x},\Phi,E^{c}]= \mathcal{L}_{\text{qp}}[\mathbf{x}]+\mathcal{L}_{\text{mix}}[\mathbf{x}]+\mathcal{L}_{\text{emb}}[\mathbf{x},\Phi,E^{c}],\label{eq:Ltot0}
%\end{equation}
%where
%\begin{align}
%\mathcal{L}_{\text{qp}}[\mathbf{x}]= & \frac{-T}{N}\frac{1}{2}\sum_{k}\text{Tr log}\Big[-i\omega_{n}+H_{\mathbf{k},ab}^{\text{qp}}\Big]e^{i\omega_{n}0^{+}},
%\end{align}
%\begin{align}
%\mathcal{L}_{\text{emb}}\big[\mathbf{x},\Phi&,E^{c}\big] =\sum_{i}\big\langle\Phi(\mathbf{x})\big|\hat{H}_{\text{emb}}(\mathbf{x})\big|\Phi(\mathbf{x})\big\rangle\nonumber\\
%&+E^{c}\big(1-\langle\Phi\big|\Phi\rangle\big),
%\end{align}
%\begin{align}
%\mathcal{L}_{\text{mix}}[\mathbf{x}]= & -\Big[\sum_{ab}\frac{1}{2}(\boldsymbol{\Lambda}_{ab}+\boldsymbol{\Lambda}_{ab}^{c})\boldsymbol{\Delta}_{ab}\nonumber\\
%&+\sum_{a\alpha c}\big(\mathcal{\mathbf{D}}_{a\alpha}\mathbf{R}_{c\alpha}[\boldsymbol{\Delta}(1-\boldsymbol{\Delta})]_{ca}^{\frac{1}{2}}+\text{c.c.}\big)\Big],
%\end{align}
%and we define the spatially homogeneous quasiparticle Hamiltonian as:
%\begin{equation}
%\big[H_{\mathbf{k}}^{\text{qp}}\big]_{ab}=\big[\mathbf{R}\big]_{a\alpha}\big[\tilde{\epsilon}_{\mathbf{k}}\big]_{\alpha\beta}\big[\mathbf{R}^{\dagger}\big]_{\beta b}+\big[\boldsymbol{\Lambda}\big]_{ab}.\label{eq:Hqp}
%\end{equation}

Performing the partial derivatives of Eq.~\eqref{eq:Lemb_const} with respect to $\mathbf{x}$, we arrive the following saddle-point equations:
\begin{gather}
\big[\boldsymbol{\Delta}\big]_{ab}=\frac{1}{N}\sum_{\mathbf{k}}\big[f_{T}(H^{\text{qp}}_{\mathbf{k}})\big]_{ba},\label{eq:sp1-1}\\
\big[\boldsymbol{\Delta}(1-\boldsymbol{\Delta})\big]_{ac}^{1/2}\mathbf{D}_{ca}=\frac{1}{N}\frac{1}{2}\sum_{\mathbf{k}}\big[\tilde{\mathbf{\epsilon}}_{\mathbf{k}}R^{\dagger}f_{T}(H^{\text{qp}}_{\mathbf{k}})\big]_{\alpha a},\label{eq:sp2-1}\\
\sum_{cb\alpha}\partial_{d_s}\big[\boldsymbol{\Delta}(1-\boldsymbol{\Delta})\big]_{cb}^{1/2}\big[\mathbf{D}\big]_{b\alpha}\big[\mathbf{R}\big]_{c\alpha}\nonumber \\
+\text{c.c.}+\frac{1}{2}\big[l_{s}+l_{s}^{c}\big]=0,\label{eq:sp3-1}\\
\hat{H}_{\text{emb}}\big|\Phi\rangle=E^{c}\big|\Phi\rangle,\label{eq:sp4-1}\\
\Big[\mathcal{F}^{(1)}\Big]_{ab}\equiv\langle\Phi|\bar{I}_{bc}\hat{\Psi}_{c}\hat{\Psi}_{d}^{\dagger}\bar{I}_{da}|\Phi\rangle-\big[\boldsymbol{\Delta}\big]_{ab}=0,\label{eq:sp5-1}\\
\Big[\mathcal{F}^{(2)}\Big]_{\alpha a}\equiv\langle\Phi|\hat{\Xi}_{\alpha}^{\dagger}\hat{\Psi}_{b}\bar{I}_{ba}|\Phi\rangle-\mathbf{R}_{c\alpha}\big[\boldsymbol{\Delta}(1-\boldsymbol{\Delta})\big]_{ca}^{1/2}=0,\label{eq:sp6-1}
\end{gather}
where $f_{T}$ is the Fermi function and $H^{\text{qp}}_\mathbf{k}=\mathbf{R}\tilde{\epsilon}_\mathbf{k}\mathbf{R}^\dagger+\boldsymbol{\Lambda}$ is the saddle-point quasiparticle Hamiltonian. Equations~\eqref{eq:sp1-1}-\eqref{eq:sp6-1}
can be solved numerically utilizing quasi-Newton methods \citep{Lanata_2015_PRX,Lanata_2017_PRL}.
%We can then compute the susceptibility around the normal-state saddle-point through the following fluctuation approach.
%is finite as opposed to the normal-state saddle-point solution. 
Note that our saddle-point equations yield consistent results compared to the formalism in Ref.~\citep{Isidori_RISB_SC_2009}. 

It is also interesting to point out that Eqs.~\eqref{eq:sp1-1}-\eqref{eq:sp6-1} are equivalent to the {\color{black}NIB-DMET} self-consistent equations when setting the renormalization matrix to unity $\mathbf{R}=I$ and enforcing the so-called ``quasiparticle constraint'' that we will introduce later in Sec.~\ref{sec:QP_sus}~\cite{Ayral_2017}. 
%The $\mathbf{R}$ in RISB allows the description of the Mott transition within the single-site approach~\cite{BrinkmanRice1970}, while in the standard DMET, one has to use at least a two-site cluster to capture the Mott transition~\cite{DMET_2012}.

Given the saddle-point solution in the normal phase, we want to compute
the corresponding susceptibilities. This will be accomplished using
the approach described below.

\subsection{Calculation of susceptibilities \label{subsec:emb_sus}}

Here we describe the formalism for calculating the susceptibilities
in multi-orbital systems within the RISB framework. For concreteness,
we focus on uniform susceptibilities in this section, where $\mathbf{x}_i$ is independent of $i$ and we suppress the $i$ index in the following derivation. The generalization
to susceptibilities with finite momentum transfer is described in Sec. \ref{sec:QP_sus}.

Let us consider the RISB Lagrange function (Eq.~\eqref{eq:Lemb}) in the presence of a local
perturbation, proportional to a generic {\color{black}operator}
$\hat{\mathcal{O}}$:
\begin{align}
\mathcal{L}[{\color{black}\xi},\mathbf{x},\Phi,E^{c}]= & \mathcal{L}_{\text{qp}}[\mathbf{x}]+\mathcal{L}_{\text{mix}}[\mathbf{x}]+\mathcal{L}_{\text{emb}}[{\color{black}\xi},\mathbf{x},\Phi,E^{c}],\label{eq:Ltot}
\end{align}
where we have modified the embedding part of the Lagrangian to
\begin{align}
\mathcal{L}_{\text{emb}}&\big[{\color{black}\xi},\mathbf{x},\Phi,E^{c}\big]  =\sum_{i}\big\langle\Phi(\mathbf{x})\big|\hat{H}_{\text{emb}}[\mathbf{x}]\nonumber\\
&-\xi\mathcal{\hat{O}}\big|\Phi(\mathbf{x})\big\rangle+E^{c}\big(1-\langle\Phi(\mathbf{x})\big|\Phi(\mathbf{x})\rangle\big),
\end{align}
which was obtained by adding a {\color{black}field $\xi$ coupled to $\mathcal{\hat{O}}$} in the embedding Hamiltonian of Eq.~\eqref{eq:Hemb} and expressing
the variational parameters in terms of the vector $\mathbf{x}$, see Eq.~\eqref{eq:X}.

To calculate the linear response of the system to the perturbation
$\mathcal{\hat{O}}$, we need to evaluate how the saddle-point
variational parameters $\mathbf{x}$ of Eq.~\eqref{eq:Ltot} evolves
as a function of $\xi$. For this purpose, it is convenient to introduce
the following functional: 
\begin{align}
\Omega[{\color{black}\xi},\mathbf{x}]&=\mathcal{L}_{\text{qp}}[\mathbf{x}]+\mathcal{L}_{\text{mix}}[\mathbf{x}]\nonumber\\
&+\mathcal{L}_{\text{emb}}[{\color{black}\xi},\Phi(\xi,\mathbf{x}),E^{c}(\xi,\mathbf{x})]\,,
\end{align}
where $|\Phi(\xi,\mathbf{x})\rangle$ and $E^{c}(\xi,\mathbf{x})$ are the ground state
of $\hat{H}_{\text{emb}}$ and its eigenvalue, respectively, see Eq.~\eqref{eq:sp4-1}.
{\color{black} Within these definitions, the saddle-point solution of $\mathbf{x}$ for a given $\xi$, that we call $\mathbf{x}(\xi)$, is defined by:
\begin{equation}
%\left.\frac{\partial\Omega}{\partial \mathbf{x}}\right|_{\mathbf{x}(\xi)}=0\,,\label{stationary-formal}
\left.\partial_\mathbf{x}\Omega[{\color{black}\xi,\mathbf{x}}]\right|_{(\xi,\mathbf{x}(\xi))}=0\,,\label{stationary-formal}
\end{equation}
and the linear response for the {\color{black}operator $\hat{\mathcal{O}}$} is given by the following
equation (see Appx.~\ref{sec:derivation_chis} for derivation): 
\begin{equation}
\chi^{}_{\mathcal{O}\mathcal{O}} =\chi^{\text{emb}}_{\mathcal{O}\mathcal{O}}+\sum_{\mu}\chi^{\text{emb}}_{\mu\mathcal{O}}\mathcal{M}^{-1}_{\mu\nu}\chi^{\text{emb}}_{\nu\mathcal{O}}\,,\label{chis}
\end{equation}
where we introduced the susceptibilities: %\begin{equation}
\begin{align}
\chi^{\text{emb}}_{\mathcal{O}\mathcal{O}}=\partial_\xi\langle\Phi(\xi,\mathbf{x})|\hat{\mathcal{O}}|\Phi(\xi,\mathbf{x})\rangle|_{(\xi=0,\mathbf{x}(\xi=0))},\label{eq:chiemb0}\\
\chi^{\text{emb}}_{\mu\mathcal{O}}=\partial_{x_{\mu}}\langle\Phi(\xi,\mathbf{x})|\hat{\mathcal{O}}|\Phi(\xi,\mathbf{x})%\rangle|_{\xi=0}.
\rangle|_{(\xi=0,\mathbf{x}(\xi=0))}.\label{eq:chimu}
\end{align}
The so-called ``fluctuation matrix'' is:
\begin{align}
%\mathcal{M}_{\mu\nu}=\left.\frac{\partial^{2}\Omega[{\color{black}\xi},\mathbf{x}]}{\partial \mathbf{x}_{\mu}\partial \mathbf{x}_{\nu}}\right|_{\xi=0}\label{eq:Mtot}
\mathcal{M}_{\mu\nu}=\left.\partial_{x_\mu} \partial_{x_\nu}\Omega[{\color{black}\xi},\mathbf{x}]\right|_{(\xi=0,\mathbf{x}(\xi=0))}.\label{eq:Mtot}
\end{align}
%=\left.\frac{\partial\mathcal{L}_{\text{emb}}[\mathbf{x}]}{\partial \mathbf{x}_{\mu}\partial \mathbf{x}_{\nu}}\right|_{\xi=0}\nonumber\\
%&+\left.\frac{\partial\mathcal{L}_{\text{qp}}[\mathbf{x}]}{\partial \mathbf{x}_{\mu}\partial \mathbf{x}_{\nu}}\right|_{\xi=0}+\left.\frac{\partial\mathcal{L}_{\text{mix}}[\mathbf{x}]}{\partial \mathbf{x}_{\mu}\partial \mathbf{x}_{\nu}}\right|_{\xi=0}\label{eq:Mtot}
Here, the indices $\mu$ and $\nu$ run through all the variational variables in Eq.~\eqref{eq:X}, i.e., $r_{s}$, $l_{s}$, $d_{s}$, $D_{s}$, $l^c_{s}$.
To keep track of the structure of the fluctuation matrix 
(where different second order derivatives are computed through different equations, see Appx.~\ref{sec:M_fluc}),
from now on we will often use these variational variables as matrix subscripts. For example, $\mathcal{M}_{D_s,l^c_{s'}}$ corresponds to the second order derivatives with respect to $D_{s}$ and $l^c_{s'}$  (see Eq.~\eqref{eq:Memb3}).}

It is important to note that 
%the solution of Eq.~\eqref{eq:fordxdxi} is not unique, as 
$\mathcal{M}$ is not invertible. The reason is
that the functional $\Omega$ is invariant with respect to the gauge
transformation (Eq.~\eqref{eq:gauge_X}), so $\mathcal{M}$ is not unique because of the would-be Goldstone modes.
As explained in Appx.~\ref{subsec:Gauge-fix}, this redundancy can
be systematically resolved by operating a gauge fixing process that
removes from the onset of the would-be Goldstone modes \citep{Fabrizio2017}.
%%Here we name $\bar{M}$ the resulting (invertible) restriction of the fluctuation matrix. 
A simpler alternative is to solve the overdetermined linear system
(Eq.~\eqref{eq:Mtot}) by introducing the Moore-Penrose pseudo-inverse
of the fluctuation matrix, which we are going to indicate as $\bar{\mathcal{M}}^{-1}$.
In terms of the pseudo-inverse, the susceptibility can be formally
expressed as follows:
{\color{black}
\begin{equation}
\chi^{}_{\mathcal{O}\mathcal{O}}=\chi^{\text{emb}}_{\mathcal{O}\mathcal{O}}+\sum_{\mu\nu}\chi^{\text{emb}}_{\mu\mathcal{O}}\bar{\mathcal{M}}_{\mu\nu}^{-1}\chi^{\text{emb}}_{\nu\mathcal{O}}.\label{eq:chi}
\end{equation}
}%\vspace{4cm}
{\color{black} Note that Eq.~\eqref{eq:chi} applies for general multiorbital Hubbard models, and the procedure for evaluating each element, Eqs.~\eqref{eq:chiemb0}, \eqref{eq:chimu}, \eqref{eq:Mtot}, is described in Appx. \ref{sec:M_fluc}.}

\begin{figure}[t]
\begin{centering}
\includegraphics[scale=0.25]{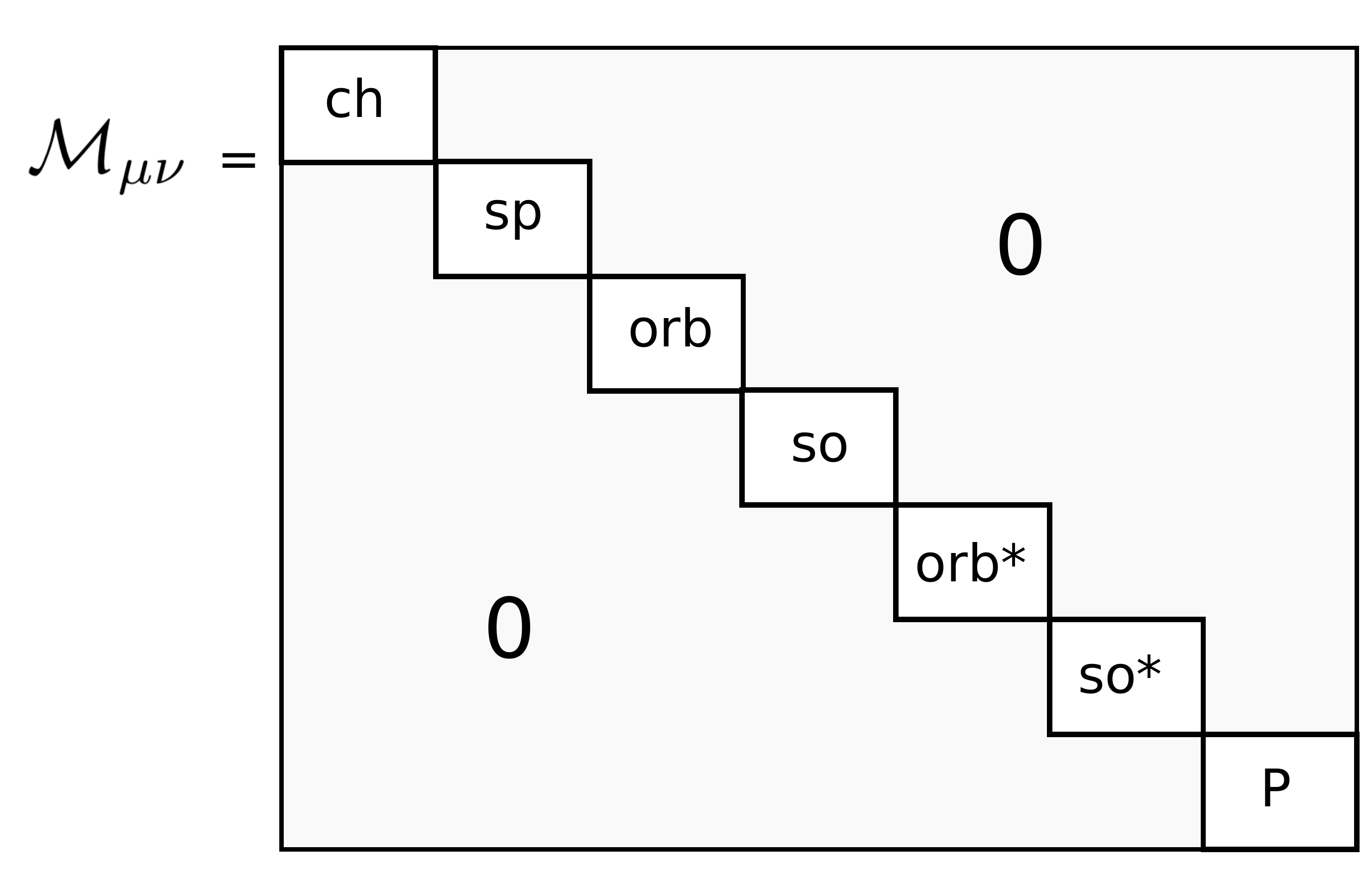}
\par\end{centering}
\caption{Schematic representation of the block-diagonalized fluctuation matrix
in the charge, spin, orbital, spin-orbital, and pairing sector for
the three-orbital degenerate Hubbard-Kanamori model {\color{black}(see Eqs.~\eqref{eq:3orb_triplet}-\eqref{eq:3orb_flucs})}. \label{fig:M_scheme}}
\end{figure}

{\color{black} We now discuss the application of our formalism to the degenerate three-orbital Hubbard-Kanamori model.} For the considered model, the fluctuation matrix $\mathcal{M}$
reduces to a block-diagonal matrix, constructed by seven $5\times5$
matrices shown schematically in Fig. \ref{fig:M_scheme} (one for
each fluctuation channel $s$), because of the orthonormality of the
fluctuation basis $\text{Tr}[\mathbf{h}_{s}\mathbf{h}_{s'}^{\dagger}]=\delta_{ss'}$.
Furthermore, for a given channel $s$, $\chi^{\text{emb}}_{\mu\mathcal{O}}$ (see Eq.~\eqref{eq:chimu})
is nonzero only for the components $\mu=D_{s}$ and $l_{s}$. Therefore,
Eq.~\eqref{eq:chi}, for a given channel $s$, can be further simplified to: 
\begin{align}
\chi^{}_{\mathcal{O}_s\mathcal{O}_s} & =\chi^{\text{emb}}_{\mathcal{O}_s\mathcal{O}_s}+\chi^{\text{emb}}_{\sm{D_{s}}\mathcal{O}_s}\bar{\mathcal{M}}_{D_{s}D_{s}}^{-1}\chi^{\text{emb}}_{\sm{D_{s}}\mathcal{O}_s}\nonumber \\
 & +2\chi^{\text{emb}}_{\sm{D_{s}}\mathcal{O}_s}\bar{\mathcal{M}}_{D_{s}l_{s}^{c}}^{-1}\chi^{\text{emb}}_{l_{s}^c\mathcal{O}_s}+\chi^{\text{emb}}_{\sm{l_{s}^{c}}\mathcal{O}_s}\bar{\mathcal{M}}_{l_{s}^{c}l_{s}^{c}}^{-1}\chi^{\text{emb}}_{\sm{l_{s}^{c}}\mathcal{O}_s},\label{eq:chi_deg}
\end{align}
%where we define
%\begin{equation}
%\begin{align}
%\chi_{\sm{D_s}} &=\frac{\partial}{\partial
%D_s}\langle\Phi(\mathbf{x},\xi)|\hat{\mathcal{O}}_{s}|\Phi(\mathbf{x},\xi)\rangle|_{\xi=0}, \label{eq:chiDs}\\
%\end{equation}
%\begin{equation}
%\chi_{\sm{l^c_s}} &=\frac{\partial}{\partial
%l^c_s}\langle\Phi(\mathbf{x},\xi)|\hat{\mathcal{O}}_{s}|\Phi(\mathbf{x},\xi)\rangle|_{\xi=0}.\label{eq:chiDs}  
%\end{equation}
%\end{align}
where $\bar{\mathcal{M}}_{D_{s}D_{s}}^{-1}$ denotes the $\mu=D_{s}$
and $\nu=D_{s}$ component of $\bar{\mathcal{M}}_{\mu\nu}^{-1}$,
and similarly applies to $\bar{\mathcal{M}}_{D_{s}l_{s}}^{-1}$ and
$\bar{\mathcal{M}}_{l_{s}^{c}l_{s}^{c}}^{-1}$. %In this case, 
We only need to evaluate the $5\times5$ fluctuation matrix and its
pseudo-inversion within each $s$ block to compute the corresponding
susceptibility. Note that the block-diagonal structure is not directly applicable to generic systems, because of effects such as orbital differentiation or spin-orbit coupling. In these cases, one has to compute the full fluctuation matrix for calculating response functions.

\section{Fermi-liquid approximation and diagrammatic approach\label{sec:QP_sus}}

The Landau Fermi-liquid theory allows one to describe the thermodynamic
properties of metals in terms of an effective non-interacting picture.
Importantly, this framework applies only to conserved quantities. In
particular, since the superconducting order parameter $\hat{\mathcal{O}}_{P}$
does not commute with Eq.~\eqref{hamiltonian}, the corresponding
susceptibility is not rigorously expressible in terms of quasiparticle
parameters. Nevertheless, as we are going to show below, within the
RISB framework, it is possible to derive an approximate (but accurate)
expression for the superconducting susceptibility in terms of the quasiparticle
Green's function and interaction vertices. Moreover, the susceptibility can be formulated in terms
of the Bethe-Salpeter equation, allowing further diagrammatic analysis
for the pairing mechanism.

From the point of view of the RISB methodology, the reason why the
superconducting susceptibility cannot be calculated in terms of quasiparticle
parameters is that:

\begin{equation}
K_{s}[\Phi_i,\boldsymbol{\Delta}_i]=\sum_{\alpha\beta}\big[\mathbf{h}_{s}\big]_{\alpha\beta}(\langle\Phi_i|\Xi_{i\alpha}^{\dagger}\Xi_{i\beta}|\Phi_i\rangle-\boldsymbol{\Delta}_{i\alpha\beta})\neq0\,
\end{equation}
for $s=\text{P}$, i.e., the physical density matrix is, in general,
not the same as the quasiparticle density matrix.

Here we propose to modify the {spatially inhomogeneous RISB Lagrange function (Eq.~\eqref{eq:Lemb})}
by imposing the constraint: 
\begin{equation}
K_{\text{s}}[\Phi_i,\boldsymbol{\Delta}_i]=0,\label{qpcondition}
\end{equation}
%which is a condition over the variational parameters. 
{\color{black} which is accomplished by introducing additional Lagrange multipliers $\zeta_{i,s}$ into Eq.~\eqref{eq:X} so the $\mathbf{x}$ vector becomes:
\begin{align}
\mathbf{x}_{i} &=( r_{i,\text{ch}},l_{i,\text{ch}},d_{i,\text{ch}},D_{i,\text{ch}},l_{i,\text{ch}}^{c},\zeta_{i,\text{ch}},...,r_{i,s},l_{i,s},\nonumber \\
 & d_{i,s},D_{i,s},l_{i,s}^{c},\zeta_{i,s},...,r_{i,\text{P}},l_{i,\text{P}},d_{i,\text{P}},D_{i,\text{P}},l_{i,\text{P}}^{c},\zeta_{i,\text{P}}).\label{eq:X_zeta}
\end{align}
{\color{black}We also introduce $\mathbf{x}_\mathbf{q}$, which is the momentum conjugate to $\mathbf{x}_i$.}

The Lagrange function now has the following form:}
\begin{align}
\mathcal{L}[{\color{black}\xi},\mathbf{x},\Phi,E^{c}]= & \mathcal{L}_{\text{qp}}[{\color{black}\xi_\mathbf{q}},\mathbf{x}_{\mathbf{q}}]+\mathcal{L}_{\text{mix}}[\mathbf{x}_i]+\mathcal{L}_{\text{emb}}[\mathbf{x}_i,\Phi_i,E^{c}_i],\label{eq:Lqptot}
\end{align}
where 
\begin{align}
\mathcal{L}_{\text{qp}}[{\color{black}\xi_\mathbf{q}},\mathbf{x}_{\mathbf{q}}] & =-\frac{T}{N}\frac{1}{2}\sum_{\omega_{n}}\sum_{\mathbf{k}_{1}\mathbf{k}_{2}}\text{Tr log}\Big[-\mathbf{G}_{\omega_{n},\mathbf{k}_{1},\mathbf{k}_{2}}^{-1}[\mathbf{x},\xi]\Big],\label{eq:Lqp_const}
\end{align}
\begin{align}
\mathcal{L}_{\text{emb}}&[\mathbf{x}_i,\Phi_i,E^{c}_i]= \sum_i\big\langle\Phi_i({\mathbf{x}_i})\big|\hat{H}_{\text{emb}}[\mathbf{x}_i]+\frac{1}{2}\sum_{\alpha\beta s}\zeta_{i,s}\mathbf{h}_{s,\alpha\beta}\nonumber\\
&\hat{\Xi}_{i\alpha}^{\dagger}\hat{\Xi}_{i\beta}\big|\Phi_i(\mathbf{x}_i)\big\rangle+E^{c}_i\big(1-\langle\Phi_i(\mathbf{x}_i)\big|\Phi_i(\mathbf{x}_i)\rangle\big),\label{eq:Lemb_const}
\end{align}
\begin{align}
\mathcal{L}_{\text{mix}} &[\mathbf{x}_i]= -\sum_i\Big[\frac{1}{2}\sum_{ab}(\boldsymbol{\Lambda}_{iab}+\boldsymbol{\Lambda}_{iab}^{c}+\sum_{s}\zeta_{is}\mathbf{h}_{s,ab})\boldsymbol{\Delta}_{iab}\nonumber\\
&+\sum_{a\alpha c}\big(\mathcal{\mathbf{D}}_{ia\alpha}\mathbf{R}_{ic\alpha}[\boldsymbol{\Delta}(1-\boldsymbol{\Delta})]_{ica}^{\frac{1}{2}}+\text{c.c.}\big)\Big],\label{eq:Lmix_const}
\end{align}
where we have introduced the physical Green's function:
\begin{equation}
\mathbf{G}_{\omega_{n},\mathbf{k}_{1},\mathbf{k}_{2}}[{\color{black}\xi},\mathbf{x}]=\mathbf{R}^{\dagger}\mathbf{G}^{\text{qp}}_{\omega_{n},\mathbf{k}_{1},\mathbf{k}_{2}}[{\color{black}\xi},\mathbf{x}]\mathbf{R}
\end{equation}
and the quasiparticle Green's function:{\color{black} 
\begin{equation}
\big[\mathbf{G}^{\text{qp}}_{\omega_{n},\mathbf{k}_{1},\mathbf{k}_{2}}[{\color{black}\xi},\mathbf{x}]\big]_{ab}^{-1}=i\omega_{n}-\big[H_{\mathbf{k}_{1}\mathbf{k}_{2}}^{\text{qp}}[\mathbf{x}]\big]_{ab}+\xi_{\mathbf{k}_{1}-\mathbf{k}_{2}}\big[\mathcal{O}\big]_{ab}.
\end{equation}
Similar to the previous section, we also introduced a field $\xi_{\mathbf{k}_1-\mathbf{k}_2}$
%and $\mathbf{x}_\mathbf{q}$ and $\mathbf{x}_i$ 
coupled to a generic quasiparticle {\color{black}operator} $\hat{\mathcal{O}}=\sum_{ab}\Psi^\dagger_{\mathbf{k}_1a} \big[\mathcal{O}\big]_{ab}\Psi_{\mathbf{k}_2b}$ into $\mathcal{L}_\text{qp}$.
%so that the quasiparticle
This modification will allow us to derive momentum dependent susceptibilities, for investigating the finite momentum (commensurate or incommensurate) instabilities. From now on, we refer to Eq.~\eqref{qpcondition}
as the ``quasiparticle constraint.''
}

Since utilizing the Lagrange equation Eqs.~\eqref{eq:Lqp_const}-\eqref{eq:Lmix_const}
amounts to solve the RISB equations Eqs.~\eqref{eq:sp1-1}-\eqref{eq:sp6-1}
within a reduced variational space, the corresponding solution is
an approximation to the original one.
% ---whose accuracy can be systematically
%verified a-posteriori, by comparison with Eq.~~\eqref{eq:chi_deg}.
%the respective solutions of Eqs.~\eqref{eq:chi_deg} and \eqref{eq:chiqp_deg}.
%%%To gain further insight into our fluctuation approach, it is important to develop a quasiparticle Fermi-liquid framework allowing the diagrammatic analysis for the quasiparticle susceptibilities, Fermi-liquid Landu parameters, and the pairing interactions. To achieve this goal, we apply Fermi-liquid approximation to the Lagrangian enforcing the equivalence between the quasiparticle order parameters and the physical order parameters, i.e., $\mathbf{h}_{s,\alpha\beta}\langle\Phi|\hat{\Xi}_{\alpha}^{\dagger}\hat{\Xi}_{\beta}\big|\Phi\rangle=\boldsymbol{\Delta}_{ab}\mathbf{h}_{s,ab}$ for a given channel $s$. 
%Consequently, $\mathcal{M}_{\mu\nu}(\mathbf{q})$ becomes a block-diagonal matrix constructed by seven $6\times6$ matrices from each channel $s$. 
In principle, enforcing the constraint (Eq.~\eqref{qpcondition}) does not affect the results for the conserving channels, where the fluctuating operator commutes with the Hamiltonian, e.g., the charge and spin channels. However, it reduces slightly the variational freedom when the constraint
is imposed on the non-conserving channel, e.g., the pairing channel. Nevertheless, as we are going to show, it is always possible to verify a-posteriori the accuracy of the approximation, by comparison to the formalism without the constraint (see also Appx.~\ref{sec:FL_approx_check}).
%, with and without the constraint, to check the differences. 
%For the pairing channel, we have verified in Appx. H that the results are essentially identical with and without enforcing the constraint.. Nevertheless, we
%can always compare a-posteriori the two approaches, with and without
%the constraint, to check the differences. 

It is also interesting to point out that Eq.~\eqref{qpcondition} corresponds to the density matrix mapping constraint in DMET~\citep{DMET_2012}. Therefore, the formalism presented in this section is also applicable to the {\color{black}NIB-DMET}, by removing the $r_s$ sector of the fluctuation basis (Eq.~\eqref{eq:X_zeta}) and setting $\mathbf{R}=I$~\citep{Ayral_2017}. This application is discussed in Appx.~\ref{sec:dmet}.

\subsection{Susceptibility: diagrammatic expression}

Here we show how the susceptibility evaluated with the quasiparticle
constraint can be expressed in terms of the Feynman diagram in perturbation
theory.

Following the procedure in Sec. \ref{subsec:emb_sus}, we introduce
the following functional: 
\begin{align}
\Omega[{\color{black}\xi},\mathbf{x}]&=\mathcal{L}_{\text{qp}}[{\color{black}\xi_\mathbf{q}},\mathbf{x}_\mathbf{q}]+\mathcal{L}_{\text{mix}}[\mathbf{x}_i]\nonumber\\
&+\mathcal{L}_{\text{emb}}[\Phi(\mathbf{x}_i),E^{c}(\mathbf{x}_i)]\,,\label{eq:Xqp}
\end{align}
where now $\mathcal{L}_{\text{qp}}$ depends on the field $\xi_{\mathbf{q}}$.
The linear response for {\color{black}a generic operator} is given by the following
equation:{\color{black}

\begin{align}
\chi^{}_{\mathcal{O}\mathcal{O}}(\mathbf{q}) & =\frac{T}{2N}\sum_{\mathbf{k}\omega_n}\left.\frac{d}{d\xi_{\mathbf{q}}}\text{Tr}\big[\mathbf{G}_{\omega_{n},\mathbf{k}+\mathbf{q},\mathbf{k}}[{\color{black}\xi},\mathbf{x}]\bar{\mathcal{O}}\big]\right|_{(\xi=0,\mathbf{x}(\xi=0))}\nonumber \\
 & =\chi^{(0)}_{\mathcal{O}\mathcal{O}}(\mathbf{q})+\sum_{\mu\nu}\chi_{\mu\mathcal{O}}(\mathbf{q})\mathcal{M}_{\mu\nu}^{-1}(\mathbf{q})\chi_{\nu\mathcal{O}}(\mathbf{q}),\label{chiqpstep1}
\end{align}
where the bare susceptibilities are 
\begin{align}
\chi^{(0)}_{\mathcal{O}\mathcal{O}}(\mathbf{q}) & =-\frac{T}{2N}\sum_{\mathbf{k}\omega_n}\text{Tr}\Big[\mathbf{G}_{\omega_n,\mathbf{k}+\mathbf{q}}\bar{\mathcal{O}}\mathbf{G}_{\omega_n,\mathbf{k}}\bar{\mathcal{O}}\Big],\label{eq:chi0qp}
\end{align}
\begin{align}
\chi_{\mu\mathcal{O}}(\mathbf{q}) & =\frac{T}{2N}\sum_{\mathbf{k}\omega_n}\left.\partial_{x_{\mu,\mathbf{q}}}\text{Tr}\big[\mathbf{G}_{\omega_n,\mathbf{k}+\mathbf{q},\mathbf{k}}[{\color{black}\xi},\mathbf{x}]\bar{\mathcal{O}}\big]\right|_{(\xi=0,\mathbf{x}(\xi=0))}.\label{eq:chiXqp}
\end{align}
Note again that $\mu$ runs through all the elements in Eq.~\eqref{eq:X_zeta}, and we use the variational parameters as subscripts. We also introduced the saddle-point Green's function $\mathbf{G}_{\omega_n,\mathbf{k}}=\mathbf{R}^\dagger [i\omega_n-H^\text{qp}_{\mathbf{k}}]^{-1} \mathbf{R}$ and $\bar{\mathcal{O}}=[\mathbf{R}]^{-1}\mathcal{O}[\mathbf{R}^{\dagger}]^{-1}$.
The fluctuation matrix $\mathcal{M}$ now depends on momentum $\mathbf{q}$
and has an additional component $\zeta_{s}$
(see Eq.~\eqref{eq:X_zeta}).} The specific form of $\mathcal{M}$ is given in Appx.~\ref{sec:M_fluc}. Furthermore, $\mathcal{M}$ is now an
invertible matrix because the quasiparticle constraint breaks the
gauge symmetry. {\color{black} Note that Eq.~\eqref{chiqpstep1} applies for generic multiorbital Hubbard models.}

{\color{black} We now discuss the application of our approach to the degenerate three-orbital Hubbard-Kanamori model.} As described in the previous section, for the degenerate model considered here,
$\mathcal{M}$ is a block-diagonal matrix shown schematically in Fig.~\ref{fig:M_scheme}. Also, from Eqs.~\eqref{eq:3orb_triplet}-\eqref{eq:3orb_flucs} and Eqs.~\eqref{eq:h_ch}-\eqref{eq:h_P}, we have $\bar{\mathcal{O}}_s=\bar{\mathbf{h}}_s=[\mathbf{R}]^{-1}\mathbf{h}_s[\mathbf{R}^{\dagger}]^{-1}$ for each fluctuation channel $s$.
Therefore, the susceptibility can be simplified to: 
\begin{align}
\chi^{}_{\mathcal{O}_s\mathcal{O}_s}(\mathbf{q}) & =\chi^{(0)}_{\mathcal{O}_s\mathcal{O}_s}(\mathbf{q})+\chi_{r_{s}\mathcal{O}_s}(\mathbf{q})\mathcal{M}_{r_{s}r_{s}}^{-1}(\mathbf{q})\chi_{r_{s}\mathcal{O}_s}(\mathbf{q})\nonumber \\
 & +2\chi_{r_{s}\mathcal{O}_s}(\mathbf{q})\mathcal{M}_{r_{s}l_{s}}^{-1}(\mathbf{q})\chi_{l_{s}\mathcal{O}_s}(\mathbf{q})\nonumber\\
 & +\chi_{l_{s}\mathcal{O}_s}(\mathbf{q})\mathcal{M}_{l_{s}l_{s}}^{-1}(\mathbf{q})\chi_{l_{s}\mathcal{O}_s}(\mathbf{q}),\label{eq:chiqp_deg}
\end{align}
where 
\begin{align}
\chi_{r_{s}\mathcal{O}_s}(\mathbf{q}) & =-\frac{T}{2N}\sum_{\mathbf{k}\omega_n}\text{Tr}\big[\mathbf{G}_{\omega_n,\mathbf{k}+\mathbf{q}}\big[\mathbf{R}\big]^{-1}\big[(\tilde{\mathbf{h}}_{s}\boldsymbol{\epsilon}_{\mathbf{k}+\mathbf{q}}\mathbf{R}^{\dagger}
 \nonumber\\
 &+\mathbf{R}\boldsymbol{\epsilon}_{\mathbf{k}}\tilde{\mathbf{h}}_{s}^{\dagger})\big[\mathbf{R}^{\dagger}\big]^{-1}\mathbf{G}_{\omega_n,\mathbf{k}}\bar{\mathbf{h}}_{s}\big]
,\label{eq:chiqp_r}\\
\chi_{l_{s}\mathcal{O}_s}(\mathbf{q}) & =-\frac{T}{2N}\sum_{\mathbf{k}\omega_n}\text{Tr}\Big[\mathbf{G}_{\omega_n,\mathbf{k}+\mathbf{q}}\bar{\mathbf{h}}_{s}\mathbf{G}_{\omega_n,\mathbf{k}}\bar{\mathbf{h}}_{s}\Big]
%=\chi^{(0)}_{\mathcal{O}_s}(\mathbf{q})\label{eq:chiqp_l}.
\end{align}
The $\mathcal{M}_{r_{s}r_{s}}^{-1}(\mathbf{q})$ denotes the $\mu=r_{s}$
and $\nu=r_{s}$ component of $\mathcal{M}_{\mu\nu}^{-1}(\mathbf{q})$
and similarly applies to $\mathcal{M}_{r_{s}l_{s}}^{-1}(\mathbf{q})$
and $\mathcal{M}_{l_{s}l_{s}}^{-1}(\mathbf{q})$. We only need to
evaluate the $6\times6$ fluctuation matrix and its inversion within
each $s$ block to compute the corresponding susceptibility.

To make a connection to perturbation theory, we compare Eq.~\eqref{eq:chiqp_deg}
with the Bethe-Salpeter representation of the susceptibility: 
\begin{eqnarray}
& \chi^{}_{\mathcal{O}_s\mathcal{O}_s}(\mathbf{q}) =\chi^{(0)}_{\mathcal{O}_s\mathcal{O}_s}(\mathbf{q})-\Big(\frac{-T}{2N}\Big)^{2}\underset{\alpha\beta\gamma\delta}{\sum}\underset{\mathbf{k}\mathbf{k}'}{\sum}\underset{\omega_{n}\omega_{n'}}{\sum}\big[\mathbf{G}_{\omega_n,\mathbf{k}}\bar{\mathbf{h}}_{s}\nonumber \\
& \mathbf{G}_{\omega_n,\mathbf{k}+\mathbf{q}}\big]_{\beta\alpha} \tilde{\Gamma}_{\alpha\beta\gamma\delta}^{s}(\mathbf{k},\mathbf{k}',\mathbf{q})\big[\mathbf{G}_{\omega_{n'},\mathbf{k}'}\bar{\mathbf{h}}_{s}\mathbf{G}_{\omega_{n'},\mathbf{k}'+\mathbf{q}}\big]_{\delta\gamma},\label{eq:chiqp_BSE}
\end{eqnarray}
where $\tilde{\Gamma}_{\alpha\beta\gamma\delta}^{s}(\mathbf{k},\mathbf{k}',\mathbf{q})$
is the (reducible) interaction vertex. To extract the $\tilde{\Gamma}_{\alpha\beta\gamma\delta}^{s}(\mathbf{k},\mathbf{k}',\mathbf{q})$
from Eq.~\eqref{eq:chiqp_deg}, we introduced the following three-leg
vertices: 
\begin{align}
\tilde{\Lambda}_{\alpha\beta r_{s}}(\mathbf{k},\mathbf{q}) & \equiv\frac{1}{2}\big[\mathbf{R}\big]_{\alpha a}^{-1}\big[\mathbf{R}\tilde{\epsilon}_{\mathbf{k}+\mathbf{q}}\tilde{\mathbf{h}}_{s}^{\dagger}+\tilde{\mathbf{h}}_{s}\tilde{\epsilon}_{\mathbf{k}}\mathbf{R}^{\dagger}\big]_{ab}\big[\mathbf{R}^{\dagger}\big]_{b\beta}^{-1},\label{eq:lamr}\\
\tilde{\Lambda}_{\alpha\beta l_{s}} & \equiv\frac{1}{2}\big[\mathbf{R}\big]_{\alpha a}^{-1}\mathbf{h}_{s,ab}\big[\mathbf{R}^{\dagger}\big]_{b\beta}^{-1},\label{eq:laml}
\end{align}
such that the susceptibilities can be written as: 
\begin{align}
\chi_{r_{s}\mathcal{O}_s}(\mathbf{q})&=-\frac{T}{N}\sum_{\mathbf{k}\omega_n}\text{Tr}\big[\mathbf{G}_{\omega_n,\mathbf{k}+\mathbf{q}}\tilde{\Lambda}_{r_{s}}(\mathbf{k},\mathbf{q})\mathbf{G}_{\omega_n,\mathbf{k}}\bar{\mathbf{h}}_{s}\big],\label{eq:chirs_BSE}\\
\chi_{l_{s}\mathcal{O}_s}(\mathbf{q})&=-\frac{T}{N}\sum_{\mathbf{k}\omega_n}\text{Tr}\big[\mathbf{G}_{\omega_n,\mathbf{k}+\mathbf{q}}\tilde{\Lambda}_{l_{s}}\mathbf{G}_{\omega_n,\mathbf{k}}\bar{\mathbf{h}}_{s}\big].\label{eq:chils_BSE}
\end{align}
%where we suppress the $\alpha$ and $\beta$ indices in the trace.
Substituting Eqs.~\eqref{eq:chirs_BSE} and \eqref{eq:chils_BSE}
into Eq.~\eqref{eq:chiqp_deg}, we obtain the interaction vertex
(see Eq.~\eqref{eq:chiqp_BSE}): 
\begin{align}
\tilde{\Gamma}_{\alpha\beta\gamma\delta}^{s}(\mathbf{k},\mathbf{k}',\mathbf{q}) & =-4\begin{pmatrix}\tilde{\Lambda}_{\alpha\beta r_{s}}(\mathbf{k},\mathbf{q}) & \tilde{\Lambda}_{\alpha\beta l_{s}}\end{pmatrix}\nonumber \\
 & \cdot\begin{pmatrix}\mathcal{M}_{r_{s}r_{s}}^{-1}(\mathbf{q}) & \mathcal{M}_{r_{s}l_{s}}^{-1}(\mathbf{q})\\
\mathcal{M}_{r_{s}l_{s}}^{-1}(\mathbf{q}) & \mathcal{M}_{l_{s}l_{s}}^{-1}(\mathbf{q})
\end{pmatrix}\begin{pmatrix}\tilde{\Lambda}_{\gamma\delta r_{s}}(\mathbf{k}',\mathbf{q})\\
\tilde{\Lambda}_{\gamma\delta l_{s}}
\end{pmatrix},\label{eq:qp_interaction-1}
\end{align}
%The interaction vertex $\tilde{\Gamma}^s$ 
describing the effective interaction between quasiparticles mediated
by the bosonic propagator $\mathcal{M}_{\mu\nu}^{-1}$ in the corresponding
channel.
\begin{figure}[t]
\begin{centering}
\includegraphics[scale=0.4]{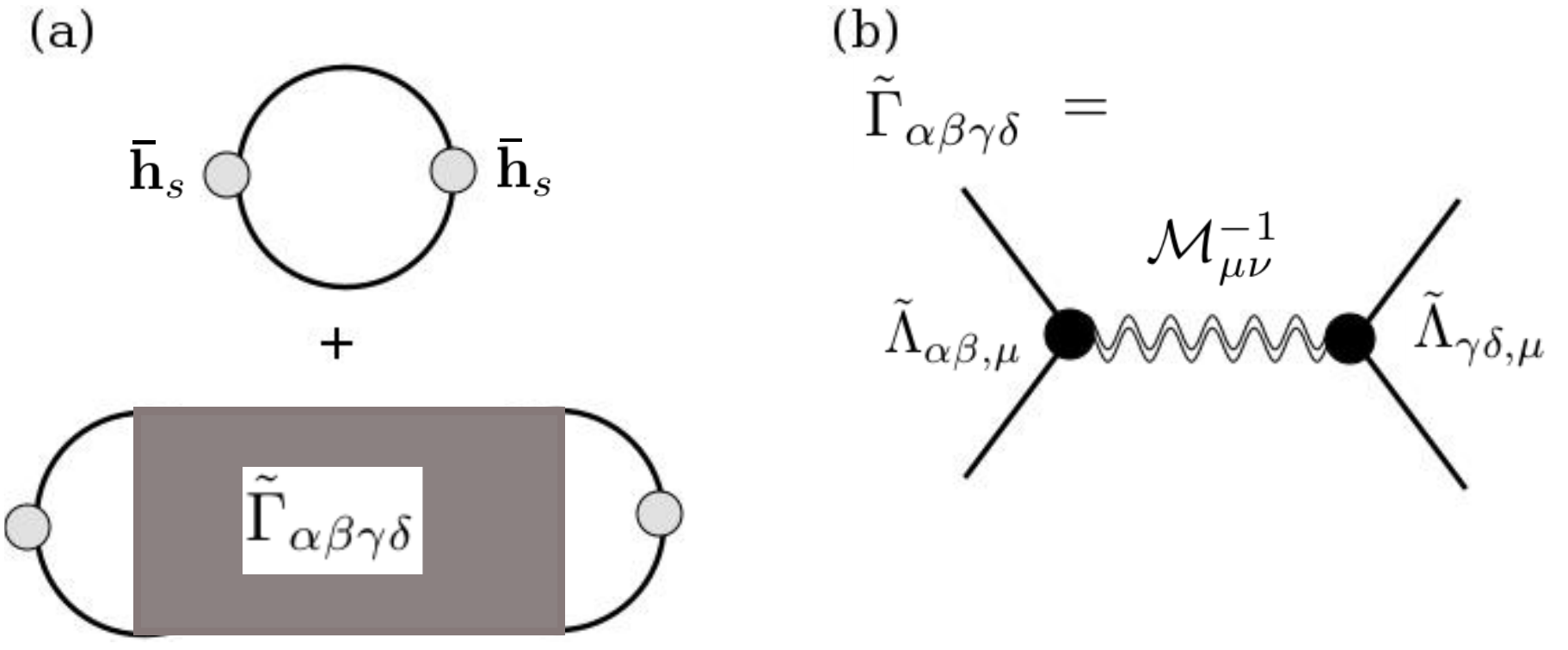}
\par\end{centering}
\caption{(a) Diagrammatic representation of the susceptibility ({\color{black}Eq.~\eqref{eq:chiqp_BSE}}). The thick solid
line indicates the Nambu fermionic propagator. The grey circle corresponds
to the fluctuation basis $\bar{\mathbf{h}}_{s}$, and the grey square
corresponds to the quasiparticle interaction vertex $\tilde{\Gamma}^s_{\alpha\beta\gamma\delta}$.
(b) Diagrammatic representation of the quasiparticle interaction vertex $\tilde{\Gamma}^s_{\alpha\beta\gamma\delta}$ ({\color{black}Eq.~\eqref{eq:qp_interaction-1}}).
The double wavy line corresponds to the dressed bosonic propagator
containing the infinite summation of the particle-particle or the
particle-hole fermionic bubbles. The black circles denotes the three-leg
vertices $\tilde{\Lambda}_{\alpha\beta\mu}$ (see main text for detail).
\label{fig:diagram-1}}
\end{figure}

The diagrammatic representation of Eq.~\eqref{eq:chiqp_BSE} is shown
in Fig. \ref{fig:diagram-1} (a), where the solid line corresponds
to the Nambu propagator, the grey circle corresponds to $\bar{\mathbf{h}}_{s}$,
and the grey rectangle corresponds to the interaction vertex $\tilde{\Gamma}_{\alpha\beta\gamma\delta}^{s}(\mathbf{k},\mathbf{k}',\mathbf{q})$.
The diagrammatic representation for the interaction vertex $\tilde{\Gamma}^s_{\alpha\beta\gamma\delta}(\mathbf{k},\mathbf{k}',\mathbf{q})$
is shown in Fig. \ref{fig:diagram-1}(b), where the solid circles
correspond to the three-leg vertices $\tilde{\Lambda}_{\alpha\beta\mu}$.
The double wavy line corresponds to $\mathcal{M}_{\mu\nu}^{-1}(\mathbf{q})$, which
can be viewed as the dressed bosonic propagator (see Appx.~\ref{sec:propagator}) summing the particle-hole
bubbles, for $s\in\{\text{ch},\text{sp},\text{orb},\text{so},\text{orb*},\text{so*}\}$,
or the particle-particle bubbles, for $s=P$, to the infinite order.

%The effective interaction for each channel $s$ can be written as
%\begin{align}
%H_{\text{eff}}^{s} & =\sum_{\mathbf{k}\mathbf{k'}}\tilde{\Gamma}_{\alpha\beta\gamma\delta}^{s}(\mathbf{k},\mathbf{k}')\hat{\Xi}_{\mathbf{k}\alpha}^{\dagger}\hat{\Xi}_{\mathbf{k}\beta}\hat{\Xi}_{\mathbf{k'}\gamma}^{\dagger}\hat{\Xi}_{\mathbf{k'}\delta},\label{eq:Veff_1}
%\end{align}
%describing the quasiparticle interaction mediated by the bosonic propagator
%$\mathcal{M}_{\mu\nu}^{-1}$ in the corresponding channel.

\subsection{Landau Fermi-liquid parameters}

We can now calculate the Landau Fermi-liquid parameters for the considered three-orbital degenerate model from Eq.~\eqref{eq:qp_interaction-1}.
For each channel $s\in\{\text{ch},\text{sp},\text{orb},\text{so},\text{orb*},\text{so*}\}$,
we have:
\begin{align}
\Gamma^{s}(\mathbf{k},\mathbf{k}',\mathbf{q}) & =-\frac{1}{2Z^{2}}\Big[Z(\epsilon_{\mathbf{k}}+\epsilon_{\mathbf{k+\mathbf{q}}})(\epsilon_{\mathbf{k}'}+\epsilon_{\mathbf{k}'+\mathbf{q}})\mathcal{M}_{r_{s}r_{s}}^{-1}(\mathbf{q})\nonumber \\
 & +R_{0}(\epsilon_{\mathbf{k}}+\epsilon_{\mathbf{k}+\mathbf{q}})\mathcal{M}_{r_{s}l_{s}}^{-1}(\mathbf{q})+R_{0}(\epsilon_{\mathbf{k}'}+\epsilon_{\mathbf{k}'+\mathbf{q}})\cdot\nonumber \\
 & \mathcal{M}_{r_{s}l_{s}}^{-1}(\mathbf{q})+\mathcal{M}_{l_{s}l_{s}}^{-1}(\mathbf{q})\Big],\label{eq:Gamma_s}
\end{align}
where we applied $\mathbf{R}=R_{0}I$ and $Z=R_{0}^{2}$ for the degenerate
model considered here$.$ The scattering amplitude for each particle-hole
channel $s$ can be evaluated from 
\begin{equation}
A_{s}(\mathbf{q})=N_{F}Z^{2}\big\langle\big\langle\Gamma^{s}(\mathbf{k},\mathbf{k}',\mathbf{q})\big\rangle_{\mathbf{k}_{F}}\big\rangle_{\mathbf{k}'_{F}},\label{eq:scattering_amplitude}
\end{equation}
where we introduce the Fermi surface average 
\begin{equation}
\langle\langle\Gamma(\mathbf{k},\mathbf{k}')\rangle_{\mathbf{k}_{F}}\rangle_{\mathbf{k}'_{F}}=\frac{\underset{\mathbf{kk}'}{\sum}\Gamma(\mathbf{k},\mathbf{k}')\delta_{\mathbf{k},\mathbf{k}_{F}}\delta_{\mathbf{k}',\mathbf{k}_{F}}}{\underset{\mathbf{k}\mathbf{k}'}{\sum}\delta_{\mathbf{k},\mathbf{k}_{F}}\delta_{\mathbf{k}',\mathbf{k}_{F}}}.\label{eq:FSavg}
\end{equation}
$N_{F}\equiv\chi^{(0)}_{\mathcal{O}_s\mathcal{O}_s}(0)$ is the density of state at the Fermi-level,
which coincides with the bare susceptibility $\chi^{(0)}_{\mathcal{O}_s\mathcal{O}_s}$. The Fermi-liquid
parameters $F_{s}$ can be extracted from the scattering amplitude
(see Appx.~\ref{sec:RPA}) 
\begin{equation}
A_{s}(\mathbf{q})=\frac{F_{s}(\mathbf{q})}{1+F_{s}(\mathbf{q})}.\label{eq:Fs}
\end{equation}
From the definition of the quasiparticle susceptibility Eq.~\eqref{eq:chiqp_BSE}
and Eq. \eqref{eq:Fs}, we obtain the random phase approximation (RPA) like expression
for the susceptibilities 
\begin{equation}
\chi_{\mathcal{O}_s\mathcal{O}_s}(\mathbf{q})=\frac{\chi^{(0)}_{\mathcal{O}_s\mathcal{O}_s}(\mathbf{q})}{1+F_{s}(\mathbf{q})},\label{eq:chis_RPA}
\end{equation}
for $s\in\{\text{ch},\ \text{sp},\ \text{orb},\ \text{so},\ \text{orb}^{*},\ \text{so}^{*}\}$.
Note that we have applied the Fermi-surface average over $\mathbf{k}$
and $\mathbf{k}'$. The divergence of the quasiparticle susceptibilities
and the scattering amplitudes can be determined from the condition
$F_{s}(\mathbf{q})=-1$. Although Eq.~\eqref{eq:chis_RPA} has an RPA-like form,
the Fermi-liquid parameters are renormalized by the correlation effect
for different $\mathbf{q}$, which provides a more accurate description
for strongly correlated systems.
\begin{figure}[t]
\begin{centering}
\includegraphics[scale=0.4]{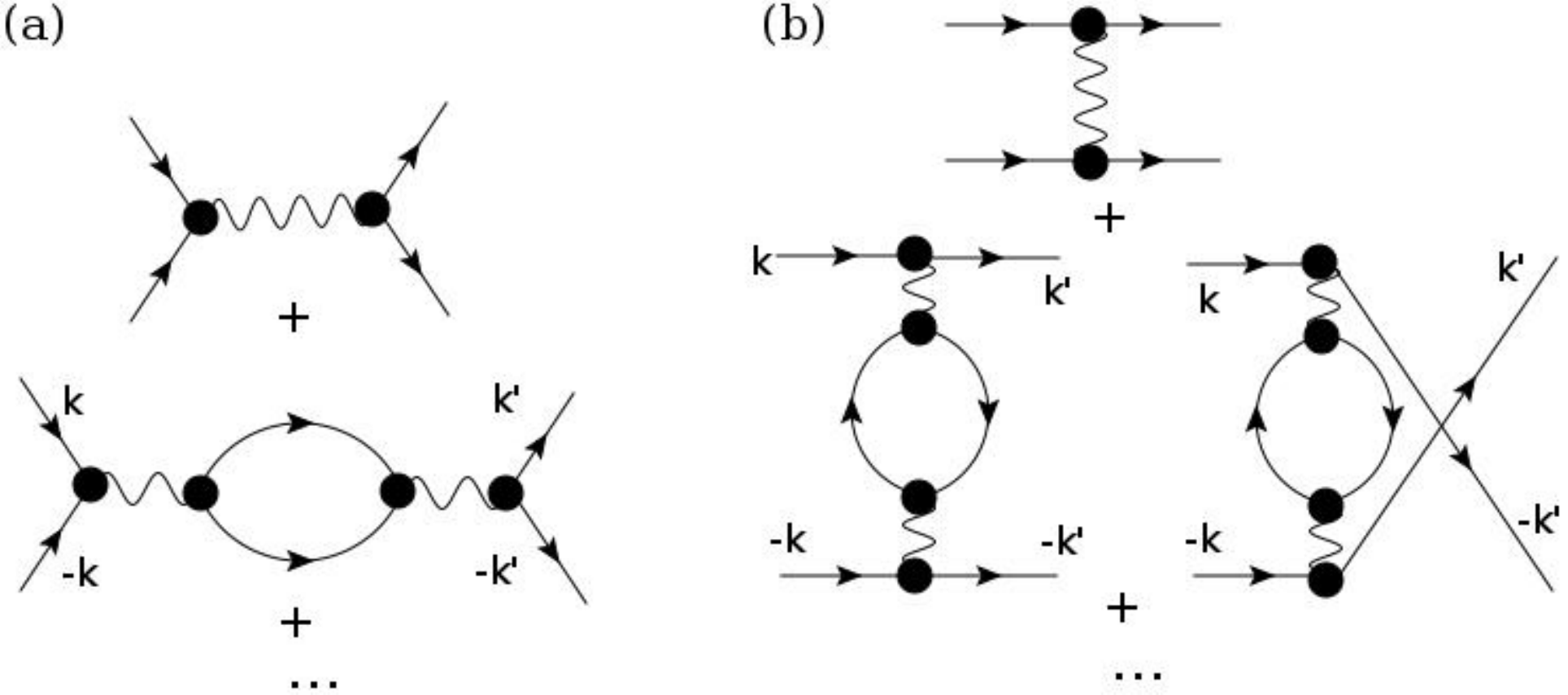}
\par\end{centering}
\caption{(a) The pairing vertex from the local particle-particle fluctuation (Eq.~\eqref{eq:Gamma_pp}). (b)
The pairing vertex from the particle-hole fluctuations (Eq.~\eqref{eq:Gamma_ph}). The bubbles
are summed to the infinite order. The arrow solid line corresponds
to the normal fermionic propagator. The wavy line corresponds to the
bare bosonic propagator. \label{fig:diagram}}
\end{figure}

\subsection{Pairing interaction from the particle--particle channel}

The reducible pairing vertex in the orbital-antisymmetric spin-triplet
pairing channel $s=\text{P}$ can be computed by projecting the particle-particle
scattering vertex $\tilde{\Gamma}^{\text{P}}$ (Eq.~\eqref{eq:qp_interaction-1})
onto the orbital-antisymmetric spin-triplet pairing basis $\mathbf{h}_{P}$
(see Appx.~\eqref{subsec:Single-particle-basis}): 
\begin{align}
\Gamma_{\text{pp}}^{\text{sc}}(\mathbf{k},\mathbf{k}',\mathbf{q}=0) & =\big[\mathbf{h}_{\text{P}}^{\dagger}\big]_{\alpha\beta}\tilde{\Gamma}_{\alpha\beta\gamma\delta}^{P}(\mathbf{k},\mathbf{k}',\mathbf{q}=0)\big[\mathbf{h}_{\text{P}}^{\dagger}\big]_{\delta\gamma}\nonumber \\
= & -\frac{1}{2Z^{4}}\Big[Z(\epsilon_{\mathbf{k}}+\epsilon_{\mathbf{k}})(\epsilon_{\mathbf{k}'}+\epsilon_{\mathbf{k}'})\mathcal{M}_{r_{P}r_{P}}^{-1}(0)\nonumber \\
 & -R_{0}(\epsilon_{\mathbf{k}}+\epsilon_{\mathbf{k}})\mathcal{M}_{r_{P}l_{P}}^{-1}(0)-R_{0}(\epsilon_{\mathbf{k}'}+\epsilon_{\mathbf{k}'})\cdot\nonumber \\
 & \mathcal{M}_{r_{P}l_{P}}^{-1}(0)+\mathcal{M}_{l_{P}l_{P}}^{-1}(0)\Big],\label{eq:Gamma_pp}
\end{align}
where we applied $\mathbf{R}=R_{0}I$ and $Z=R_{0}^{2}$ for the degenerate
model considered here and restrict the pairing at $\mathbf{q}=0$.
The diagrammatic representation for Eq.~\eqref{eq:Gamma_pp} is shown
in Fig. \ref{fig:diagram}(a). In this scattering process, only the
particle-particle fermionic bubbles and the local multiplets fluctuation
between different particle number sectors in $\mathcal{M}_{\mu\nu}^{-1}$
are involved (the fluctuation basis $\mathbf{h}_{\text{P}}$ and $\tilde{\mathbf{h}}_{\text{P}}$
in Eqs.~\eqref{eq:Memb1}-\eqref{eq:Memb2-1-1-1} selects the fluctuation
that does not conserve the particle number.)

We can now derive the RPA-like form for the quasiparticle
susceptibility. From $\Gamma_{\text{pp}}^{\text{sc}}$, we compute
the reducible pairing interaction by averaging the $\mathbf{k}$ and
$\mathbf{k}'$ over the Fermi surface 
\begin{equation}
\Gamma_{\text{pp}}^{\text{sc}}=Z^{2}\big\langle\big\langle\Gamma_{\text{pp}}^{\text{sc}}(\mathbf{k},\mathbf{k}')\big\rangle_{\mathbf{k}_{F}}\big\rangle_{\mathbf{k}'_{F}}.\label{eq:scattering_amplitude_P}
\end{equation}
The irreducible pairing interaction $\Gamma_{\text{pp}}^{\text{irr}}$
can be extracted from (see Appx.~\ref{sec:RPA}) 
\begin{equation}
\Gamma_{\text{pp}}^{\text{sc}}=\frac{\Gamma_{\text{pp}}^{\text{irr}}}{1+\Gamma_{\text{pp}}^{\text{irr}}\chi^{(0)}_{\mathcal{O}_P\mathcal{O}_P}}.\label{eq:Vsc}
\end{equation}
From the definition of the quasiparticle susceptibility Eq.~\eqref{eq:chiqp_BSE}
and Eq.~\eqref{eq:Vsc} , we obtain the RPA-like expression for the
pairing susceptibility 
\begin{equation}
\chi_{P}=\frac{\chi^{(0)}_{\mathcal{O}_P\mathcal{O}_P}}{1+\Gamma_{\text{pp}}^{\text{irr}}\chi^{(0)}_{\mathcal{O}_P\mathcal{O}_P}}.\label{eq:chiP_RPA}
\end{equation}
The divergence of the pairing susceptibilities and vertex can be determined
from the condition $\Gamma_{\text{pp}}^{\text{irr}}\chi^{(0)}_{\mathcal{O}_P\mathcal{O}_P}=-1$.

\subsection{Pairing interaction from the particle-hole channel \label{subsec:Gamma_ph_irr}}

Besides the $s$-wave pairing induced from the particle-particle vertex,
the particle-hole vertices can also induce the local and the non-local
pairing through the charge and spin-fluctuation mechanism \citep{Gabi1988,Grilli1990,Grilli1991,Houghton1990}.
To compute the irreducible pairing vertex for the orbital-antisymmetric
spin-triplet pairing, we again project the particle-hole vertices
onto the pairing basis $\mathbf{h}_{P}$:

\begin{align}
\Gamma_{\text{ph}}^{\text{\text{irr}}} & (\mathbf{k},\mathbf{k}')=\sum_{\substack{s\in\{\text{ch},\text{sp},\text{orb},\\
\text{so},\text{orb}^{*},\text{so}^{*}\}
}
}\big[\mathbf{h}_{\text{P}}^{\dagger}\big]_{\alpha\gamma}\tilde{\Gamma}_{\alpha\beta\gamma\delta}^{s}(\mathbf{k},\mathbf{k}')\big[\mathbf{h}_{\text{P}}^{\dagger}\big]_{\beta\delta}\nonumber \\
 & =\frac{1}{8}\Big[\Gamma^{\text{ch}}(\mathbf{k},\mathbf{k}',\mathbf{q}=\mathbf{k}-\mathbf{k}')+\Gamma^{\text{sp}}(\mathbf{k},\mathbf{k}',\mathbf{q}=\mathbf{k}-\mathbf{k}')\nonumber \\
 & -\Gamma^{\text{orb}}(\mathbf{k},\mathbf{k}',\mathbf{q}=\mathbf{k}-\mathbf{k}')-\Gamma^{\text{so}}(\mathbf{k},\mathbf{k}',\mathbf{q}=\mathbf{k}-\mathbf{k}')\nonumber \\
 & -\frac{5}{3}\Gamma^{\text{orb*}}(\mathbf{k},\mathbf{k}',\mathbf{q}=\mathbf{k}-\mathbf{k}')-\frac{5}{3}\Gamma^{\text{so*}}(\mathbf{k},\mathbf{k}',\mathbf{q}=\mathbf{k}-\mathbf{k}')\nonumber \\
 & +(\mathbf{k}'\rightarrow-\mathbf{k}')\Big],\label{eq:Gamma_ph}
\end{align}
where the charge, spin, orbital, and spin-orbital scattering vertices
$\tilde{\Gamma}^{s}$ are defined in Eq.~\eqref{eq:Gamma_s}. The diagrammatic
representation for Eq.~\eqref{eq:Gamma_ph} is shown in Fig. \ref{fig:diagram}(b),
where the $\mathcal{M}_{r_{s}r_{s}}^{-1}$, $\mathcal{M}_{r_{s}l_{s}}^{-1}$,
and $\mathcal{M}_{l_{s}l_{s}}^{-1}$ contain the summation of the
particle-hole bubbles to the infinite order (see Appx.~\ref{sec:propagator}), and we include both the
direct and the exchange (crossing) diagrams. The irreducible pairing
interaction from the particle-hole channel can be computed from:
\begin{equation}
\Gamma_{\text{ph}}^{\text{irr}}=Z^{2}\big\langle\big\langle\Gamma_{\text{ph}}^{\text{irr}}(\mathbf{k},\mathbf{k}')\big\rangle_{\mathbf{k}_{F}}\big\rangle_{\mathbf{k}'_{F}},
\end{equation}
where we assume an $s$-wave pairing to compare with the local pairing
fluctuation mechanism in the previous section. 

\section{Results and discussion}

\subsection{Superconducting phase diagram}

In this subsection, we apply our RISB saddle-point approximation and
fluctuation approach to the degenerate three-orbital Hubbard-Kanamori
model with Hund's coupling $J=U/4$, which serves as an effective
model for Hund's metals. We will focus on the order parameter $\langle\hat{\mathcal{O}}_{\text{P}}\rangle$
computed from Eq.~\eqref{eq:Op} and the pairing susceptibility $\chi_{\text{P}}$
computed from Eq.~\eqref{eq:chi_deg}.

\begin{figure}[t]
\begin{centering}
\includegraphics[scale=0.31]{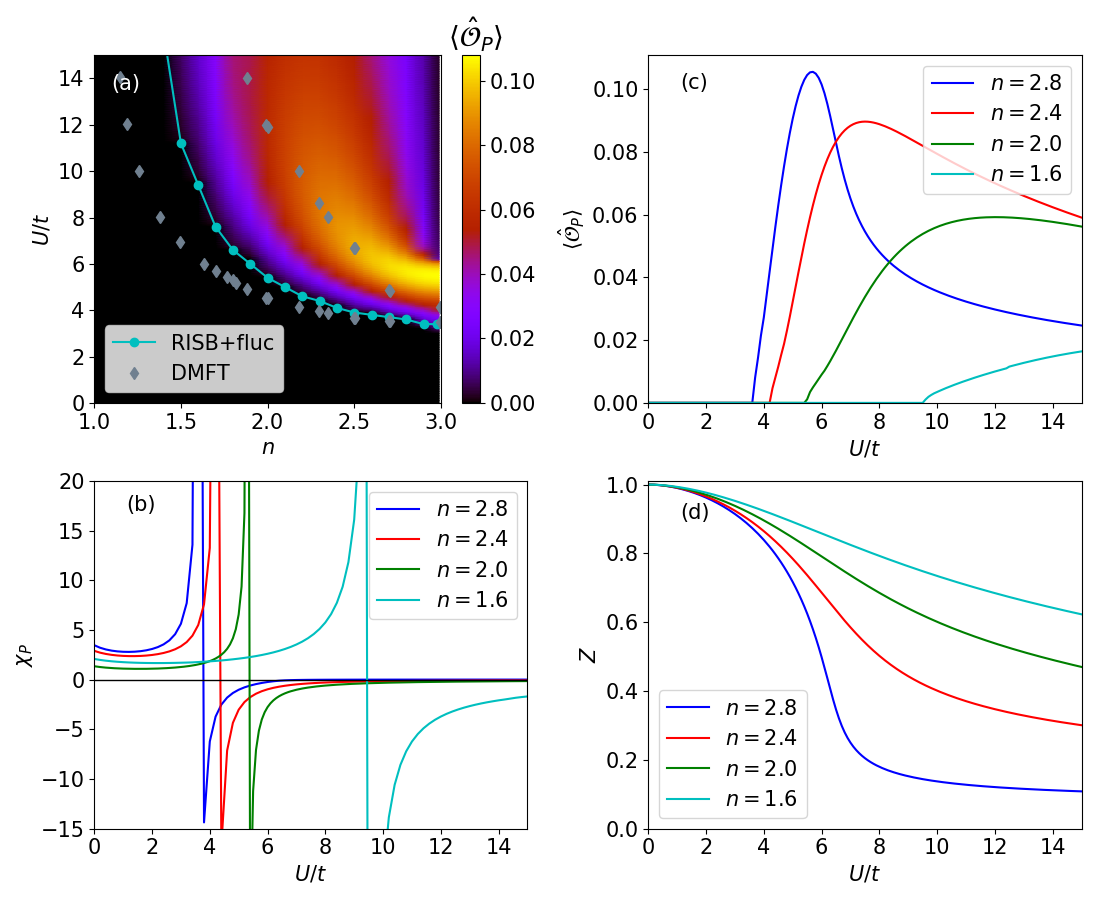}
\par\end{centering}
\caption{(a) The density plot of the $s$-wave spin-triplet superconducting
order parameter $\langle\hat{\mathcal{O}}_{P}\rangle$ as a function
of electron filling $n$ and Coulomb interaction $U$ with $J=U/4$
at $T=0.0005t$. The cyan line is the phase boundary determined from
the instability in the pairing susceptibility $\chi_{\text{P}}$.
(b) The uniform pairing susceptibility $\chi_{\text{P}}$ for $n=2.8,\ 2.4,\ 2.0,\ 1.6$.
(c) The spin-triplet superconducting order parameters $\langle\hat{\mathcal{O}}_{P}\rangle$
for $n=2.8,\ 2.4,\ 2.0,\ 1.6$. (d) The quasiparticle weight $Z$
for $n=2.8,\ 2.4,\ 2.0,\ 1.6$.\label{fig:PD_T0}}
\end{figure}
\begin{figure}[t]
\begin{centering}
\includegraphics[scale=0.31]{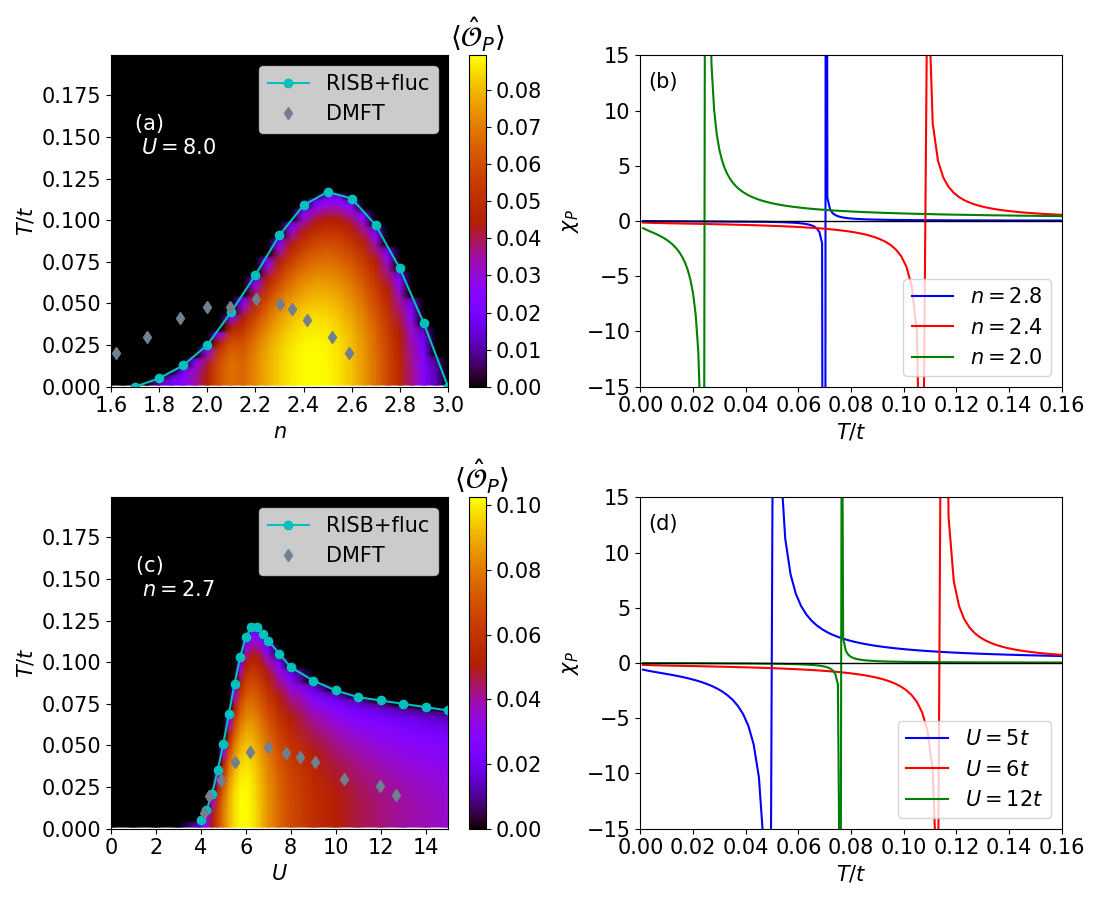}
\par\end{centering}
\caption{(a) The density plot of the $s$-wave spin-triplet superconducting
order parameter $\langle\hat{\mathcal{O}}_{P}\rangle$ as a function
of electron filling $n$ and temperature $T$ at $U=8$ and $J=U/4$.
The cyan line is the phase boundary determined from the instability
of the pairing susceptibility $\chi_{\text{P}}$. (b) The uniform
pairing susceptibility $\chi_{\text{P}}$ for $n=2.8,\ 2.4,\ 2.0$.
(c) The density plot of the spin-triplet superconducting order parameter
$\langle\hat{\mathcal{O}}_{P}\rangle$ as a function of Coulomb interaction
$U$ and temperature $T$ with $n=2.7$ and $J=U/4$. The cyan line
is the phase boundary determined from the instability of the pairing
susceptibility $\chi_{\text{P}}$. (d) The uniform pairing susceptibility
$\chi_{\text{P}}$ for $U=5t,\ 16t,\ 12t$.\label{fig:PD_T}}
\end{figure}

Figure \ref{fig:PD_T0} (a) shows the intensity plot of the spin-triplet
pairing order parameter $\langle\hat{\mathcal{O}}_{\text{P}}\rangle$
at $T=0.0005t$. The peak of the order parameters locates %along a particular trace, which is known as
at the so-called Hund's metal crossover, where the quasiparticle weights
$Z$ decrease significantly, as shown in Fig. \ref{fig:PD_T0} (c)
and (d) for selected fillings $n=1.6,\ 2.0,\ 2.4,\ \text{and}\ 2.8$.
The faster the decrease in $Z$, the stronger the enhancement in the
pairing order parameters $\langle\hat{\mathcal{O}}_{\text{P}}\rangle$.
The normal state in the superconducting regime can be viewed as Hund's
metals, where the quasiparticle weight is small, and the local multiplet
is populated with high spin states, favoring the local spin-triplet
pairing \citep{Georges_Hunds_review,Medici_Janus_PRL,Isidori_2019_PRL,Han_2004,Hoshino_Werner_2015_PRL}.

We also show the uniform pairing susceptibility $\chi_{\text{P}}$
evaluated from the fluctuation technique in Fig. \ref{fig:PD_T0}
(b). The pairing susceptibility is initially positive at small Coulomb
interaction $U$ and diverges at the critical point. Then, the pairing
susceptibility turns negative, indicating the instability towards
the $s$-wave spin-triplet ordering state. The phase boundary determined
from the divergence of the pairing susceptibility is shown in Fig.
\ref{fig:PD_T0} (a), which agrees with the onset of the mean-field
order parameters indicating the consistency of our approach. We also
compare our phase diagram with the DMFT results on a Bethe lattice
at $T=0.04t$ rescaled to the 2D bandwidth $W=8t$ in Fig. \ref{fig:PD_T0}
(a). While the RISB superconducting regime is broader than the DMFT
results, the overall phase diagram agrees qualitatively with the DMFT~\citep{Hoshino_Werner_2015_PRL}. 
%Note that the quantitative differences may also arise from the Ising anisotropic Coulomb interaction used in the DMFT study or the choice of the lattice.

We now turn to the finite-temperature phase diagram for the $s$-wave
spin-triplet pairing state. Figure \ref{fig:PD_T} (a) shows the intensity
plot of the $s$-wave spin-triplet order parameters $\langle\hat{\mathcal{O}}_{\text{P}}\rangle$
at $U=8t$ as a function of electron filling $n$ and temperature
$T$. The superconducting region has a dome shape structure, where
the maximum $T_{c}$ locates around $n=2.5$. Figure \ref{fig:PD_T}
(b) shows the uniform pairing susceptibility $\chi_{\text{P}}$ computed
from the fluctuation approach for filling $n=2.0,\ 2.4,\ \text{and}\ 2.8$
as a function of temperature $T$. With decreasing $T$, the pairing
susceptibility increases and diverges at the critical temperature
$T_{c}$. The critical temperature obtained from the divergence of
the pairing susceptibility agrees with the onset of the mean-filed
order parameters, as shown in Fig. \ref{fig:PD_T} (a). We also compare
our phase diagram with the DMFT results on a Bethe lattice in Fig.
\ref{fig:PD_T} (a) corresponding to $U=6t$ rescaled to the 2D bandwidth
$W=8t$ considered here. \citep{Hoshino_Werner_2015_PRL}. Both methods
generate a dome shape structure where the peak in RISB is closer to
half-filling.

\begin{figure}[t]
\begin{centering}
\includegraphics[scale=0.38]{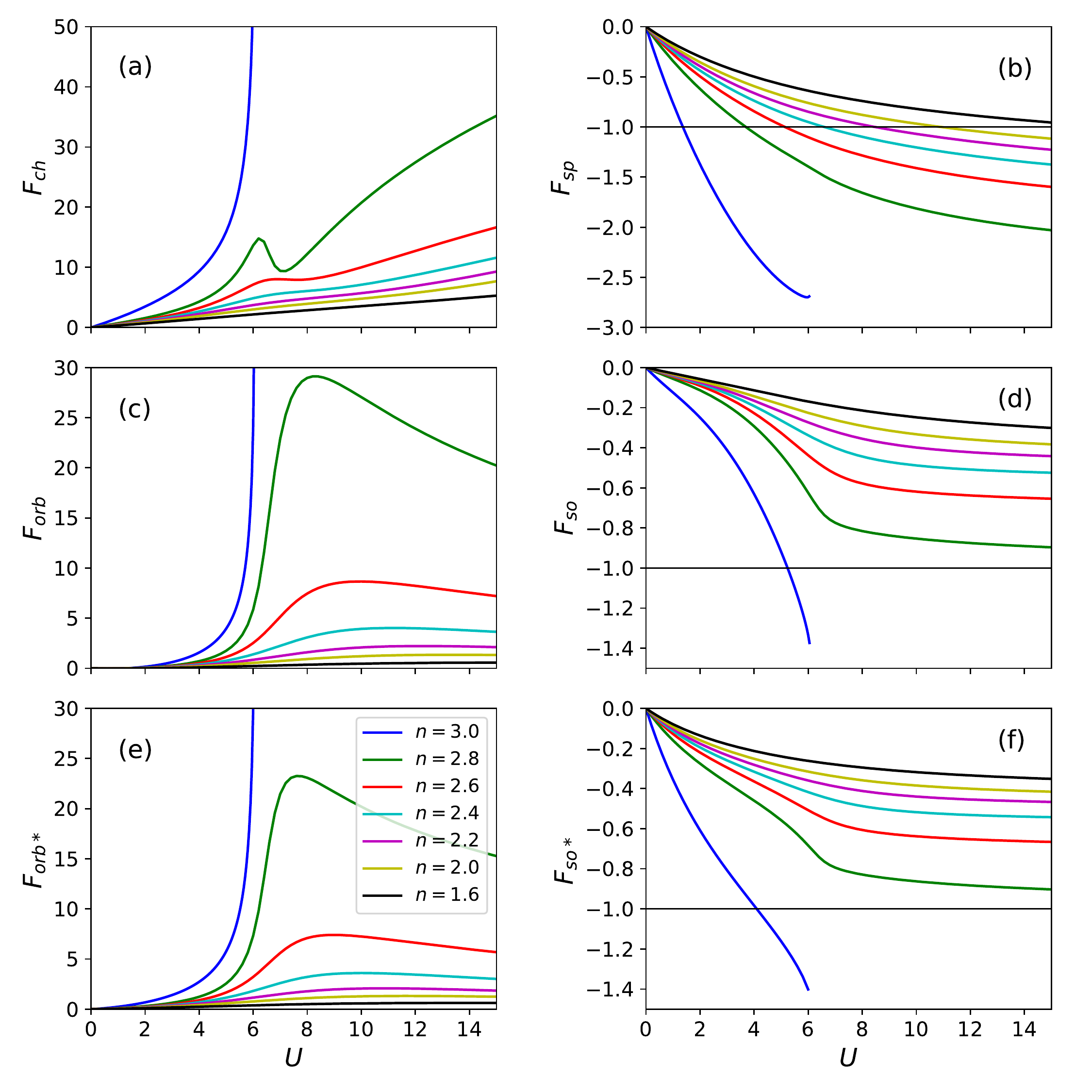}
\par\end{centering}
\caption{The Landau parameters in the (a) charge, (b) spin, (c) orbital, (d),
spin-orbital, (e) orbital{*}, and (f) spin-orbital{*} channel defined
in Eq.~\eqref{eq:3orb_flucs} as a function of coulomb interaction $U$  and $J=U/4$ 
for filling $n=3.0,\ 2.8,\ 2.6,\ 2.4,\ 2.2,\ 2.0,\ 1.6$ and $T=0.0005t$.
\label{fig:FL_params}}
\end{figure}

Figure \ref{fig:PD_T} (c) shows the intensity plot of the $s$-wave
spin-triplet pairing order parameters $\langle\hat{\mathcal{O}}_{\text{P}}\rangle$
as a function of Coulomb interaction $U$ and $J=U/4$  at filling $n=2.7$. The
critical temperature $T_{c}$ peaks around $U=6t$, which is around
the Hund's metal crossover. Figure \ref{fig:PD_T} (d) shows the corresponding
uniform pairing susceptibility $\chi_{\text{P}}$ computed from the
fluctuation approach for $U=5t,\ 6t,\ \text{and}\ 12t$. The pairing
susceptibility diverges at $T_{c}$ and turns negative, indicating
the instability towards the $s$-wave spin-triplet pairing states.
The $T_{c}$ obtained from the divergence of the susceptibility again
agrees with the onset of the mean-field order parameters, as shown
in Fig. \ref{fig:PD_T} (c). We also compare our phase diagram with
the DMFT results on a Bethe lattice in Figure \ref{fig:PD_T} (c)
at $n=2.0$ to match with our critical $U_{c}$ at $T=0.0005t$. The
phase diagrams obtained from both methods are again similar with a
dome shape structure where the $T_{c}$ peaks around the Hund's crossover.
% Note that the overestimation of the critical temperature $T_c$ may originated from
%the fact that RISB cannot capture the non-Fermi liquid factional power-law self-energy at finite temperature~
%\cite{Werner_power_law}, due to the linearized self-energy in RISB~\citep{Lecherman_2007}. 
%%The detailed discussion for the connection between RISB and DMFT can be found in Refs.~\citep{Lanata_2015_PRX,Ayral_2017}.
%

Note that there are two main reasons for expecting qualitative agreement (but quantitative agreement) between our RISB results and the DMFT results of Ref.~\citep{Hoshino_Werner_2015_PRL}. The first reason is that RISB (equivalently GA) is essentially a variational approximation to DMFT, in the sense that it is variational in the limit of infinite dimension~\citep{Metzner_1989}, where DMFT is exact. %, where DMFT is exact in the infinite dimension. Also, 
Also, RISB can be viewed as an approximation to DMFT, from a quantum embedding perspective, where the uncorrelated bath has the same number of orbitals as the impurity (while the bath is infinite in DMFT). Hence, RISB is expected to be less accurate (but more efficient) compared to DMFT. 
Nevertheless, we note that, in this work, we assumed a 2D square lattice, while a Bethe lattice was used in Ref.~\citep{Hoshino_Werner_2015_PRL}. In fact, it is known that different lattice structures can lead to quantitative differences in the results, but the qualitative behaviors are generally similar~\citep{Bak1998}.
%We now discuss the origin of the quantitative differences between our RISB results and the DMFT results in Ref.~\citep{Hoshino_Werner_2015_PRL}. One of the factors is the formal differences between the two methods. From the formal standpoint, RISB (equivalently GA) can be interpreted as a variational approximation to DMFT, where DMFT is exact in the infinite dimension. One can also view RISB as an approximation to DMFT, where the continuous bath is replaced by a single multiorbital atom. Hence, the two methods can have quantitative differences where DMFT is more accurate than RISB. The other factor is the underlying difference in the chosen lattice structure. We used 2D square lattice in our calculations, while in Ref.~\citep{Hoshino_Werner_2015_PRL}, the Bethe lattice was utilized. The different lattice structures can lead to quantitative variations in $T_c$, but the qualitative behaviors are similar. 
%Given the above reasons,  our RISB results has qualitative but not quantitative agreement with the DMFT results in Ref.~\citep{Hoshino_Werner_2015_PRL}.

\subsection{Landau parameter and pairing interaction}

For studying the pairing mechanism, it is instructive to investigate
the quasiparticle interaction vertex in the spin, charge, orbital,
spin-orbital, and pairing channel. To obtain these quantities,
we applied the Fermi-liquid approximation in Sec. \ref{sec:QP_sus}, which reproduces the exact physical
susceptibility, as shown in Appx. \ref{sec:FL_approx_check}. 

Let us first discuss the charge, spin, orbital, and spin-orbital fluctuation, encoded in the Landau parameters $F_{s}$. The Landau parameters $F_{s}$ in
each channel are shown in Fig. \ref{fig:FL_params}. We found that
the Landau parameters in the charge $F_{\text{ch}}$ and orbital $F_{\text{orb (orb*)}}$
channels show a peak around the Hund's crossover and diverges at
the Mott transition at $n=3$. The kink in $F_{\text{ch}}$ corresponds
to the possible phase separation instability found in the previous
slave-spin study \citep{Medici_instability_2017}. Moreover, we found
the instability towards the ferromagnetic ordering $F_{\text{sp}}=-1$
for a wide range of electron filling. Consequently, $F_{\text{sp}}$
is the dominant fluctuation in the particle-hole channel. %As we are going to show later, the strong ferromagnetic fluctuation mechanism induces the -wave superconducting instability around the ferromagnetic instability. 
In addition, the spin-orbital channel $F_{\text{so(so*)}}$ also shows a subleading
instability %compared to $F_{\text{sp}}$ 
at $n=3$.

\begin{figure}[t]
\begin{centering}
\includegraphics[scale=0.42]{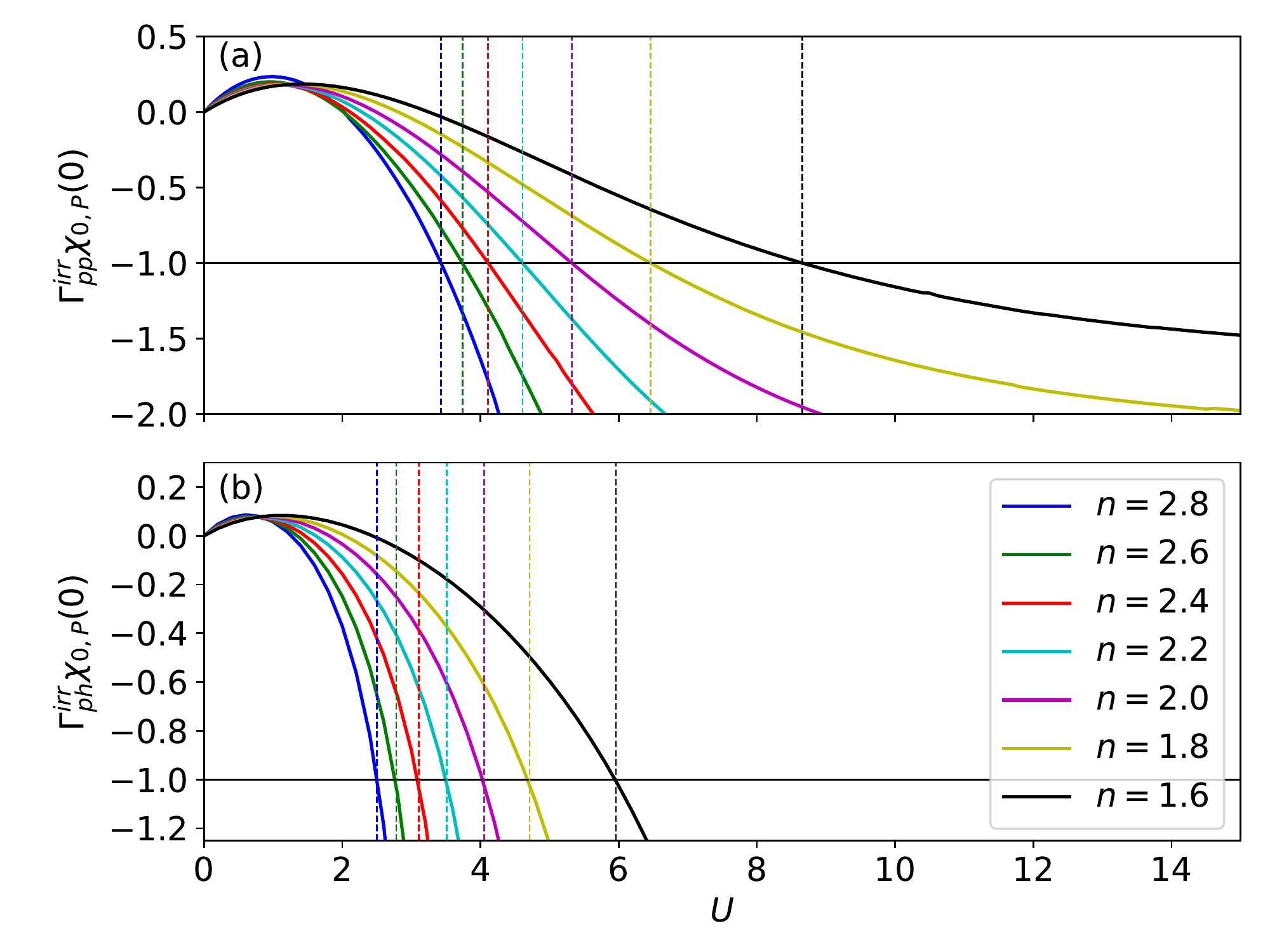}
\par\end{centering}
\caption{ (a) The irreducible particle-particle $s$-wave spin-triplet pairing vertex
$\Gamma_{\text{pp}}^{\text{\text{irr}}}\chi^{(0)}_{O_P}$ as a function
of Coulomb interaction $U$ and {\color{black}$J=U/4$} for filling $n=2.8,\ 2.6,\ 2.4,\ 2.2,\ 2.0,\ 1.8,\ 1.6$
and temperature $T=0.0005t$.  (b) The irreducible particle-hole $s$-wave
spin-triplet pairing vertex $\Gamma_{\text{ph}}^{\text{\text{irr}}}\chi^{(0)}_{O_P}$
with the same parameters setting. The vertical dashed lines indicate the
critical $U_{c}$ determined from $\Gamma_{\text{pp}}^{\text{irr}}\chi_{0}^{P}=-1$,
signalizing the divergence of the superconducting susceptibility and
the scattering amplitude. \label{fig:Ueff}}
\end{figure}

We now turn to the irreducible pairing vertex in the particle-particle
channel $\Gamma_{\text{pp}}^{\text{irr}}$ originated purely from
the local pairing fluctuation describing the superconducting instability.
Figure \ref{fig:Ueff} (a) shows the behavior of the pairing interaction
$\Gamma_{\text{pp}}^{\text{irr}}$ in the particle-particle channel
as a function of Coulomb interaction $U$. The condition $\Gamma_{\text{pp}}^{\text{irr}}\chi^{(0)}_{O_P}=-1$
indicates the divergence in the pairing susceptibility. In the weak-coupling
limit, i.e., $U\ll t$, $\Gamma_{\text{pp}}^{\text{\text{irr}}}$
follows the bare pairing interaction $U-3J$ for all the electron
filling $n$. With increasing $U$, the effective interactions for
different electronic filling are renormalized to smaller values and
eventually become negative signalizing the instability towards the
pairing states. The pairing instability determined from $\Gamma_{\text{pp}}^{\text{irr}}$
locates around the Hund's metal crossover as discussed in the previous
subsection. On the other hand, as shown in Fig. \ref{fig:Ueff} (b),
the pairing instability determined from the particle-hole scattering
channel $\Gamma_{\text{ph}}^{\text{irr}}$ takes place at a much lower
$U$ below the Hund's metal crossover. Consequently, the particle-hole
spin-fluctuation mechanism cannot explain the pairing instability
around Hund's metal crossover. The strong attraction in $\Gamma_{\text{ph}}^{\text{irr}}$
is, however, related to the ferromagnetic instability, as shown in
Fig. \ref{fig:FL_params}(b).

\section{Conclusions}

On the basis of the fluctuation approach around the RISB normal state
saddle-point, we developed an efficient method to compute general
susceptibilities, quasiparticle interaction vertex, Fermi-liquid parameters,
and pairing interaction for the multiorbital Hubbard model. The method has an RPA-like efficiency and a similar accuracy compared to DMFT for correlated systems.

We applied our method to the degenerate three-orbital Hubbard-Kanamori model
to investigate the origin of the $s$-wave orbital-antisymmetric spin-triplet
pairing in Hund's metal, previously found in the DMFT studies \citep{Hoshino_Werner_2015_PRL}.
We showed that, in agreement with DMFT, the pairing susceptibility
of the $s$-wave spin-triplet pairing states diverges around the Hund's
metal crossover. The phase diagram is in good qualitative agreement with DMFT. By computing the pairing interaction considering the particle-particle
and the particle- hole scattering channel, we identified that the
origin of the superconducting pairing around Hund's crossover arises from the particle-particle
channel, containing the local electron pair fluctuation between different
particle-number sectors of the local Hilbert space. The pairing interaction
is strongly renormalized in the incoherent Hund's metal regime and
becomes negative. On the other hand, the particle-hole spin-fluctuation mechanism induces an $s$-wave pairing instability already for a smaller value of Coulomb interaction, before entering the Hund's regime.%This behavior is similar to the Fulleride superconductor where the superconducting state also emerges from the incoherent metal near the Mott transition .  

The local interorbital pairing mechanisms revealed in this work can be applied to the
$s$-wave orbital-antisymmetric spin-triplet pairing states proposed
for Sr$_{2}$RuO$_{4}$~\citep{Puetter_2012_EPL,Cheung_2019,Agterberg_2020_PRR,Kee_2020_PRR,Kee_2020_shadow}
and KFe$_{2}$As$_{2}$~\citep{Vafek_Chubukov_2017,Piers_TRVB},
where the interplay between the Hund's rule coupling and the spin-orbital
coupling leads to intriguing gap structures on the Fermi surface. Our approach
provides an efficient route for investigating the pairing mechanism
for these materials, with the combination of density functional theory.  
The general formalism that we presented is also applicable for different purposes. For example, it could be utilized for investigating the response functions in the correlation-induced topological materials, e.g., the topological Kondo and topological Mott insulators~\cite{Raghu_TMI_2008,Dzero2010,Dai_Topological,Klett2020}, and the recently proposed topological iron-based superconductors~\cite{FeTeSe_2018,CaKFe4As4_2020}.
%Note that, for the topological Mott insulators, one has to apply the cluster-RISB approach or the ghost-orbital extension of RISB (equivalently GA) to capture the incoherent Hubbard bands~\citep{Lecherman_2007,RISB_DMET_Lee_2019,gGA_2017}.
%Our finding shows the importance of the particle-particle vertex contribution to the pairing interaction for Hund's metals, which generates additional renormalization to the pairing interaction that cannot be captured from the spin-fluctuation mechanism. Consequently, it is important to include this vertex contribution when performing the pairing instability analysis with the linearized Eliashberg equations, where the particle-particle vertex contribution was not considered \citep{Graser_2009,Yin_pairing,Gingras_2019}. Our formalism is not restricted to zero momentum transfer. Therefore, it can describe the non-local pairing by projecting the pairing vertex onto the momentum dependent symmetry basis\citep{Gabi1988,Grilli1990,Grilli1991,Houghton1990}. With the combination of density functional theory, our approach paves a way to investigate the pairing mechanism for realistic correlated materials. 
%[Review the recent articles on this type of local $s$-wave pairing, which leads to $d$-wave or $p$-wave like form factor when the SOC is considered. The methods and the finding here  will be important to analysis the competition between the local pairing and the spin-fluctuation effect.]
In addition, the diagrammatic approach proposed in this work may serve as a basis for the non-local extensions beyond RISB, similarly to the diagrammatic approaches beyond DMFT~\cite{Rohringer2018}.
Finally, our formalism can be applied to the NIB-DMET and other similar quantum embedding methods~\cite{SOET_2018,DET_2014,sekaran2021householder}. 
%The benchmark of the DMET and RISB susceptibilities will be an interesting future topic.

%Regarding the efficiency of our approach, the most time-consuming part of our method is the diagonalization
%of the embedding Hamiltonian $H_{\text{emb}}$. With the current state of the art, we can easily study the $f$-electron materials, containing 7 correlated orbitals, using exact-diagonalization and machine learning techniques~\citep{ML_RISB_2021}. In addition, one could also utilize the density matrix renormalization group or auxiliary-field quantum Monte Carlo to study the systems with more correlated orbitals~\citep{zheng1155}.
%Therefore, our method will be useful for investigating the response functions in heavy-fermion systems, and for extrapolating the two-particle response functions to the thermodynamic limit with the cluster approaches~\citep{RISB_DMET_Lee_2019}.

%Finally, we comment on the possible applications of our approach to DMET \cite{DMET_2012}. Since DMET can be interpreted as a special case of RISB by setting the renormalization matrix to unity $\mathbf{R}=I$ , we expect that our formalism can be directly applied to DMET by removing the $\mathbf{R}$ sector of the fluctuation matrix. The benchmark of the DMET and RISB susceptibilities will be an interesting future topic. Furthermore, our derivation based on the linear response of the embedding system can also be applied to other wavefunction-based embedding methods \cite{SOET_2018,DET_2014,sekaran2021householder}.

\begin{acknowledgments}
This work was supported by the Computational Materials Sciences Program funded by the US Department of Energy, Office of Science, Basic Energy Sciences, Materials Sciences and Engineering division. N.L. gratefully acknowledges support from the Novo Nordisk Foundation through the Exploratory Inter- disciplinary Synergy Programme project NNF19OC0057790, and from the VILLUM FONDEN through the Villum Experiment project 00028019 and the Centre of Excellence for Dirac Materials (Grant. No. 11744).
\end{acknowledgments}

\appendix
%dummy comment inserted by tex2lyx to ensure that this paragraph is not empty

\section{Gell-Mann matrices\label{sec:Gell-mann-matrices}}

We use the following convention for the Gell-Mann matrices 
\begin{eqnarray}
\lambda^{1}=\begin{pmatrix}0 & 1 & 0\\
1 & 0 & 0\\
0 & 0 & 0
\end{pmatrix},\  & \lambda^{4}=\begin{pmatrix}0 & -i & 0\\
i & 0 & 0\\
0 & 0 & 0
\end{pmatrix},\  & \lambda^{7}=\begin{pmatrix}1 & 0 & 0\\
0 & -1 & 0\\
0 & 0 & 0
\end{pmatrix},\nonumber \\
\lambda^{2}=\begin{pmatrix}0 & 0 & 1\\
0 & 0 & 0\\
1 & 0 & 0
\end{pmatrix},\  & \lambda^{5}=\begin{pmatrix}0 & 0 & -i\\
0 & 0 & 0\\
i & 0 & 0
\end{pmatrix},\  & \lambda^{8}=\frac{1}{\sqrt{3}}\begin{pmatrix}1 & 0 & 0\\
0 & 1 & 0\\
0 & 0 & -2
\end{pmatrix},\nonumber \\
\lambda^{3}=\begin{pmatrix}0 & 0 & 0\\
0 & 0 & 1\\
0 & 1 & 0
\end{pmatrix},\  & \lambda^{6}=\begin{pmatrix}0 & 0 & 0\\
0 & 0 & -i\\
0 & i & 0
\end{pmatrix},\  & \lambda^{0}=\begin{pmatrix}1 & 0 & 0\\
0 & 1 & 0\\
0 & 0 & 1
\end{pmatrix},\label{eq:-1}
\end{eqnarray}
{\color{black} where $\lambda^{1}$, $\lambda^{2}$, $\lambda^{3}$ describe the symmetric interorbital interactions or pairings; $\lambda^{4}$, $\lambda^{5}$, $\lambda^{6}$ describe the anti-symmetric interorbital interactions or pairings; $\lambda^{7}$, $\lambda^{8}$, $\lambda^{0}$ describe the intraorbital interactions or pairings. This set of matrices is the most general basis that parameterizes the $3\times3$ quadratic operators in the orbital space for three-orbital models. In the degenerate three-orbital Hubbard-Kanamori model, the $O(3)$ symmetry implies that the order parameters corresponds to the symmetric interorbital fluctuations $\lambda_{1}$, $\lambda_{2}$, and $\lambda_{3}$ are identical to each other. Similarly, the order parameters corresponds to the antisymmetric interorbital fluctuations $\lambda_{4}$, $\lambda_{5}$, and $\lambda_{6}$ are identical to each other.}

\section{Rotationally-invariant slave-boson Nambu formalism \label{sec:RISB_Nambu}}

In this section, we outline the basis of the RISB Nambu formalism. We start from a generic multiorbital Hubbard model in the Nambu notation:
 
\begin{equation}
\hat{H}=\frac{1}{2}\sum_{\mathbf{k}}\Xi_{\mathbf{k}\alpha}^{\dagger}\tilde{\epsilon}_{\mathbf{k},\alpha\beta}\Xi_{\mathbf{\mathbf{k}}\beta}+\sum_{i}\hat{H}_{\text{loc}}\big[\{d_{i\alpha\sigma}^{\dagger},d_{i\alpha\sigma}\}\big],\label{eq:Hph}
\end{equation}
where
\begin{equation}
\tilde{\mathbf{\epsilon}}_{\mathbf{k},\alpha\beta}=\begin{pmatrix}\epsilon_{\mathbf{k}} & 0\\
0 & -\epsilon_{\mathbf{-k}}^{*}
\end{pmatrix}
\end{equation}
is the energy dispersion in the Nambu basis. We also define the Nambu spinor $\Xi_{\mathbf{k}}^{\dagger}=(d_{\mathbf{k}1\uparrow}^{\dagger}d_{\mathbf{k}1\downarrow}^{\dagger}...d_{\mathbf{k}M\uparrow}^{\dagger}d_{\mathbf{k}M\downarrow}^{\dagger}d_{\mathbf{-k}1\uparrow}d_{\mathbf{-k}1\downarrow}...d_{\mathbf{-k}M\uparrow}d_{\mathbf{-k}M\downarrow})$, where $M$ is the total number of orbitals.
The $H_{\text{loc}}$ contains the generic local one-body and two-body interactions.

Within RISB framework, the physical operators $\Xi_{i\alpha}$ is mapped to the product
of a renormalization matrix and a quasiparticle Nambu spinor:

\begin{equation}
\Xi_{i\alpha}^{\dagger}=\sum_{a}\mathbf{R}_{ia\alpha}[\Phi_i^\dagger,\Phi_i]\Psi_{ia}^{\dagger},\label{eq:RISB_map}
\end{equation}
where the quasiparticle spinor is $\Psi_{i}^{\dagger}=(f_{i1\uparrow}^{\dagger}f_{i1\downarrow}^{\dagger}...,f_{iM\uparrow}^{\dagger}f_{iM\downarrow}^{\dagger},f_{i1\uparrow}f_{i1\downarrow}...f_{iM\uparrow}f_{iM\downarrow})$, and the renormalization matrix has the following form~\citep{Isidori_RISB_SC_2009,Fabrizio_SC,Nicola_SC}:
\begin{equation}
\mathbf{R}_{ia\alpha}[\Phi_i^\dagger,\Phi_i]=\sum_{b}\text{Tr}\big[\Phi_i^{\dagger}\Xi_{i\alpha}\Phi_i\Psi_{ib}^{\dagger}\big]\big[\boldsymbol{\Delta}_i(1-\boldsymbol{\Delta}_i)\big]_{ba}^{-\frac{1}{2}},\label{eq:R_mat}
\end{equation}
where
\begin{equation}
\big[\boldsymbol{\Delta}_i\big]_{ab}= \text{Tr}\big[\Phi_i^{\dagger}\Phi_i\Psi_{ia}^{\dagger}\Psi_{ib}\big]
\end{equation}
corresponds to the local quasiparticle Nambu density matrix, and $\big[\Phi_i\big]_{An}$ is the slave-boson amplitude matrix.
We also define the matrices $\big[\Xi_{i\alpha}\big]_{AB}=\langle A|\hat{\Xi}_{i\alpha}|B\rangle$ and $\big[\Psi_{i\alpha}\big]_{nm}=\langle n|\hat{\Psi}_{i\alpha}|m\rangle$
for the fermionic operator in the arbitrary local many-body basis $|A\rangle$ and the local Fock basis $|n\rangle$, respectively~\citep{Lecherman_2007,Lanata_2015_PRX}.
%Note that differ from the renormalization matrix in Ref. \citep{Isidori_RISB_SC_2009}, we further allow the gauge transformation to mix the particle and hole blocks so the anomalous part of $\boldsymbol{\Delta}_i$ is non-zero in the superconducting state. 
%The $\mathbf{R}$ is related to the quasiparticle weight through $Z=\mathbf{R}^\dagger\mathbf{R}$. 
The local interactions can be expressed in terms of the bosonic amplitudes as~\citep{Lecherman_2007}
\begin{equation}
\hat{H}_{\text{loc}}=\sum_{ABn}[\Phi]^{}_{Bn}[\Phi]^\dagger_{nA} [H_\text{loc}]^{}_{AB},
\end{equation}
where $\big[H_\text{loc}\big]_{AB}=\langle A|\hat{H}_{\text{loc}}|B\rangle$.

In order to select the physical states out of the enlarged boson and quasiparticle Hilbert space, one has to enforce the following RISB constraints~\citep{Lecherman_2007,Isidori_RISB_SC_2009}
\begin{align}
\text{\text{Tr}\ensuremath{\big[}\ensuremath{\ensuremath{\Phi_i\Phi_i^{\dagger}}}\ensuremath{\big]}} & =1,\label{eq:const1_sup}
\end{align}
\begin{align}
\big[\boldsymbol{\Delta}_i\big]_{ab} =\langle\hat{\Psi}_{ia}^{\dagger}\hat{\Psi}_{ib}\rangle &=\text{Tr}\big[\Phi_i^{\dagger}\Phi_i\Psi_{ia}^{\dagger}\Psi_{ib}\big].\label{eq:const2_sup}
\end{align}
The first constraint limits the Hilbert space to the single-boson states, while
the second constraint ensures the rotational invariance of the quasiparticle
density matrix under the gauge transformation (see Appx.~\ref{subsec:Gauge}).

With the RISB representations and constraints (Eqs~\eqref{eq:RISB_map}-\eqref{eq:const2_sup}), the RISB Lagrangian for the generic Hubbard model (Eq.~\eqref{eq:Hph}) can be expressed as: \begin{widetext}

\begin{align}
\mathcal{L}[\Phi,\mathcal{\mathbf{R}},\boldsymbol{\mathbf{\mathbf{\Lambda}}};\mathcal{\mathbf{D}},\boldsymbol{\Lambda}^{c} &,E^{c},\Delta] =\frac{-T}{N}\frac{1}{2}\sum_{\mathbf{k}_{1}\mathbf{k}_{2}\omega_{n}}\text{Tr log}\Big[-i\omega_{n}+H_{\mathbf{k}_{1}\mathbf{k}_{2}}^{\text{qp}}\Big]e^{i\omega_{n}0^{+}}+\sum_{i}\text{Tr}\Big[\Phi_{i}\Phi_{i}^{\dagger}H_{\text{loc}}\nonumber \\
 & +\big(\sum_{a\alpha}\big[\mathbf{D}_i]_{a\alpha}\Phi_{i}^{\dagger}\Xi_{i\alpha}^{\dagger}\Phi_{i}\Psi_{ia}+\text{H.c.}\big)+\sum_{ab}\frac{1}{2}\big[\boldsymbol{\Lambda}^{c}_i\big]_{ab}\Phi_{i}^{\dagger}\Phi_{i}\Psi_{ia}^{\dagger}\Psi_{ib}\Big]+E^{c}_i\big(\text{Tr}\Big[\Phi_{i}\Phi_{i}^{\dagger}\Big]-1\big)\nonumber \\
 & -\sum_{i}\Big[\sum_{ab}\frac{1}{2}\big(\big[\boldsymbol{\Lambda}_i\big]_{ab}+\big[\boldsymbol{\Lambda}^{c}_i\big]_{ab}\big)\big[\boldsymbol{\Delta}_i\big]_{ab}+\sum_{ca\alpha}\big(\big[\mathcal{\mathbf{D}}_i\big]_{a\alpha}\big[\mathbf{R}_i\big]_{c\alpha}\big[\boldsymbol{\Delta}_i(1-\boldsymbol{\Delta}_i)\big]_{ca}^{1/2}+\text{c.c.}\big)\Big],\label{eq:L_phi}
\end{align}
\end{widetext}
where the original kinetic hopping term in Eq.~\eqref{eq:Hph} is described by the quasiparticle Hamiltonian:
\begin{equation}
\big[H_{\mathbf{k}_{1}\mathbf{k}_{2}}^{\text{qp}}\big]_{ab}=\frac{1}{N}\sum_{\mathbf{k}}\big[\mathbf{R}_{\mathbf{k}_{1}-\mathbf{k}}\tilde{\epsilon}_{\mathbf{k}}\mathbf{R}_{\mathbf{k}_2-\mathbf{k}}^{\dagger}\big]_{ab}+\big[\boldsymbol{\mathbf{\mathbf{\Lambda}}}_{\mathbf{k}_{1}-\mathbf{k}_{2}}\big]_{ab},\label{eq:Hqp}
\end{equation}
while the local interaction $\hat{H}_{\text{loc}}$ in Eq.~\eqref{eq:Hph} is mapped to the slave-boson representation $\text{Tr}[\Phi_i \Phi_i^\dagger H_{\text{loc}}]$.
The $\boldsymbol{\Lambda}_i$, $\boldsymbol{\Lambda}^{c}_i$, $\mathbf{D}_i$, $E^{c}_i$
are the Lagrange multipliers enforcing the RISB constraints (Eqs.~\eqref{eq:const1_sup} and \eqref{eq:const2_sup}) and the
structure of the $\mathbf{R}_i$ matrix (Eq.~\eqref{eq:R_mat}). Note that all these single-particle
matrices contains the particle, hole, and anomalous sector defined as follows:

\begin{equation}
\mathbf{R}_i=\begin{bmatrix}R_i & Q^{*}_i\\
Q_i & R^{*}_i
\end{bmatrix},\label{eq:R_sup}
\end{equation}
\begin{equation}
\boldsymbol{\Lambda}_i=\begin{bmatrix}\Lambda_i & \Lambda'_i\\
{\Lambda_i'^{\dagger}} & -\Lambda^{*}_i
\end{bmatrix},\label{eq:Lam_sup}
\end{equation}

\begin{equation}
\boldsymbol{\Lambda}^{c}_i=\begin{bmatrix}\Lambda^{c}_i & {\Lambda^{c}_i}'\\
{{\Lambda^{c}_i}'}^{\dagger} & -{\Lambda^{c}_i}^{*}
\end{bmatrix},\label{eq:Lamc_sup}
\end{equation}

\begin{equation}
\mathbf{D}_i=\begin{bmatrix}D_i & D^{'*}_i\\
D^{'}_i & D^{*}_i
\end{bmatrix},\label{eq:D_sup}
\end{equation}	

\begin{equation}
\boldsymbol{\Delta}_i=\begin{bmatrix}\Delta_i & \Delta^{'}_i\\
\Delta^{'\dagger}_i & (\mathbf{1}-\Delta_i)
\end{bmatrix}.\label{eq:Delta_p_sup}
\end{equation}
The $\mathbf{\Lambda}_i$, $\mathbf{\Lambda}^{c}_i$ and $\mathbf{\Delta}_i$
are Hermitian matrices, and the $\mathbf{R}_i$ and $\mathbf{D}_i$ are non-Hermitian
matrices. {\color{black} These single-particle matrices are parameterized by Eqs.~\eqref{eq:R}-\eqref{eq:Lamc} utilizing the matrix basis $\mathbf{h}_{s}$ and $\tilde{\mathbf{h}}_s$, whose structure (for the three-orbital degenerate Hubbard-Kanamori model) is discussed in Sec.~\ref{subsec:Single-particle-basis}.}

The slave-boson amplitude can be constructed from the symmetry adaptive
basis $\phi_{ip}$:

\begin{equation}
\big[\Phi_{i}\big]_{An}=\sum_{p}c_{p}\big[\phi_{ip}\big]_{An}\label{eq:phi}
\end{equation}
where 
\begin{equation}
\text{Tr}\big[\phi_{ip}^{\dagger}\phi_{ip'}\big]=\delta_{p,p'}\ \ \ p,p'=1,...,N_{\phi},\label{eq:phi_constraint}
\end{equation}
and the matrix basis commutes with all the symmetry operation in the
group $G$ of the given problem, i.e., $[\phi_{ip},R(g)]=0\ \forall g\in G$. The procedure for determining $\phi_{ip}$ is discussed in Appx.~\ref{subsec:Many-body-basis}.

\subsection{Embedding mapping\label{subsec:Embedding-mapping}}

We now introduce the embedding wavefunction \citep{Lanata_2015_PRX}
\begin{equation}
|\Phi_i\rangle=\sum_{An}e^{i(\pi/2)N_{n}(N_{n}-1)}[\Phi_i]_{An}U_{\text{PH}}|A\rangle|n\rangle,\label{eq:}
\end{equation}
where $U_{\text{PH}}$ is the particle-hole transformation on the
bath site and $N_{n}$ is the particle number of Fock state $|n\rangle$.
Substituting the following identities to Eq.~\eqref{eq:L_phi}: 
\begin{align}
\text{Tr}\big[\Phi_i\Phi_i^{\dagger}H_{\text{loc}}\big] & =\langle\Phi_i|\hat{H}_{\text{loc}}[\hat{d}_{i\alpha}^{\dagger},\hat{d}_{i\alpha}]|\Phi_i\rangle\label{eq:-10}\\
\text{Tr}[\Phi_i^{\dagger}\Xi_{i\alpha}^{\dagger}\Phi\Psi_{ia}] & =\sum_{b}\langle\Phi_i|\hat{\Xi}_{i\alpha}^{\dagger}\hat{\Psi}_{ib}|\Phi_i\rangle\bar{I}_{ba}\label{eq:-11}\\
\text{Tr}[\Phi_i^{\dagger}\Phi_i\Psi_{ia}^{\dagger}\Psi_{ib}] & =\sum_{cd}\bar{I}_{bc}\langle\Phi_i|\hat{\Psi}_{ic}\hat{\Psi}_{id}^{\dagger}|\Phi_i\rangle\bar{I}_{da},\label{eq:-12}
\end{align}
where
\begin{equation}
\bar{I}=\begin{pmatrix}\mathbf{1} & 0\\
0 & \mathbf{-1}
\end{pmatrix},\label{eq:-13}
\end{equation}
and $\mathbf{1}$ is the identity matrix, we obtained the RISB Lagrangian in terms of $|\Phi_i\rangle$ in Eq.~\eqref{eq:Lemb} in the main text.

\section{Variational basis\label{sec:Variational-basis}}

In this section, we describe the construction of our variational many-body
basis $\phi_{p}$ and the single-particle basis $\mathbf{h}_{s}$
and $\tilde{\mathbf{h}}_{s}$ of our fluctuation approach to the degenerate
three-orbital Hubbard-Kanamori model.

\subsection{Many-body basis\label{subsec:Many-body-basis}}

\begin{table}[p]
\begin{centering}
{\tiny{}{}}%
\begin{tabular}{|c|c|c|c|}
\hline 
{\small{}{}$(N,L,S)$} & {\small{}{}Degeneracy} & {\small{}{}$E_{\Gamma}$} & {\small{}{}$\Phi_{\Gamma n}$}\tabularnewline
\hline 
\hline 
\multirow{3}{*}{{\small{}{}$(0,0,0)$}} & \multirow{3}{*}{{\small{}{}$1$}} & \multirow{3}{*}{{\small{}{}$0$}} & {\tiny{}{}$\Phi(E_{000})$}\tabularnewline
 &  &  & {\tiny{}{}$\Phi(E_{000};2)$}\tabularnewline
 &  &  & {\tiny{}{}$\Phi(E_{000};4)$}\tabularnewline
\hline 
\multirow{3}{*}{{\small{}{}$(1,1,\frac{1}{2})$}} & \multirow{3}{*}{{\small{}{}$6$}} & \multirow{3}{*}{{\small{}{}$0$}} & {\tiny{}{}$\Phi(E_{11\frac{1}{2}})$}\tabularnewline
 &  &  & {\tiny{}{}$\Phi(E_{11\frac{1}{2}};2)$}\tabularnewline
 &  &  & {\tiny{}{}$\Phi(E_{11\frac{1}{2}};4)$}\tabularnewline
\hline 
\multirow{4}{*}{{\small{}{}$(2,2,0)$}} & \multirow{4}{*}{{\small{}{}$5$}} & \multirow{4}{*}{{\small{}{}$U-J$}} & {\tiny{}{}$\Phi(E_{220})$}\tabularnewline
 &  &  & {\tiny{}{}$\Phi(E_{220};-2)$}\tabularnewline
 &  &  & {\tiny{}{}$\Phi(E_{220};2)$}\tabularnewline
 &  &  & {\tiny{}{}$\Phi(E_{220};4)$}\tabularnewline
\hline 
\multirow{3}{*}{{\small{}{}$(2,1,1)$}} & \multirow{3}{*}{{\small{}{}$9$}} & \multirow{3}{*}{{\small{}{}$U-3J$}} & {\tiny{}{}$\Phi(E_{211})$}\tabularnewline
 &  &  & {\tiny{}{}$\Phi(E_{211};-2)$}\tabularnewline
 &  &  & {\tiny{}{}$\Phi(E_{211};2)$}\tabularnewline
\hline 
\multirow{3}{*}{{\small{}{}$(2,0,0)$}} & \multirow{3}{*}{{\small{}{}$1$}} & \multirow{3}{*}{{\small{}{}$U+2J$}} & {\tiny{}{}$\Phi(E_{200})$}\tabularnewline
 &  &  & {\tiny{}{}$\Phi(E_{200};2)$}\tabularnewline
 &  &  & {\tiny{}{}$\Phi(E_{200};4)$}\tabularnewline
\hline 
\multirow{3}{*}{{\small{}{}$(3,2,\frac{1}{2})$}} & \multirow{3}{*}{{\small{}{}$10$}} & \multirow{3}{*}{{\small{}{}$3U-6J$}} & {\tiny{}{}$\Phi(E_{32\frac{1}{2}})$}\tabularnewline
 &  &  & {\tiny{}{}$\Phi(E_{32\frac{1}{2}};-2)$}\tabularnewline
 &  &  & {\tiny{}{}$\Phi(E_{32\frac{1}{2}};2)$}\tabularnewline
\hline 
\multirow{3}{*}{{\small{}{}$(3,1,\frac{1}{2})$}} & \multirow{3}{*}{{\small{}{}$6$}} & \multirow{3}{*}{{\small{}{}$3U-4J$}} & {\tiny{}{}$\Phi(E_{31\frac{1}{2}})$}\tabularnewline
 &  &  & {\tiny{}{}$\Phi(E_{31\frac{1}{2}};-2)$}\tabularnewline
 &  &  & {\tiny{}{}$\Phi(E_{31\frac{1}{2}};2)$}\tabularnewline
\hline 
\multirow{3}{*}{{\small{}{}$(3,0,\frac{3}{2})$}} & \multirow{3}{*}{{\small{}{}$4$}} & \multirow{3}{*}{{\small{}{}$3U-9J$}} & {\tiny{}{}$\Phi(E_{30\frac{3}{2}})$}\tabularnewline
 &  &  & {\tiny{}{}$\Phi(E_{30\frac{3}{2}};-2)$}\tabularnewline
 &  &  & {\tiny{}{}$\Phi(E_{30\frac{3}{2}};2)$}\tabularnewline
\hline 
\multirow{4}{*}{{\small{}{}$(4,2,0)$}} & \multirow{4}{*}{{\small{}{}$5$}} & \multirow{4}{*}{{\small{}{}$6U-11J$}} & {\tiny{}{}$\Phi(E_{420})$}\tabularnewline
 &  &  & {\tiny{}{}$\Phi(E_{420};-4)$}\tabularnewline
 &  &  & {\tiny{}{}$\Phi(E_{420};-2)$}\tabularnewline
 &  &  & {\tiny{}{}$\Phi(E_{420};2)$}\tabularnewline
\hline 
\multirow{4}{*}{{\small{}{}$(4,1,1)$}} & \multirow{4}{*}{{\small{}{}$9$}} & \multirow{4}{*}{{\small{}{}$6U-13J$}} & {\tiny{}{}$\Phi(E_{411})$}\tabularnewline
 &  &  & {\tiny{}{}$\Phi(E_{411};-4)$}\tabularnewline
 &  &  & {\tiny{}{}$\Phi(E_{411};-2)$}\tabularnewline
 &  &  & {\tiny{}{}$\Phi(E_{411};2)$}\tabularnewline
\hline 
\multirow{4}{*}{{\small{}{}$(4,0,0)$}} & \multirow{4}{*}{{\small{}{}$1$}} & \multirow{4}{*}{{\small{}{}$U+2J$}} & {\tiny{}{}$\Phi(E_{400})$}\tabularnewline
 &  &  & {\tiny{}{}$\Phi(E_{400};-4)$}\tabularnewline
 &  &  & {\tiny{}{}$\Phi(E_{400};-2)$}\tabularnewline
 &  &  & {\tiny{}{}$\Phi(E_{400};2)$}\tabularnewline
\hline 
\multirow{3}{*}{{\small{}{}$(5,1,\frac{1}{2})$}} & \multirow{3}{*}{{\small{}{}$6$}} & \multirow{3}{*}{{\small{}{}$10U-20J$}} & {\tiny{}{}$\Phi(E_{51\frac{1}{2}})$}\tabularnewline
 &  &  & {\tiny{}{}$\Phi(E_{51\frac{1}{2}};-4)$}\tabularnewline
 &  &  & {\tiny{}{}$\Phi(E_{51\frac{1}{2}};-2)$}\tabularnewline
\hline 
\multirow{3}{*}{{\small{}{}$(6,0,0)$}} & \multirow{3}{*}{{\small{}{}$1$}} & \multirow{3}{*}{{\small{}{}$15U-30J$}} & {\tiny{}{}$\Phi(E_{600})$}\tabularnewline
 &  &  & {\tiny{}{}$\Phi(E_{600};-4)$}\tabularnewline
 &  &  & {\tiny{}{}$\Phi(E_{600};-2)$}\tabularnewline
\cline{4-4} 
\end{tabular}
\par\end{centering}
\caption{Quantum numbers $(N,L,S)$, degeneracy, Eigenvalues, and the corresponding
slave-bosons $\Phi(E_{\Gamma};2q)$ for each local multiplets $|\Gamma\rangle$.\label{tab:boson_multiplets}}
\end{table}

For the charge, spin, orbital, and spin-orbital fluctuations, we construct
the many-body basis in Eq.~\eqref{eq:phi} using the symmetry adapted
basis. The procedure can be found in Ref. \citep{Lanata_2012}. On
the other hand, for the pairing state, we construct the many-body
variational basis following the procedure in Ref. \citep{Isidori_RISB_SC_2009}.
First, since the Hubbard-Kanamori interaction (Eq.~\eqref{eq:Kanamorai_int}) can be written into
\begin{equation}
H_{\text{loc}}=(U-3J)\frac{\hat{N}(\hat{N}-1)}{2}-J\Big[2\hat{\mathbf{S}}^{2}+\frac{1}{2}\hat{\mathbf{L}}^{2}\Big]+\frac{5}{2}J\hat{N}\label{eq:Hint_NLS}
\end{equation}
{\color{black}with 
\begin{align}
\hat{L}_{\alpha}&=\sum_{\beta\gamma\sigma}\hat{d}^\dagger_{i\beta\sigma}[-i\epsilon_{\alpha\beta\gamma}] \hat{d}_{i\gamma\sigma}\\
\hat{\mathbf{S}}&=\frac{1}{2}\sum_{\alpha\sigma\sigma'}\hat{d}_{i\alpha\sigma}^\dagger \boldsymbol{\sigma}_{\sigma\sigma'} \hat{d}_{i\alpha\sigma'}\\
\hat{N}&=\sum_{\alpha\sigma} \hat{d}^\dagger_{\alpha\sigma}\hat{d}_{\alpha \sigma},
\end{align}}the local Hamiltonian is diagonalized in the $\Gamma=(N,L,S)$ basis. {\color{black}The $\boldsymbol{\sigma}$ is a vector of Pauli matrices, and $\epsilon_{\alpha\beta\gamma}$ is the Levi-Civita symbol, which can be expressed in terms of Gell-Mann matrices $\lambda^4$, $\lambda^5$, and $\lambda^6$.}
Therefore, the slave-boson amplitude can be significantly reduced
to 
\begin{align}
\Phi_{\Gamma n} & =\langle\Gamma|n\rangle\Phi(E_{\Gamma})+\sum_{q=1}^{3}\Bigg[\frac{\langle n|(\mathcal{\hat{O}}_{\text{P}})^{q}|\Gamma\rangle}{\sqrt{\langle\Gamma|(\hat{\mathcal{O}}_{\text{P}}^{\dagger})^{q}(\hat{\mathcal{O}}_{\text{P}})^{q}|\Gamma\rangle}}\Phi(E_{\Gamma};2q),\nonumber \\
 & +\frac{\langle n|(\mathcal{\hat{O}}_{\text{P}}^{\dagger})^{q}|\Gamma\rangle}{\sqrt{\langle\Gamma|(\hat{\mathcal{O}}_{\text{P}})^{q}(\hat{\mathcal{O}}_{\text{P}}^{\dagger})^{q}|\Gamma\rangle}}\Phi(E_{\Gamma};-2q)\Bigg]\label{eq:Phi_Gn}
\end{align}
where $E_{\Gamma}$ and $|\Gamma\rangle$ is the eigenvalue and the
eigenstate of Eq.~\eqref{eq:Hint_NLS}, respectively. Comparing Eq.~\eqref{eq:Phi_Gn} to Eq.~\eqref{eq:phi}, we identify that the many-body basis for the normal
state part is 
\[
\phi_{p}=\langle\Gamma|n\rangle,
\]
with the corresponding slave-boson $c_p=\Phi(E_{\Gamma})$, and the pairing
part are 
\[
\phi_{p}=\frac{\langle n|(\mathcal{\hat{O}}_{\text{P}})^{q}|\Gamma\rangle}{\sqrt{\langle\Gamma|(\hat{\mathcal{O}}_{\text{P}}^{\dagger})^{q}(\hat{\mathcal{O}}_{\text{P}})^{q}|\Gamma\rangle}}
\]
and 
\[
\phi_{p}=\frac{\langle n|(\mathcal{\hat{O}}_{\text{P}}^{\dagger})^{q}|\Gamma\rangle}{\sqrt{\langle\Gamma|(\hat{\mathcal{O}}_{\text{P}})^{q}(\hat{\mathcal{O}}_{\text{P}}^{\dagger})^{q}|\Gamma\rangle}}
\]
with the corresponding slave-boson amplitudes $c_p=\Phi(E_{\Gamma};2q)$
and $c_p=\Phi(E_{\Gamma};-2q)$, respectively. In the end, we have $43$
bosonic amplitudes listed in Tab. \ref{tab:boson_multiplets}.

\subsection{Single-particle basis\label{subsec:Single-particle-basis}}

The single-particle basis $\mathbf{h}_{s}$ and $\tilde{\mathbf{h}}_{s}$, parameterizing Eqs.~\eqref{eq:R_sup}-\eqref{eq:Delta_p_sup},
are block matrices, 
\begin{equation}
\mathbf{h}_{s}=\begin{pmatrix}h_{s} & h'_{s}\\
{h'_{s}}^{\dagger} & -h_{s}^{*}
\end{pmatrix}\quad\tilde{\mathbf{h}}_{s}=\begin{pmatrix}h_{s} & {h'_{s}}^{*}\\
{h'_{s}} & h_{s}^{*}
\end{pmatrix},\label{eq:hs}
\end{equation}
where the component $h_{s}$ corresponds to the normal part and $h'_{s}$
corresponds to the anomalous part of the matrix. The components for
each fluctuation channel, in the degenerate three-orbital model, are as follow: 
\begin{equation}
h_{\text{ch}}=\lambda_{0}\otimes\sigma_{0},\label{eq:h_ch}
\end{equation}
\begin{equation}
h_{\text{sp}}=\lambda_{0}\otimes\sigma_{z},
\end{equation}
\begin{equation}
h_{\text{orb}}=\lambda_{4}\otimes\sigma_{0},
\end{equation}
\begin{equation}
h_{\text{so}}=\lambda_{4}\otimes\sigma_{z},
\end{equation}
\begin{equation}
h_{\text{orb*}}=\lambda_{1}\otimes\sigma_{0},
\end{equation}
\begin{equation}
h_{\text{so*}}=\lambda_{1}\otimes\sigma_{z},
\end{equation}
\begin{equation}
h_{\text{P}}=0,
\end{equation}
for the normal part, and 
\begin{gather}
h'_{\text{ch}}=h'_{\text{sp}}=h'_{\text{orb}}=h'_{\text{so}}=h'_{\text{orb*}}=h'_{\text{so*}}=0\\
h_{\text{P}}'=\lambda_{6}\otimes[-i\sigma_{y}\sigma_{z}]\label{eq:h_P}
\end{gather}
for the anomalous part, where the basis is chosen to be normalized,
i.e., $\text{Tr}\big[\mathbf{h}_{s}\mathbf{h}_{s}^{\dagger}\big]=1$.
We see that $\mathbf{h}_{\text{P}}$ describes the pairing fluctuation,
while $\mathbf{h}_{\text{ch}}$, $\mathbf{h}_{\text{sp}}$, $\mathbf{h}_{\text{orb}}$,
$\mathbf{h}_{\text{so}}$, $\mathbf{h}_{\text{orb}*}$, and $\mathbf{h}_{\text{so*}}$
describes the charge, spin, orbital, and spin-orbital fluctuations, respectively.

\section{Derivation of Eq.~\eqref{chis} \label{sec:derivation_chis}}

The linear response for {\color{black}a generic operator} is given by the following
equation: 
\begin{align}
\chi^{}_{\mathcal{O}\mathcal{O}} & =\frac{d}{d\xi}\langle\Phi(\xi,\mathbf{x})|\hat{\mathcal{O}}|\Phi(\xi,\mathbf{x})\rangle|_{(\xi=0,\mathbf{x}(\xi=0))}\nonumber \\
 & =\chi^{(0)}_{\mathcal{O}\mathcal{O}}+\sum_{\mu}\left.\frac{dx_{\mu}}{d\xi}\right|_{\xi=0}\chi_{\mu\mathcal{O}}\,.\label{chistep1}
\end{align}
{\color{black} Note again that $\mu$ runs through all the variational variables in $\mathbf{x}$ (Eq.~\eqref{eq:X}), and we use the variational parameters as the subscripts.}
% such that $x_\mu\equiv\mu$, e.g., $x_{r_s}\equiv r_s$.}
To evaluate Eq.~\eqref{chistep1}, it is necessary to calculate $\frac{dx_\mu}{d\xi}\big|_{\xi=0}$,
which can be determined by taking the total derivative of Eq.~\eqref{stationary-formal}
with respect to $\xi$, as follows: 
\begin{equation}
\sum_{\nu}\mathcal{M}_{\mu\nu}\left.\frac{dx_{\nu}}{d\xi}\right|_{\xi=0}-\chi_{\mu\mathcal{O}}=0\,,\label{eq:fordxdxi}
\end{equation}
where $\mathcal{M}$ is the fluctuation matrix defined in Eq.~\eqref{eq:Mtot}. Substituting Eq.~\eqref{eq:fordxdxi} into Eq.~\eqref{chistep1}, we obtain Eq.~\eqref{chis} in the main text. Since physical susceptibilities in Eq.~\eqref{chiqpstep1} are gauge invariant, all solutions of Eq.~\eqref{eq:fordxdxi}, connected by the gauge transformations (Eq.~\eqref{eq:gauge_X}),
are equivalent.

\section{Fluctuation Matrix \label{sec:M_fluc}}
The fluctuation matrix can be separated into three parts: 
\begin{align}
\mathcal{M}_{\mu\nu}(\mathbf{q}) & = \mathcal{M}_{\mu\nu}^{\text{mix}}+\mathcal{M}_{\mu\nu}^{\text{qp}}(\mathbf{q})+\mathcal{M}_{\mu\nu}^{\text{emb}}%\nonumber \\
% & =\big[\partial_{x_{\mu,-\mathbf{q}}}\partial_{x_{\nu,\mathbf{q}}}\mathcal{L}_{\text{emb}}[{\color{black}\xi},\mathbf{x}]+\partial_{x_{\mu,-\mathbf{q}}}\partial_{x_{\nu,\mathbf{q}}}\mathcal{L}_{\text{qp}}[\mathbf{x}]\nonumber \\
% & +\partial_{x_{\mu,-\mathbf{q}}}\partial_{x_{\nu\mathbf{q}}}\mathcal{L}_{\text{mix}}[\mathbf{x}]\big]\Big|_{(\xi=0,\mathbf{x}(\xi=0))}.
\end{align}
The first part $\mathcal{M}^{\text{mix}}$, which involves the partial derivatives of the mixing term of the Lagrangian $\mathcal{L}_{\text{mix}}$ with respect to $r_s$, $l_s$, $d_s$, $D_s$, $l^c_s$, and $\zeta_s$, is computed from the following equations:
\begin{align}
\mathcal{M}_{r_{s}d_{s'}}^{\text{mix}} & \equiv \partial_{r_s}\partial_{d_{s'}}\mathcal{L}_{\text{mix}}[\mathbf{x}]\Big|_{(\xi=0,\mathbf{x}(\xi=0))}=-\sum_{a\alpha c}\big(\mathbf{D}_{a\alpha}\tilde{\mathbf{h}}_{s,c\alpha}\nonumber \\
 & \partial_{d_{s'}}[\boldsymbol{\Delta}(1-\boldsymbol{\Delta})]_{ca}^{\frac{1}{2}}+\text{c.c}\big),
\end{align}
\begin{align}
\mathcal{M}_{r_{s}D_{s'}}^{\text{mix}} & \equiv\partial_{r_s}\partial_{D_{s'}}\mathcal{L}_{\text{mix}}[\mathbf{x}]\Big|_{(\xi=0,\mathbf{x}(\xi=0))}=-\sum_{a\alpha c}\big(\tilde{\mathbf{h}}_{s,a\alpha}\tilde{\mathbf{h}}_{s',c\alpha}\nonumber \\
 & [\boldsymbol{\Delta}(1-\boldsymbol{\Delta})]_{ca}^{\frac{1}{2}}+\text{c.c}\Big),
\end{align}
\begin{equation}
\mathcal{M}_{l_{s}d_{s'}}^{\text{mix}}=\mathcal{M}_{d_{s}l^{c}_{s'}}^{\text{mix}}=\mathcal{M}_{d_{s}\zeta_{s'}}^{\text{mix}}=-\frac{1}{2}\sum_{ab}\mathbf{h}_{ab}^{s}[\mathbf{h}^{s'}]_{ab}^{t},
\end{equation}
\begin{align}
\mathcal{M}_{d_{s}d_{s'}}^{\text{mix}} & \equiv\partial_{d_{s}}\partial_{d_{s'}}\mathcal{L}_{\text{mix}}[\mathbf{x}]\Big|_{(\xi=0,\mathbf{x}(\xi=0))}=-\sum_{a\alpha c}\big(\mathbf{D}_{a\alpha}\mathbf{R}_{c\alpha}\nonumber \\
 & \partial_{d_{s}}\partial_{d_{s'}}[\boldsymbol{\Delta}(1-\boldsymbol{\Delta})]_{ca}^{\frac{1}{2}}+\text{c.c.}\big),
\end{align}
\begin{align}
\mathcal{M}_{d_{s}D_{s'}}^{\text{mix}} & \equiv\partial_{d_{s}}\partial_{D_{s'}}\mathcal{L}_{\text{mix}}[\mathbf{x}]\Big|_{(\xi=0,\mathbf{x}(\xi=0))}=-\sum_{a\alpha c}\big(\tilde{\mathbf{h}}_{s',a\alpha}\mathbf{R}_{c\alpha}\nonumber \\
 & \partial_{d_{s}}[\boldsymbol{\Delta}(1-\boldsymbol{\Delta})]_{ca}^{\frac{1}{2}}+\text{c.c}\big),
\end{align}
and the other unlisted components of $\mathcal{M}^{\text{mix}}$ are zero.

The second part  $\mathcal{M}^{\text{qp}}$, which involves the partial derivatives of the quasiparticle term of the Lagrangian $\mathcal{L}_{\text{qp}}$ with respect to $r_s$ and $l_s$, is computed from the following equations:
\begin{align}
\mathcal{M}_{r_{s}r_{s'}}^{\text{qp}}&(q)= \partial_{r_{s,-q}}\partial_{r_{s',q}}\mathcal{L}_{\text{qp}}[\mathbf{x}]\Big|_{(\xi=0,\mathbf{x}(\xi=0))}\nonumber\\
&=\frac{1}{2N}\sum_{\mathbf{k}}\text{Tr}\Big\{ n_{F}(H_{\mathbf{k}}^{\text{qp}})\Big[\tilde{\mathbf{h}}_{s}\tilde{\boldsymbol{\epsilon}}_{\mathbf{k}+\mathbf{q}}\tilde{\mathbf{h}}_{s'}^{\dagger} + \tilde{\mathbf{h}}_{s'}\tilde{\boldsymbol{\epsilon}}_{\mathbf{k}-\mathbf{q}}\tilde{\mathbf{h}}_{s}^{\dagger}\Big]\nonumber \\
 &+T\sum_{\omega_{n}}\mathbf{G}_{k}\big[\mathbf{R}\big]^{-1}\big[\mathbf{R}\tilde{\boldsymbol{\epsilon}}_{\mathbf{k}}\tilde{\mathbf{h}}_{s}^{\dagger}+\tilde{\mathbf{h}}_{s}\tilde{\boldsymbol{\epsilon}}_{\mathbf{k}+\mathbf{q}}\mathbf{R}^{\dagger}\big]\big[\mathbf{R}^{\dagger}\big]^{-1}\mathbf{G}_{k+q}\nonumber \\
 & \cdot\big[\mathbf{R}\big]^{-1}\big[\mathbf{R}\tilde{\boldsymbol{\epsilon}}_{\mathbf{k}+\mathbf{q}}\tilde{\mathbf{h}}_{s'}^{\dagger} +\tilde{\mathbf{h}}_{s'}\tilde{\boldsymbol{\epsilon}}_{\mathbf{k}}\mathbf{R}^{\dagger}\Big]\big[\mathbf{R}^{\dagger}\big]^{-1}\Big\},\label{eq:Mrr}
\end{align}
\begin{align}
\mathcal{M}_{r_{s}l_{s'}}^{\text{qp}}&(q) \equiv\partial_{r_{s,-q}}\partial_{l_{s',q}}\mathcal{L}_{\text{qp}}[\mathbf{x}]\Big|_{(\xi=0,\mathbf{x}(\xi=0))}\nonumber \\
 &=\frac{T}{2N}\sum_{k}\text{Tr}\Big\{\mathbf{G}_{k}\big[\mathbf{R}\big]^{-1}\Big[\mathbf{R}\tilde{\boldsymbol{\epsilon}}_{\mathbf{k}}\tilde{\mathbf{h}}_{s}^{\dagger} +\tilde{\mathbf{h}}_{s}\tilde{\boldsymbol{\epsilon}}_{\mathbf{k}+\mathbf{q}}\mathbf{R}^{\dagger}\Big]\nonumber\\
 &\cdot\big[\mathbf{R}^{\dagger}\big]^{-1}\mathbf{G}_{k+q}\big[\mathbf{R}\big]^{-1}\mathbf{h}_{s'}\big[\mathbf{R}^{\dagger}\big]^{-1}\Big\},\label{eq:Mrl}
\end{align}
\begin{align}
\mathcal{M}_{l_{s}l_{s'}}^{\text{qp}}&(q)  \equiv\partial{l_{s,-q}}\partial{l_{s',q}}\mathcal{L}_{\text{qp}}[\mathbf{x}]\Big|_{(\xi=0,\mathbf{x}(\xi=0))}\nonumber \\
 &=\frac{T}{2N}\sum_{k}\text{Tr}\Big\{\mathbf{G}_{k} \big[\mathbf{R}\big]^{-1}\mathbf{h}_{s}\big[\mathbf{R}^{\dagger}\big]^{-1}\nonumber\\
 &\cdot\mathbf{G}_{k+q}\big[\mathbf{R}\big]^{-1}\mathbf{h}_{s'}\big[\mathbf{R}^{\dagger}\big]^{-1}\Big\},\label{eq:Mll}
\end{align}
and the other unlisted components of $\mathcal{M}^{\text{qp}}$ are zero. We also defined $k=(\omega_n,\mathbf{k})$ and $\underset{k}{\sum}\equiv\underset{\mathbf{k}}{\sum}\underset{\omega_{n}}{\sum}$.
Note that since we consider degenerate three-orbital model, at the
normal-state saddle-point, the renormalization matrix, the local potential,
the quasiparticle energy dispersion, and the Green's functions are
all degenerate and diagonal matrices, i.e., 
\begin{align}
\mathbf{R} & =R_{0}\begin{pmatrix}I & 0\\
0 & I
\end{pmatrix},
\end{align}
\begin{equation}
\boldsymbol{\Lambda}=l_{0}\begin{pmatrix}I & 0\\
0 & -I
\end{pmatrix},
\end{equation}
\begin{align}
H_{\mathbf{k}}^{\text{qp}} & =E_{\mathbf{k}}^{\text{qp}}\begin{pmatrix}I & 0\\
0 & -I
\end{pmatrix}
\end{align}
\begin{equation}
\mathbf{G}^{\text{qp}}(k)=\begin{pmatrix}\frac{1}{i\omega_{n}-E_{\mathbf{k}}^{\text{qp}}}I & 0\\
0 & \frac{1}{-i\omega_{n}-E_{\mathbf{k}}^{\text{qp}}}I
\end{pmatrix},
\end{equation}
where $E_{\mathbf{k}}^{\text{qp}}=R_{0}^{2}\epsilon_{\mathbf{k}}+l_{0}$
and $I$ is the $6\times6$ identity matrix. The Matsubara summation
for the fermionic Green's function convolutions in $\mathcal{M}^{\text{qp}}_{rr}$, $\mathcal{M}^{\text{qp}}_{rl}$,
and $\mathcal{M}^{\text{qp}}_{ll}$ can be evaluated analytically from the Lindhard
function.
For example, the particle-hole convolution:
\begin{eqnarray}
T&&\sum_{\omega_{m}}\frac{1}{i\omega_{m}-E^\text{qp}_{\mathbf{k}}}\frac{1}{i\omega_{m}+i\Omega_{n}-E^\text{qp}_{\mathbf{k}+\mathbf{q}}} \nonumber \\
&& =\frac{n_{F}(E^\text{qp}_{\mathbf{k}})-n_{F}(E^\text{qp}_{\mathbf{k}+\mathbf{q}})}{i\Omega_{n}-E^\text{qp}_{\mathbf{k}+\mathbf{q}}+E^\text{qp}_{\mathbf{k}}}.
\end{eqnarray}
and the particle-particle convolution:
\begin{eqnarray}
T&&\sum_{\omega_{m}}\frac{1}{i\omega_m+i\Omega_n-E^\text{qp}_{\mathbf{k}+\mathbf{q}}}\frac{1}{-i\omega_m-E^\text{qp}_{-\mathbf{k}}}\nonumber \\ 
&&= \frac{n_{F}(E^\text{qp}_{\mathbf{k}+\mathbf{q}})-n_{F}(-E^\text{qp}_{\mathbf{-k}})}{i\Omega_{n}-E^\text{qp}_{\mathbf{k}+\mathbf{q}}-E^\text{qp}_{\mathbf{-k}}},
\end{eqnarray}
The analytical continuation to real frequency can be achieved by replacing $i\Omega_n\rightarrow \omega+i0^+$.

The third part  $\mathcal{M}^{\text{emb}}$ involves the partial derivatives of the embedding term of the Lagrangian $\mathcal{L}_{\text{emb}}$ with respect to $D_s$, $l^c_s$, and $\zeta_s$, which can be evaluated as follows.
First, we evaluate the first order derivatives using the Hellmann-Feynman theorem:
\begin{align}
\partial_{l_{s}^{c}}\mathcal{L}_{\text{emb}}[\xi,\mathbf{x}]=\sum_{abcd}\frac{1}{2}\mathbf{h}_{ab}^{s}\langle\Phi(\xi,\mathbf{x})|\bar{I}_{bc}\hat{\Psi}_{c}
\hat{\Psi}_{d}^{\dagger}\bar{I}_{da}|\Phi(\xi,\mathbf{x})\rangle,\label{eq:dLembdlc}
\end{align}
\begin{align}
\partial_{D_{s}}\mathcal{L}_{\text{emb}}[\xi,\mathbf{x}]=2\sum_{a\alpha b}\tilde{\mathbf{h}}_{a\alpha}^{s}\langle\Phi(\xi,\mathbf{x})|\hat{\Xi}_{\alpha}^{\dagger}\hat{\Psi}_{b}\bar{I}_{ba}|\Phi(\xi,\mathbf{x})\rangle,\label{eq:dLembdD}
\end{align}
\begin{equation}
\partial_{\zeta_s}\mathcal{L}_{\text{emb}}[\xi,\mathbf{x}]=\frac{1}{2}\sum_{\alpha\beta}\mathbf{h}^s_{\alpha\beta}\langle\Phi(\xi,\mathbf{x})|\hat{\Xi}_{\alpha}^{\dagger}\hat{\Xi}{}_{\beta}|\Phi(\xi,\mathbf{x})\rangle.\label{eq:dLembdz}
\end{equation}
Then, we can compute the second order derivatives from the following equations:
\begin{widetext}
\begin{align}
\mathcal{M}_{l_{s}^{c}l_{s'}^{c}}^{\text{emb}} & =\partial_{l^c_{s}}\partial_{l^c_{s'}}\mathcal{L}_{\text{emb}}[\xi,\mathbf{x}]\Big|_{(\xi=0,\mathbf{x}(\xi=0))}=\partial_{l_{s}^{c}}\sum_{abcd}\frac{1}{2}\mathbf{h}_{ab}^{s'}\langle\Phi(\xi,\mathbf{x})|\bar{I}_{bc}\hat{\Psi}_{c}\hat{\Psi}_{d}^{\dagger}\bar{I}_{da}|\Phi(\xi,\mathbf{x})\rangle\Big|_{(\xi=0,\mathbf{x}(\xi=0))},\label{eq:Memb1}
\end{align}
\begin{align}
\mathcal{M}_{l_{s}^{c}D_{s'}}^{\text{emb}} & =\partial_{l_{s}^{c}}\partial_{D_{s'}}\mathcal{L}_{\text{emb}}[\xi,\mathbf{x}]\Big|_{(\xi=0,\mathbf{x}(\xi=0))}=\partial_{l_{s}^{c}}2\sum_{a\alpha b}\tilde{\mathbf{h}}_{a\alpha}^{s'}\langle\Phi(\xi,\mathbf{x})|\hat{\Xi}_{\alpha}^{\dagger}\hat{\Psi}_{b}\bar{I}_{ba}|\Phi(\xi,\mathbf{x})\rangle\Big|_{(\xi=0,\mathbf{x}(\xi=0))},\label{eq:Memb2}
\end{align}
\begin{align}
\mathcal{M}_{D_{s},l_{s'}^{c}}^{\text{emb}} & =\partial_{D_{s}}\partial_{l_{s'}^{c}}\mathcal{L}_{\text{emb}}[\xi,\mathbf{x}]\Big|_{(\xi=0,\mathbf{x}(\xi=0))}=\partial_{D_{s}}\frac{1}{2}\sum_{abcd}\mathbf{h}_{ab}^{s'}\langle\Phi(\xi,\mathbf{x})|\bar{I}_{bc}\hat{\Psi}_{c}\hat{\Psi}_{d}^{\dagger}\bar{I}_{da}|\Phi(\xi,\mathbf{x})\rangle\Big|_{(\xi=0,\mathbf{x}(\xi=0))},\label{eq:Memb3}
\end{align}
\begin{align}
\mathcal{M}_{D_{s}D_{s'}}^{\text{emb}} & =\partial_{D_{s}}\partial_{D_{s'}}\mathcal{L}_{\text{emb}}[\xi,\mathbf{x}]\Big|_{(\xi=0,\mathbf{x}(\xi=0))}=\partial_{D_{s}}2\sum_{a\alpha b}\tilde{\mathbf{h}}_{a\alpha}^{s'}\langle\Phi(\xi,\mathbf{x})|\hat{\Xi}_{\alpha}^{\dagger}\hat{\Psi}_{b}\bar{I}_{ba}|\Phi(\xi,\mathbf{x})\rangle\Big|_{(\xi=0,\mathbf{x}(\xi=0))},\label{eq:Memb4}
\end{align}

\begin{align}
\mathcal{M}_{D_{s}\zeta_{s'}}^{\text{emb}} & =\partial_{l^c_{s}}\partial_{\zeta_{s'}}\mathcal{L}_{\text{emb}}[\xi,\mathbf{x}]\Big|_{(\xi=0,\mathbf{x}(\xi=0))}=\partial_{D_{s}}\frac{1}{2}\sum_{\alpha\beta}\mathbf{h}_{\alpha\beta}^{s'}\langle\Phi(\xi,\mathbf{x})|\hat{\Xi}_{\alpha}^{\dagger}\hat{\Xi}_{\beta}|\Phi(\xi,\mathbf{x})\rangle\Big|_{(\xi=0,\mathbf{x}(\xi=0))},\label{eq:Memb2-1-1}
\end{align}
\begin{align}
\mathcal{M}_{\zeta_{s},l_{s'}^{c}}^{\text{emb}} & =\partial_{\zeta_{s}}\partial_{l_{s'}^{c}}\mathcal{L}_{\text{emb}}[\xi,\mathbf{x}]\Big|_{(\xi=0,\mathbf{x}(\xi=0))}=\partial_{\zeta_s}\frac{1}{2}\sum_{abcd}\mathbf{h}_{ab}^{s'}\langle\Phi(\xi,\mathbf{x})|\bar{I}_{bc}\hat{\Psi}_{c}\hat{\Psi}_{d}^{\dagger}\bar{I}_{da}|\Phi(\xi,\mathbf{x})\rangle\Big|_{(\xi=0,\mathbf{x}(\xi=0))},\label{eq:Memb3-1}
\end{align}
\begin{align}
\mathcal{M}_{\zeta_{s}D_{s'}}^{\text{emb}} & =\partial_{\zeta_{s}}\partial_{D_{s'}}\mathcal{L}_{\text{emb}}[\xi,\mathbf{x}]\Big|_{(\xi=0,\mathbf{x}(\xi=0))}=\partial_{\zeta_s}2\sum_{a\alpha b}\tilde{\mathbf{h}}_{a\alpha}^{s'}\langle\Phi(\xi,\mathbf{x})|\hat{\Xi}_{\alpha}^{\dagger}\hat{\Psi}_{b}\bar{I}_{ba}|\Phi(\xi,\mathbf{x})\rangle\Big|_{(\xi=0,\mathbf{x}(\xi=0))},\label{eq:Memb4-1}
\end{align}

\begin{align}
\mathcal{M}_{\zeta_{s}\zeta_{s'}}^{\text{emb}} & =\partial_{\zeta_{s}}\partial_{\zeta_{s'}}\mathcal{L}_{\text{emb}}[\xi,\mathbf{x}]\Big|_{(\xi=0,\mathbf{x}(\xi=0))}=\partial_{\zeta_s}\frac{1}{2}\sum_{\alpha\beta}\mathbf{h}_{\alpha\beta}^{s'}\langle\Phi(\xi,\mathbf{x})|\hat{\Xi}_{\alpha}^{\dagger}\hat{\Xi}_{\beta}|\Phi(\xi,\mathbf{x})\rangle\Big|_{(\xi=0,\mathbf{x}(\xi=0))},\label{eq:Memb2-1-1-1}
\end{align}
%\begin{align}
%\chi_{\nu}^{\text{emb}} & =\frac{\partial^{2}\mathcal{L}_{\text{emb}}}{\partial \mathbf{x}_{\nu}\partial\xi}\Big|_{\xi=0}=\frac{\partial}{\partial \mathbf{x}_{\nu}}\langle\Phi(\mathbf{x})|\mathcal{\hat{O}}_{s}|\Phi(\mathbf{x})\rangle\Big|_{\xi=0}.\label{eq:d2Lembd2z}
%\end{align}
%\begin{align}
%\chi_{0}^{\text{emb}} & =\frac{\partial^{2}\mathcal{L}_{\text{emb}}}{\partial\xi\partial\xi}\Big|_{\xi=0}=\frac{\partial}{\partial\zeta}\langle\Phi(\mathbf{x})|\mathcal{\hat{O}}_{s}|\Phi(\mathbf{x})\rangle\Big|_{\xi=0}.\label{eq:d2Lembd2z-1}
%\end{align}
\end{widetext}
where the other unlisted components of $\mathcal{M}^{\text{emb}}$ are zero.

{\color{black}The above second-order derivatives and Eqs.~\eqref{eq:chiemb0}-\eqref{eq:chimu} can be evaluated using the linear response
theory. We apply a perturbation to the embedding Hamiltonian 
\begin{equation}
\hat{H}_{\text{emb}}(\eta)=\hat{H}_{\text{emb}}+\eta \hat{A},
\end{equation}
where $\hat{A}=\underset{abcd}{\sum}\mathbf{h}_{ab}^{s}\bar{I}_{bc}\hat{\Psi}_{c}\hat{\Psi}_{d}^{\dagger}\bar{I}_{da}$, $\underset{a\alpha b}{\sum}\tilde{\mathbf{h}}_{a\alpha}^{s}\hat{\Xi}_{\alpha}^{\dagger}\hat{\Psi}_{b}\bar{I}_{ba}$, or $\hat{\mathcal{O}}$
corresponding to the perturbation in $\eta=l^c_{s}$, $D_{s}$, or $\xi$, respectively. %Note that, in the following derivations, the Hamiltonians and the operators are expressed in their many-body matrix representation. Let us expand $\langle \hat{A}\rangle_{\eta}$ to the linear order in $\eta$
%\begin{equation}
%\langle \hat{A}\rangle_{\eta}=\langle \hat{A}\rangle_{\eta=0}+\eta A_{1}+O(\eta^{2}),
%\end{equation}
We want to compute the change in the average of $\langle\hat{B}\rangle_{\eta}$ in the limit $\eta\rightarrow0$, where $\hat{B}=\underset{abcd}{\sum}\mathbf{h}_{ab}^{s}\bar{I}_{bc}\hat{\Psi}_{c}\hat{\Psi}_{d}^{\dagger}\bar{I}_{da}$, $\underset{a\alpha b}{\sum}\tilde{\mathbf{h}}_{a\alpha}^{s} \hat{\Xi}_{\alpha}^{\dagger}\hat{\Psi}_{b}\bar{I}_{ba}$, or $\hat{\mathcal{O}}$. This response function can be computed from the spectral representation of the static susceptibility at zero temperature:
\begin{align}
&\left.\frac{\partial\langle \hat{B}\rangle_{\eta}}{\partial\eta}\right|_{\eta=0}=\chi_{\hat{A}\hat{B}}\nonumber\\
=\underset{\epsilon\rightarrow0^+}{\text{lim}}\sum_{n}&\Big[\frac{\langle 0|\hat{A}|n\rangle \langle n|\hat{B}|0\rangle}{E_n-E_0+i\epsilon}-\frac{\langle 0|\hat{B}|n\rangle\langle n|\hat{A}|0\rangle}{E_0-E_n+i\epsilon}\Big],\label{eq:A1}
\end{align}
where $E_n$ is the $n$-th excited state energy of $\hat{H}_{\text{emb}}$ and $|n\rangle$ is the $n$-th excited state wavefunction of $\hat{H}_{\text{emb}}$.
}

Beside the method proposed in Eqs.~\eqref{eq:A1}, one can also use the finite difference method to evaluate the partial derivatives in Eqs.~\eqref{eq:Memb1}-\eqref{eq:Memb2-1-1-1}. Note that both methods requires the diagonalization of the embedding Hamiltonian $\hat{H}_{\text{emb}}$, which is the most time-consuming part of the linear-response calculations. With the current state-of-the-art, we can easily study the $f$-electron materials, containing 7 correlated orbitals, using exact-diagonalization and machine learning techniques~\citep{ML_RISB_2021}.
For the systems with more correlated orbital, one may also utilize the density matrix renormalization group or auxiliary-field quantum Monte Carlo methods~\citep{zheng1155}.
%Regarding the efficiency of our approach, the most time-consuming part of our method is the diagonalization
%of the embedding Hamiltonian $H_{\text{emb}}$. With the current state of the art, we can easily study the $f$-electron materials, containing 7 correlated orbitals, using exact-diagonalization and machine learning techniques~\citep{ML_RISB_2021}. 
%In addition, one could also utilize the density matrix renormalization group or auxiliary-field quantum Monte Carlo to study the systems with more correlated orbitals~\citep{zheng1155}. }
%Therefore, our method will be useful for investigating the response functions in heavy-fermion systems, and for extrapolating the two-particle response functions to the thermodynamic limit with the cluster approaches~\citep{RISB_DMET_Lee_2019}.}

\section{Fluctuation matrix as a bosonic propagator \label{sec:propagator}}

Here we discuss how the fluctuation matrix can be interpreted as the
propagator for the fluctuations of the bosonic variables $\mathbf{x}_{i}$.
Let us expand the Lagrangian,
Eq.~\eqref{eq:Lqptot}, to the second order in 
\begin{align}
\delta \mathbf{x}_{i}^t=(\delta r_{\text{ch}},\delta l_{\text{ch}},\delta d_{\text{ch}},\delta D_{\text{ch}},\delta l_{s}^{c},\delta\zeta_{\text{ch}},...,\delta r_{s},\delta l_{s},\delta d_{s},\nonumber \\
\delta D_{s},\delta l_{s}^{c},\delta\zeta_{s},...,\delta r_{\text{P}},\delta l_{\text{P}},\delta d_{\text{P}},\delta D_{\text{P}},,\delta l_{\text{P}}^{c},\delta\zeta_{\text{P}})
\end{align}
around the normal-state saddle-point: 
\begin{align}
\mathcal{L}^{s}[\delta \mathbf{x} & ,\Xi,\Xi^{\dagger}]=\frac{T}{2N}\sum_{k}\sum_{\alpha\beta}\Xi_{\mathbf{k}\alpha}^{\dagger}\big[\mathbf{G}(k)\big]_{\alpha\beta}^{-1}\Xi_{\mathbf{k}\beta}\nonumber \\
 & +\frac{1}{2}\sum_{i}\delta \mathbf{x}_{i}^t\Big[\mathcal{M}^{\text{mix}}+\mathcal{M}^{\text{emb}}\Big]\delta \mathbf{x}_{i,}\nonumber \\
 & +\sum_{\mathbf{k},\mathbf{q}}\sum_{\alpha\beta}\Big[(\tilde{\Lambda}_{\alpha\beta,r_{s}}^{\mathbf{k},\mathbf{q}}\delta r_{s,\mathbf{q}}\Xi_{\mathbf{k}+\mathbf{q}\alpha}^{\dagger}\Xi_{\mathbf{k}\beta}\nonumber \\
 & +\text{h.c.})+\tilde{\Lambda}_{\alpha\beta,l_{s}}\delta l_{s,\mathbf{q}}\Xi_{\mathbf{k}+\mathbf{q}\alpha}^{\dagger}\Xi_{\mathbf{k}\beta}\Big]\nonumber\\
 & + \sum_{\mathbf{k},\mathbf{k}',\mathbf{q}}\sum_{\alpha\beta} \tilde{\gamma}_{\alpha\beta,r_s r_{s'}}^{\mathbf{k},\mathbf{k}',\mathbf{q}}
 \delta r_{s,\mathbf{q}} \delta r_{s',-\mathbf{q}} \Xi_{\mathbf{k}\alpha}^{\dagger} \Xi_{\mathbf{k}'\beta},\label{eq:L_fluc}
\end{align}
where $\tilde{\Lambda}_{\alpha\beta\mu}$ are the three-leg
vertices defined in Eq.~\eqref{eq:lamr} and \eqref{eq:laml} and $\mathbf{G}(k)$
is the Nambu propagator. We also introduce the four-leg vertex:
\begin{equation}
\tilde{\gamma}_{\alpha\beta r_s r_{s'}}^{\mathbf{k},\mathbf{k}',\mathbf{q}}=\frac{1}{2}\Big[\mathbf{R}^{-1}\big(\tilde{\mathbf{h}}_{s}\tilde{\boldsymbol{\epsilon}}_{\mathbf{k}+\mathbf{q}}\tilde{\mathbf{h}}_{s'}^{\dagger}+\tilde{\mathbf{h}}_{s'}\tilde{\boldsymbol{\epsilon}}_{\mathbf{k}'-\mathbf{q}}\tilde{\mathbf{h}}_{s}^{\dagger}\big)\big(\mathbf{R}^\dagger\big)^{-1}\Big]_{\alpha\beta}.
\end{equation}
We immediately see that the $\mathbf{q}$
independent part of the fluctuation matrix:
\begin{equation}
\mathcal{M}^{\text{mix}}+\mathcal{M}^{\text{emb}}\equiv\mathcal{D}_{0}^{-1}
\end{equation}
can be viewed, in the Gaussian fluctuation sense \citep{Lavagna1990},
as the inverse of the bare bosonic propagator $\mathcal{D}_{0}^{-1}$.
It is important to note that $\mathcal{D}_{0}$ describes the local
multiplet fluctuations because it contains the embedding susceptibilities
shown in Eqs.~\eqref{eq:Memb1}-\eqref{eq:Memb2-1-1-1}. We see that,
for the pairing channel $s=\text{P}$ in Eqs.~\eqref{eq:Memb1}-\eqref{eq:Memb2-1-1-1},
the multiplet fluctuation selects the basis $\mathbf{h}_{\text{P}}$
that increases and removes electron pairs from the saddle-point wavefunction.
Therefore, it describes the local fluctuation with pair excitations.
On the other hand, for channel $s\in\{\text{ch},\text{sp},\text{orb},\text{so},\text{orb*},\text{so*}\}$,
the particle number is conserved. Consequently, they describe the
corresponding local charge, orbital, and spin fluctuations.
\begin{figure}[t]
\begin{centering}
\includegraphics[scale=0.5]{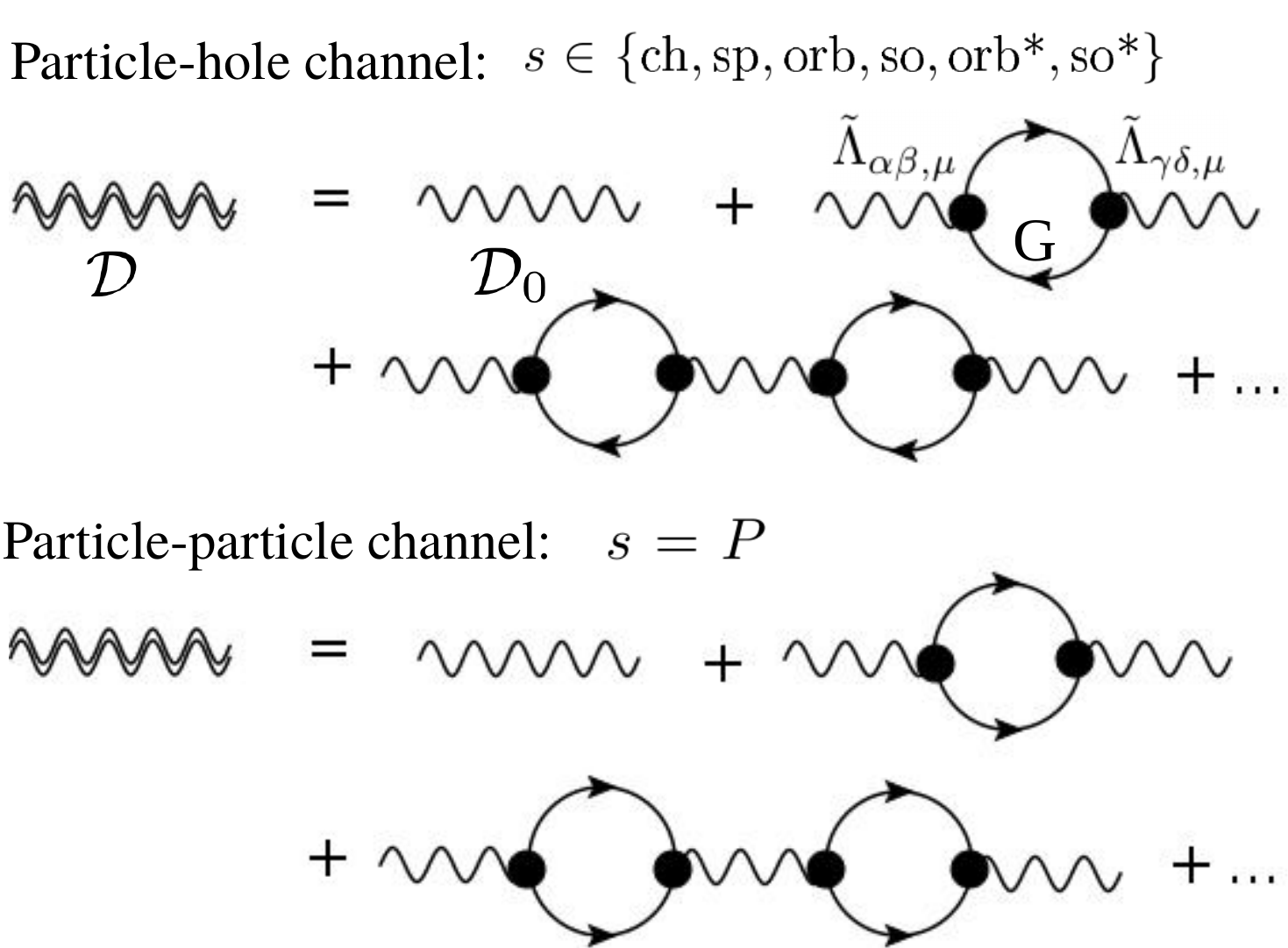}
\par\end{centering}
\caption{Diagrammatic representation of the Dyson equation in Eq.~\eqref{eq:dyson-1}.
The double wavy line and the wavy line denotes the dressed bosonic
propagator $\mathcal{D}$ and the bared bosonic propagator $\mathcal{D}_{0}$.
The solid line denotes the Nambu propagator $\mathbf{G}$. The circle
denotes the three-leg vertices $\tilde{\Lambda}$. \label{fig:dyson-1}}
\end{figure}

We now discuss the role of $\mathcal{M}^{\text{qp}}(\mathbf{q})$.
By integrating out the fermionic field $\Xi_{\mathbf{k}\alpha}$ in Eq.~\eqref{eq:L_fluc}
to the one-loop order, we found the self-energy correction is related
to the fluctuation matrix through $\pi(\mathbf{q})\equiv-\mathcal{\mathcal{M}^{\text{qp}}}(\mathbf{q})$.
Therefore, we can write the total fluctuation matrix in terms of the
Dyson equation:
\begin{align}
\mathcal{M}(\mathbf{q}) & \equiv\mathcal{D}^{-1}(\mathbf{q})=\mathcal{D}_{0}^{-1}-\pi(\mathbf{q}).\label{eq:dyson-1}
\end{align}
The total fluctuation matrix corresponds to the dressed bosonic propagator
with the self-energy correction summing the fermionic bubbles to the
infinite order. From Eq.~\eqref{eq:Mrr}-\eqref{eq:Mll}, we see
that $\mathcal{M}^{\text{qp}}$ contains only the particle-particle
bubbles for the pairing channel $s=P$ , and the particle-hole bubbles
for the other channels $s\in\{\text{ch},\text{sp},\text{orb},\text{so},\text{orb*},\text{so*}\}$.
Figure \ref{fig:dyson-1} shows the diagrammatic representation of
the Dyson equation for the particle-hole and the particle-particle
channels.

\section{Random phase approximation for the interaction vertex \label{sec:RPA}}

In this section, we derive the random phase approximation for the
interaction vertex at $\mathbf{q}=0$. Therefore, we suppress the $\mathbf{q}$ dependent of $\tilde{\Gamma}$, $\tilde{\lambda}$, $\mathcal{D}$, and $\pi$ in the following derivation. The interaction vertex has the following form
(see Eqs.~\eqref{eq:qp_interaction-1} and \eqref{eq:dyson-1}): 
\begin{align}
\tilde{\Gamma} & _{\alpha\beta\gamma\delta}^{s}(\mathbf{k},\mathbf{k}')=-4\Big[\tilde{\Lambda}_{\alpha\beta r_{s}}(\mathbf{k})\mathcal{D}_{r_{s}r_{s}}\tilde{\Lambda}_{\gamma\delta r_{s}}(\mathbf{k}')\nonumber \\
 & +2\tilde{\Lambda}_{\alpha\beta r_{s}}(\mathbf{k})\tilde{\Lambda}_{\gamma\delta l_{s}}(\mathbf{k}')\mathcal{D}_{r_{s}l_{s}}+\tilde{\Lambda}_{\alpha\beta l_{s}}(\mathbf{k})\tilde{\Lambda}_{\gamma\delta l_{s}}(\mathbf{k}')\mathcal{D}_{l_{s}l_{s}}\Big],\label{eq:gamma_sup}
\end{align}
where $\tilde{\Lambda}_{\mu}$ is the three-leg vertex and $\mathcal{D}\equiv\mathcal{M}^{-1}$
is the bosonic Green's function defined in Eq.~\eqref{eq:dyson-1}.
We want to obtain an RPA like form for the vertex: 
\begin{equation}
\tilde{\Gamma}_{\alpha\beta\gamma\delta}^{s}\equiv\langle\langle\tilde{\Gamma}_{\alpha\beta\gamma\delta}^{s}(\mathbf{k},\mathbf{k}')\rangle_{\mathbf{k}_{F}}\rangle_{\mathbf{k}'_{F}}=\frac{F_{s}}{1+F_{s}\chi^{(0)}_{\mathcal{O}_s\mathcal{O}_s}}[\mathbf{h}_{s}]_{\alpha\beta}[\mathbf{h}_{s}]_{\gamma\delta},\label{eq:gamma_sup_RPA}
\end{equation}
after averaging $\mathbf{k}$ and $\mathbf{k}'$ over the Fermi surface, where $F_s$ is the Landau parameter.

We know that the bosonic Green's function has the following Dyson
form for each sector $s$ (see Eq.~\eqref{eq:dyson-1}): 
\begin{align}
\mathcal{D}_{s} & =\Bigg[\big[\mathcal{D}_{0,s}\big]^{-1}-\pi_{s}\Bigg]^{-1}\\
 & =\Big[1-\mathcal{D}_{0,s}\pi\Big]^{-1}\mathcal{D}_{0,s},
\end{align}
where $\mathcal{D}_{0}$ is the bare bosonic propagator, and the
self-energy in each sector $s$ has the form 
\begin{equation}
\pi_{s}=\begin{pmatrix}\pi_{r_{s}r_{s}} & \pi_{r_{s}l_{s}} & 0 & 0 & 0 & 0\\
\pi_{r_{s}l_{s}} & \pi_{l_{s}l_{s}} & 0 & 0 & 0 & 0\\
0 & 0 & 0 & 0 & 0 & 0\\
0 & 0 & 0 & 0 & 0 & 0\\
0 & 0 & 0 & 0 & 0 & 0\\
0 & 0 & 0 & 0 & 0 & 0
\end{pmatrix}.
\end{equation}
The divergence of $\mathcal{D}$ can be determined from 
\begin{align}
\text{Det}\Big[1-\mathcal{D}_{0,s}\pi\Big] & =1-\mathcal{D}_{0,l_{s}l_{s}}\pi_{l_{s}l_{s}}-2\mathcal{D}_{0,r_{s}l_{s}}\pi_{r_{s}l_{s}}\nonumber \\
 & -\mathcal{D}_{0,r_{s}r_{s}}\pi_{r_{s}r_{s}}+(\mathcal{D}_{0,r_{s}l_{s}}\pi_{r_{s}l_{s}})^{2}\nonumber \\
 & -\mathcal{D}_{0,l_{s}l_{s}}\mathcal{D}_{0,r_{s}r_{s}}\pi_{r_{s}l_{s}}^{2}-(\mathcal{D}_{0,r_{s}l_{s}})^{2}\pi_{l_{s}l_{s}}\pi_{r_{s}r_{s}}\nonumber \\
 & +\mathcal{D}_{0,l_{s}l_{s}}\mathcal{D}_{0,r_{s}r_{s}}\pi_{l_{s}l_{s}}\pi_{r_{s}r_{s}}=0.
\end{align}
The interaction vertex can be expressed in terms of $\mathcal{D}_{0}$,
$\tilde{\Lambda}$, and $\pi$ as 
\begin{align}
\tilde{\Gamma}_{\alpha\beta\gamma\delta}^{s} & (\mathbf{k},\mathbf{k}')=-\frac{4}{\text{Det}\Big[1-\mathcal{D}_{0,s}\pi\Big]}\bigg[\tilde{\Lambda}_{\alpha\beta r_{s}}(\mathbf{k})\tilde{\Lambda}_{\gamma\delta r_{s}}(\mathbf{k}')\nonumber \\
 & \Big(\mathcal{D}_{0,r_{s}r_{s}}+(\mathcal{D}_{0,r_{s}l_{s}})^{2}\pi_{l_{s}l_{s}}-\mathcal{D}_{0,l_{s}l_{s}}\mathcal{D}_{0,r_{s}r_{s}}\pi_{l_{s}l_{s}}\Big)\nonumber \\
 & +2\tilde{\Lambda}_{\alpha\beta r_{s}}(\mathbf{k})\tilde{\Lambda}_{\gamma\delta l_{s}}(\mathbf{k}')\Big(\mathcal{D}_{0,r_{s}l_{s}}-(\mathcal{D}_{0,r_{s}l_{s}})^{2}\pi_{r_{s}l_{s}}\nonumber \\
 & +\mathcal{D}_{0,l_{s}l_{s}}\mathcal{D}_{0,r_{s}r_{s}}\pi_{r_{s}l_{s}}\Big)+\tilde{\Lambda}_{\alpha\beta l_{s}}(\mathbf{k})\tilde{\Lambda}_{\gamma\delta l_{s}}(\mathbf{k}')\nonumber \\
 & \Big(\mathcal{D}_{0,l_{s}l_{s}}+(\mathcal{D}_{0,r_{s}l_{s}})^{2}\pi_{r_{s}r_{s}}-\mathcal{D}_{0,l_{s}l_{s}}\mathcal{D}_{0,r_{s}r_{s}}\pi_{r_{s}r_{s}}\Big)\bigg].\label{eq:Gamma_sup2}
\end{align}
We can make further approximation that 
\begin{align}
\pi_{r_{s}r{}_{s'}} & =\frac{-T}{2N}\sum_{k}\text{tr}\Bigg\{\mathbf{G}_{k}\big[\mathbf{R}\big]^{-1}\big[\mathbf{R}\mathbf{\mathbf{\boldsymbol{\epsilon}}}_{\mathbf{k}}\tilde{\mathbf{h}}_{s}^{\dagger}+\tilde{\mathbf{h}}_{s}\mathbf{\mathbf{\boldsymbol{\epsilon}}}_{\mathbf{k}}\mathbf{R}^{\dagger}\big]\nonumber \\
 & \big[\mathbf{R}^{\dagger}\big]^{-1}\mathbf{G}_{k}\big[\mathbf{R}\big]^{-1}\big[\mathbf{R}\mathbf{\mathbf{\boldsymbol{\epsilon}}}_{\mathbf{k}}\tilde{\mathbf{h}}_{s'}^{\dagger}+\tilde{\mathbf{h}}_{s'}\mathbf{\mathbf{\boldsymbol{\epsilon}}}_{\mathbf{k}}\mathbf{R}^{\dagger}\Big]\big[\mathbf{R}^{\dagger}\big]^{-1}\Bigg\}\Bigg|_{\xi=0}\nonumber \\
 & \approx4\langle\tilde{\Lambda}_{r}\rangle^{2}\chi^{(0)}_{\mathcal{O}_s\mathcal{O}_s},\\
\pi_{r_{s}l_{s'}} & =\frac{-T}{2N}\sum_{k}\text{tr}\Bigg\{\mathbf{G}_{k}\big[\mathbf{R}\big]^{-1}\Big[\mathbf{R}\mathbf{\boldsymbol{\epsilon}}_{\mathbf{k}}\tilde{\mathbf{h}}_{s}^{\dagger}+\tilde{\mathbf{h}}_{s}\mathbf{\boldsymbol{\epsilon}}_{\mathbf{k}}\mathbf{R}^{\dagger}\Big]\nonumber \\
 & \big[\mathbf{R}^{\dagger}\big]^{-1}\mathbf{G}_{k}\big[\mathbf{R}\big]^{-1}\mathbf{h}_{s'}\big[\mathbf{R}^{\dagger}\big]^{-1}\Bigg\}\Bigg|_{\xi=0}\nonumber \\
 & \approx4\langle\tilde{\Lambda}_{r}\rangle\langle\tilde{\Lambda}_{l}\rangle\chi^{(0)}_{\mathcal{O}_s\mathcal{O}_s},\\
\pi_{l_{s}l_{s'}} & =\frac{-T}{2N}\sum_{k}\text{tr}\Bigg\{\mathbf{G}_{k}\big[\mathbf{R}\big]^{-1}\mathbf{h}_{s}\big[\mathbf{R}^{\dagger}\big]^{-1}\nonumber \\
 & \mathbf{G}_{k}\big[\mathbf{R}\big]^{-1}\mathbf{h}_{s'}\big[\mathbf{R}^{\dagger}\big]^{-1}\Bigg\}\Bigg|_{\xi=0}\nonumber \\
 & =4\langle\tilde{\Lambda}_{l}^{2}\rangle\chi^{(0)}_{\mathcal{O}_s\mathcal{O}_s},
\end{align}
where we average the vertex over the Fermi surface: 
\begin{align}
\langle\tilde{\Lambda}_{r_{s}}\rangle & =\frac{R_{0}^{2}}{2}\langle2\epsilon_{\mathbf{k}}\rangle_{\mathbf{k}_{F}},\\
\langle\tilde{\Lambda}_{l_{s}}\rangle & =\frac{1}{2}.
\end{align}
We see that, after averaging all the vertices $\tilde{\Lambda}$ and
self-energy $\pi$ in Eq.~\eqref{eq:Gamma_sup2} over the Fermi surface,
there are further cancellation in the denominator and the numerator
of $\tilde{\Gamma}^{s}$ in Eq.~\eqref{eq:Gamma_sup2}. %In the end, we have $O(\chi_{0})$ in the denominator and $O(1)$ in the numerator:
%\begin{equation}
%\tilde{\Gamma}^{s}_{\alpha\beta\gamma\delta}\approx\frac{-4\tilde{\Lambda}_{r_{s}}^{2}D_{r_{s}r_{s}}^{(0)}-8\tilde{%\Lambda}_{r_{s}}D_{r_{s}l_{s}}^{(0)}-4D_{l_{s}l_{s}}^{(0)}}{1+(-4\tilde{\Lambda}_{r_{s}}^{2}D_{r_{s}r_{s}}^{(0)}-8D%_{r_{s}l_{s}}^{(0)}\tilde{\Lambda}_{r_{s}}-4D_{l_{s}l_{s}}^{(0)})\chi^{(0)}_{\mathcal{O}_s\mathcal{O}_s}}[\mathbf{h}]_{\alpha\beta}[\mathbf{h}]_%{\gamma\delta}.
%\end{equation}
%Comparing with Eq.~\eqref{eq:gamma_sup}, 
Recasting Eq.~\eqref{eq:Gamma_sup2} in the form of Eq.~\eqref{eq:gamma_sup_RPA},
we identify that the irreducible interaction (Landau parameter) $F_{s}$ in Eq.~\eqref{eq:gamma_sup_RPA}
for each channel $s$ is 
\begin{equation}
F_{s}=-4\langle\tilde{\Lambda}_{r_{s}}\rangle^{2}\mathcal{D}_{0,r_{s}r_{s}}-8\mathcal{D}_{0,r_{s}l_{s}}\langle\tilde{\Lambda}_{r_{s}}\rangle-4\mathcal{D}_{0,l_{s}l_{s}}.
\end{equation}

%\begin{equation}
%V_{\text{pp}}^{\text{sc}}=-4\tilde{\Lambda}_{r_{P}}^{2}D_{r_{P}r_{P}}^{(0)}-8D_{r_{P}l_{P}}^{(0)}\tilde{\Lambda}_{r_{P}}-4D_{l_{P}l_{P}}^{(0)},
%\end{equation}
%and the Landau parameters
%\begin{align}
%f_{s}&=F^s/N_0=-4\tilde{\Lambda}_{r_{s}}^{2}D_{r_{s}r_{s}}^{(0)}-8D_{r_{s}l_{s}}^{(0)}\tilde{\Lambda}_{r_{s}}-4D_{l_{s}l_{s}}^{(0)},
%\end{align}
%for $s\in\{\text{ch},\ \text{sp},\ \text{orb},\ \text{so},\ \text{orb*},\ \text{so*}\}$, and $N_0=\chi^{(0)}_{\mathcal{O}_s\mathcal{O}_s}$ is %the density of state at the Fermi level.

\section{Gauge invariance\label{subsec:Gauge}}

The RISB Lagrangian is invariant under the following gauge transformation
\citep{Isidori_RISB_SC_2009,Lanata_2017_PRL}:

\begin{align}
\Phi & \rightarrow\Phi U(\theta),\ \boldsymbol{\Delta}\rightarrow u^{t}(\theta)\boldsymbol{\Delta}[u^{\dagger}]^{t}(\theta),\label{eq:gauge_1}\\
\mathbf{R} & \rightarrow u^{\dagger}(\theta)\mathbf{R},\ \boldsymbol{\Lambda}\rightarrow u^{\dagger}(\theta)\boldsymbol{\Lambda}u(\theta),\label{eq:gauge_2}\\
\mathbf{D} & \rightarrow u^{t}(\theta)\mathbf{D},\ \boldsymbol{\Lambda}^{c}\rightarrow u^{\dagger}(\theta)\boldsymbol{\Lambda}^{c}u(\theta),\label{eq:gauge_3}
\end{align}
where 
\begin{align}
u(\theta) & =e^{i\underset{s}{\sum}\theta_{s}\mathbf{T}_{s,ab}},\\
U(\theta) & =e^{i\underset{s}{\sum}\theta_{s}\mathbf{T}_{s,ab}\Psi_{a}^{\dagger}\Psi_{b}}.
\end{align}
$\mathbf{T}_{s}$ are the generators for the gauge group, and $\theta_{s}$
are the Lie parameters. The specific form of $\mathbf{T}_{s}$ corresponding
to our variational setup is shown in Appx. \ref{subsec:Gauge-fix}.

We also define the corresponding gauge transformation for $\mathbf{x}$, see
Eq.~\eqref{eq:X}: 
\begin{equation}
\mathbf{x}\rightarrow\mathcal{G}_{\theta}[\mathbf{x}],\label{eq:gauge_X}
\end{equation}
where the operator $\mathcal{G}_{\theta}$ transforms each element
in Eq.~\eqref{eq:X} according to Eqs.~\eqref{eq:gauge_1}-\eqref{eq:gauge_3}.
%(see Eqs.~\eqref{eq:gauge_4}-\eqref{eq:gauge_6}). 

\section{Gauge-fixing procedure\label{subsec:Gauge-fix}}

In this section, we describe the gauge-fixing procedure for the fluctuation matrix $\mathcal{M}$. We define a gauge transformation (see Eq.~\eqref{eq:gauge_X}): 
\begin{equation}
\mathbf{x}'=\mathcal{G}_{\theta}[\mathbf{x}(\xi)]=\mathbf{x}(\xi,\theta(\xi))
\end{equation}
where each component of $\mathbf{x}$ transform as 
\begin{align}
r_{s} & =\text{Tr}\left.[\tilde{\mathbf{h}}_{s}^{\dagger}u^{\dagger}(\theta)\mathbf{R}\right],\ l_{s}=\text{Tr}\left.[\mathbf{h}_{s}^{\dagger}u^{\dagger}(\theta)\boldsymbol{\Lambda}u(\theta)\right],\label{eq:gauge_4}\\
D_{s} & =\text{Tr}\left.[\tilde{\mathbf{h}}_{s}^{\dagger}u^{t}(\theta)\mathbf{D}\right],\ l^{c}=\text{Tr}\left.[\mathbf{h}_{s}^{\dagger}u^{\dagger}(\theta)\boldsymbol{\Lambda}^{c}u(\theta)\right],\label{eq:gauge_5}\\
d_{s} & =\text{Tr}\left.[\mathbf{h}_{s}^{\dagger}u^{t}(\theta)\boldsymbol{\Delta}[u^{\dagger}]^{t}(\theta)\right],\label{eq:gauge_6}
\end{align}
and $\theta(0)=0$ at $\xi=0$.

Given that $\mathbf{x}$ is a solution of (see Eq.~\eqref{eq:fordxdxi})
\begin{equation}
\sum_{\nu}\mathcal{M}_{\mu\nu}\left.\frac{\partial x_\nu}{\partial\xi}\right|_{\xi=0}=\chi_{\mu},
\end{equation}
$\mathbf{x}'$ is also a solution of 
\begin{equation}
\sum_{\nu}\mathcal{M}_{\mu\nu}\left.\frac{\partial x_\nu'}{\partial\xi}\right|_{\xi=0}=\chi_{\mu}.
\end{equation}
{\color{black} Note again that here $\mu$ and $\nu$ runs through all the elements in $\mathbf{x}$ (Eq.~\eqref{eq:X}), and we use the variational parameters as the subscripts.}
%such that $x_\mu\equiv\mu$, e.g., $x_{r_s}\equiv r_s$.}
Also, we have 
\begin{equation}
\left.\frac{\partial x_{\mu}'}{\partial\xi}\right|_{\xi=0}=\left.\frac{\partial {x_{\mu}}}{\partial\xi}\right|_{\xi=0}+\sum_s\left.\frac{\partial x_{\mu}}{\partial\theta_{s}}\right|_{\theta=0}\left.\frac{\partial\theta_{s}}{\partial\xi}\right|_{\xi=0}.
\end{equation}
Consequently, we show that 
\begin{equation}
\mathcal{M}_{\mu\nu}\left.\frac{\partial x_\nu}{\partial\theta_{s}}\right|_{\theta=0}\left.\frac{\partial\theta_{s}}{\partial\xi}\right|_{\xi=0}=0,
\end{equation}
which implies that $\mathcal{M}_{\mu\nu}$ has zero eigenvalues, and the
kernels are defined as 
\begin{equation}
K_{s,\mu}:=\left\{ \left.\frac{\partial {x_\mu}}{\partial\theta_{s}}\right\vert _{\theta=0}\right\} 
\end{equation}
such that: 
\begin{equation}
M_{\mu\nu}K_{s,\nu}=0\ \forall K_{s}\in K.
\end{equation}
We can fix the gauge by projecting the matrices onto the vector space
$v_{i,\mu}$ perpendicular to $K$, where $v_{i,\mu}$ can be constructed
from the Gram-Schmidt process. The reduced fluctuation matrix and
the embedding susceptibilities becomes: 
\begin{align}
\bar{\mathcal{M}}_{ij} & =v_{i,\mu}\mathcal{M}_{\mu\nu}v_{j,\nu},\label{eq:Mbar}\\
\bar{\chi}_{i\mathcal{O}} & =v_{i,\mu}\chi_{\mu\mathcal{O}}.
\end{align}
Consequently, we have the physical susceptibility 
\begin{align}
\chi^{}_{\mathcal{O}\mathcal{O}} & =\chi^{(0)}_{\mathcal{O}\mathcal{O}}+\bar{\chi}_{i\mathcal{O}}\bar{\mathcal{M}}_{ij}^{-1}\bar{\chi}_{j\mathcal{O}}.\label{eq:chiemb}
\end{align}
Now the $\bar{\mathcal{M}}_{ij}^{-1}$ does not contain zero modes
and the matrix inversion is well defined.

For the model considered in this work, where we restricted the variational variables $\mathbf{x}$ to
real numbers (Eqs.~\eqref{eq:R}-\eqref{eq:Lamc}), the $U(1)$ gauge degrees of freedom in the charge, spin, orbital, and spin-orbital 
channels are fixed. However, we are left with one gauge degree of freedom relating to the Nambu
pseudo-spin rotation generator:
%Since we restrict $\mathbf{x}$ to be real variables (Eqs.~\eqref{eq:R}-\eqref{eq:Lamc}) and the fluctuating basis $\mathbf{h}_{s}$ to a subset of symmetry adapted basis (Eqs.~\eqref{eq:h_ch}-\eqref{eq:h_P}), this variational setup removes most of the gauge redundancy and leaves us with only one gauge transformation generator:

\begin{equation}
\mathbf{T}=\tau_{1}\otimes\lambda_{6}\otimes(i\sigma_{y}\sigma_{z}),
\end{equation}
where $\tau_{i}$ is the Pauli matrix corresponding to Nambu pseudospin.
From the definition of the gauge transformation (Eqs.~\eqref{eq:gauge_1}-\eqref{eq:gauge_3}),
we derive the kernel $K$:

\begin{align}
\left.\frac{\partial r_{s}}{\partial\theta}\right|_{\theta=0} & =-i\text{Tr}\big[(\tilde{\mathbf{h}}_{s})^{\dagger}\mathbf{T}\mathbf{R}\big]=\frac{r_{0}}{2\sqrt{3}}\delta_{s,P}\\
\left.\frac{\partial l_{s}}{\partial\theta}\right|_{\theta=0} & =-i\text{Tr}\big[(\mathbf{h}_{s})^{\dagger}[\mathbf{T},\boldsymbol{\Lambda}]\big]=-\frac{l_{0}}{\sqrt{3}}\delta_{s,P}\\
\left.\frac{\partial d_{s}}{\partial\theta}\right|_{\theta=0} & =i\text{Tr}\big[(\mathbf{h}_{s})^{\dagger}[\mathbf{T}^{t},\boldsymbol{\Delta}]\big]=-\frac{d_{0}}{\sqrt{3}}\delta_{s,P}\\
\left.\frac{\partial D_{s}}{\partial\theta_{j}}\right|_{\theta=0} & =i\text{Tr}\big[(\tilde{\mathbf{h}}_{s})^{\dagger}\mathbf{T}^{t}\mathbf{D}\big]=\frac{D_{0}}{2\sqrt{3}}\delta_{s,P}\\
\left.\frac{\partial l_{s}^{c}}{\partial\theta}\right|_{\theta=0} & =-i\text{Tr}\big[(\mathbf{h}_{s})^{\dagger}[\mathbf{T},\boldsymbol{\Lambda}^{c}]\big]=-\frac{l_{0}^{c}}{\sqrt{3}}\delta_{s,P},
\end{align}
where the $K$ vector is only non-zero in the pairing channel. We
can then construct the vector space $v_{i,\mu}$ using the Gram-Schmidt
process and compute the susceptibilities through Eq.~\eqref{eq:Mbar}-\eqref{eq:chiemb}.

\begin{figure}[t]
\begin{centering}
\includegraphics[scale=0.4]{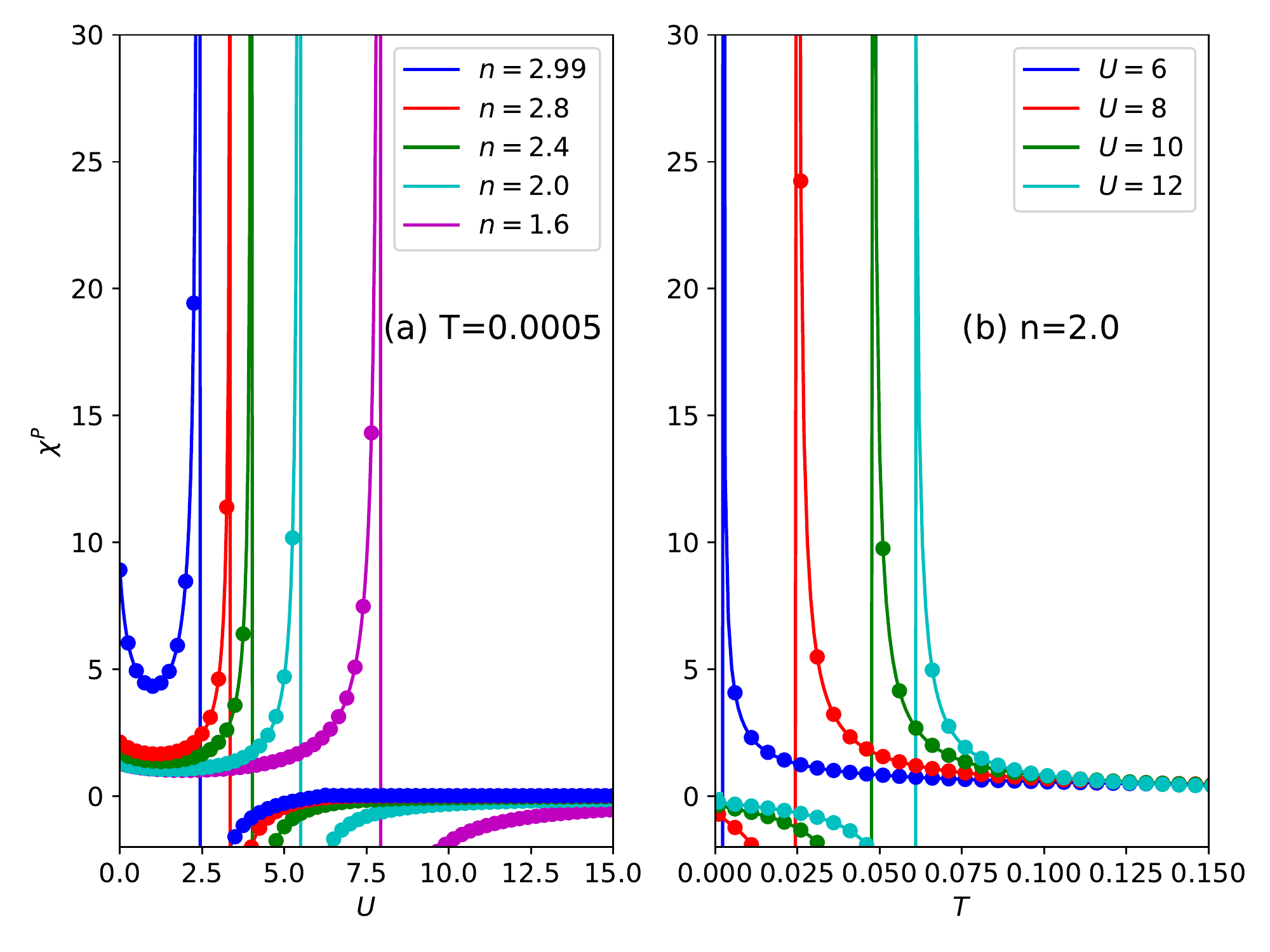}
\par\end{centering}
\caption{Comparison of the pairing susceptibilities $\chi^{P}$ computed {\color{black}from  Eq.~\eqref{eq:chi_deg} (solid line, without ``quasiparticle constraint'') and Eq.~\eqref{eq:chiqp_deg} (filled circles, with ``quasiparticle constraint'')} for (a) $T=0.0005$ and $n=2.99,\ 2.8,\ 2.4,\ 2.0,\ 1.6$
as a function of $U$  and $J=U/4$ , and (b) $n=2.0$ and $U=6,\ 8,\ 10,\ 12$ as
a function of $T$ and $J=U/4$ . \label{fig:FL_approx}}
\end{figure}

\section{Validity of the Fermi-liquid approximation\label{sec:FL_approx_check}}

In this section, we show the pairing susceptibility $\chi^{P}$ computed
from the equation without enforcing the ``quasiparticle constraint'' (Eq.~\eqref{eq:chi_deg}) and the equation with the ``quasiparticle constraint'' (Eq.~\eqref{eq:chiqp_deg}) in Fig. \ref{fig:FL_approx}.
The $\chi^{P}$ obtained from the two approaches are identical for
all the parameter regime, indicating the validity of the Fermi-liquid
approximation described in Sec. \ref{sec:QP_sus}.

\begin{figure}[t]
\begin{centering}
\includegraphics[scale=0.4]{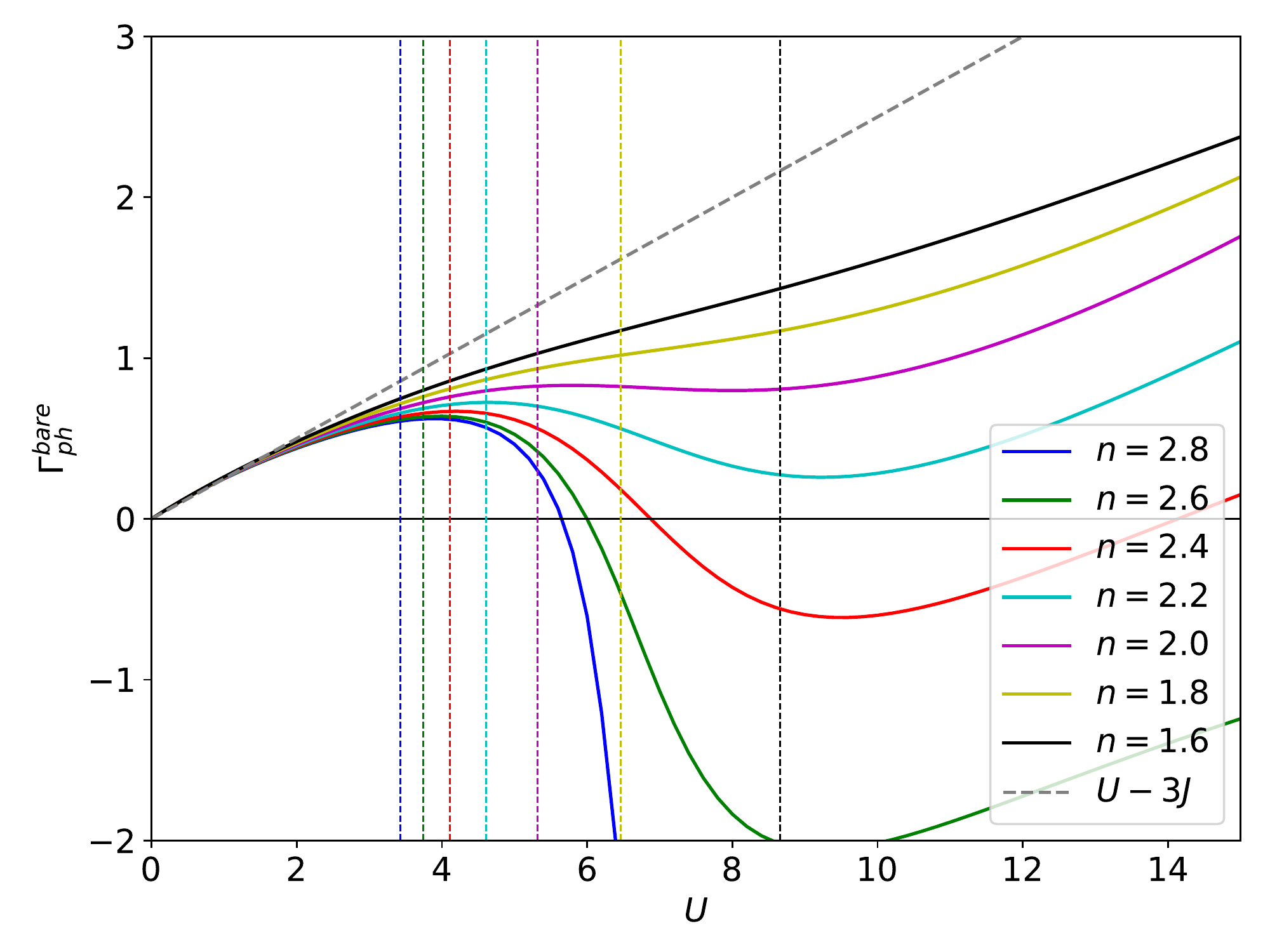}
\par\end{centering}
\caption{The bare pairing interaction in the particle-hole channel $\Gamma_{\text{ph}}^{\text{bare}}$ (Eq.~\eqref{eq:Gamma_ph_bare}) as a function of Coulomb interaction $U$ and {\color{black}$J=U/4$} for filling $n=2.8,\ 2.6,\ 2.4,\ 2.2,\ 2.0,\ 1.8,\ 1.6$ at temperature $T=0.0005t$. \label{fig:Vbare}}
\end{figure}

\section{Bare pairing interaction in the particle-hole channel \label{sec:Vbare}}

We also compute the bare pairing interaction in the particle-hole channel defined as follows:{\color{black}
\begin{align}
\Gamma_{\text{ph}}^{\text{bare}}=\frac{1}{4}\Big[F_{\text{ch}}+F_{\text{sp}}
-F_{\text{orb}}-F_{\text{so}}
-\frac{5}{3}F_{\text{orb*}}-\frac{5}{3}F_{\text{so*}}\Big],\label{eq:Gamma_ph_bare}
\end{align}}In this case, the summation of the fermionic particle-hole bubbles in Fig. \ref{fig:diagram} (b) are ignored and only the bare interaction (Landau parameters $F_s$ at $\mathbf{q}=0$) is considered. Figure \ref{fig:Vbare} shows the bare pairing interaction in the particle-hole channel. We found that the bare pairing interaction only turns negative (signalizing the pairing instability) for filling $n<2.3$.

\begin{figure}[t]
\begin{centering}
\includegraphics[scale=0.4]{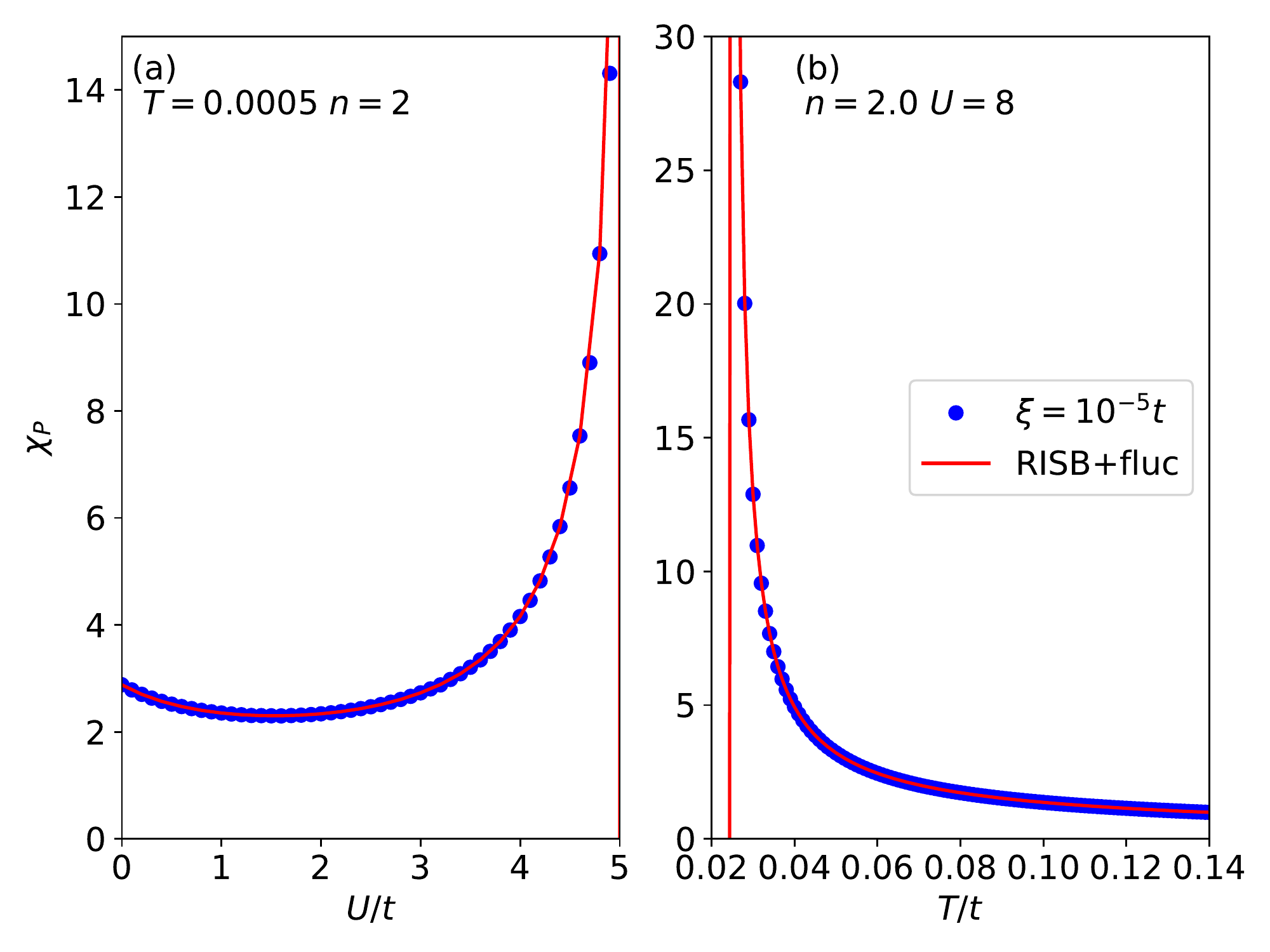}
\par\end{centering}
\caption{Comparison of the uniform pairing susceptibility $\chi^{\text{P}}(\mathbf{q}=0,\omega=0)$
evaluated from fluctuation approach with the pairing susceptibility
evaluated from the mean-field solution $\chi^{\text{P}}=\frac{d\langle\mathcal{O}_{\text{P}}\rangle}{d\zeta}$
with small pairing field $\zeta=10^{-5}$ for (a) temperature $T=0.0005t$
and filling $n=2.0$. and (b) Coulomb interaction $U=8t$ and filling
$n=2.0$. We fix the Hund's coupling interaction at $J=U/4$. \label{fig:MF_check}}
\end{figure}

\section{Consistency check for susceptibility}

We perform the consistency check for the pairing susceptibility between
the one computed from RISB fluctuation approach and the one computed from
RISB mean-field theory with a small pairing field $\zeta$. The
definition of the pairing susceptibility in the RISB self-consistent mean-field
theory is $\chi^{P}=\frac{\partial\langle\mathcal{O}_{P}\rangle}{\partial\zeta}\Big|_{\zeta\rightarrow0}$.
The results from the two approaches are shown in Fig. \ref{fig:MF_check}
(a) as function of Coulomb interaction $U$ for $T=0.0005t$ and filling
$n=2.0$ and (b) as a function of temperature $T$ for $U=8t$ and
filling $n=2.0$. We confirm that the $\chi^{\text{P}}(\mathbf{q}=0,\omega=0)$
computed from the fluctuation approach (red line) agrees excellently
with the $\chi^{\text{P}}$ computed from the mean-field theory with a small
pairing field $\zeta=10^{-5}t$ (blue dots). The agreement between
the two approaches indicates the consistency of our fluctuation approach within the RISB framework. 

\begin{figure}[t]
\begin{centering}
\includegraphics[scale=0.4]{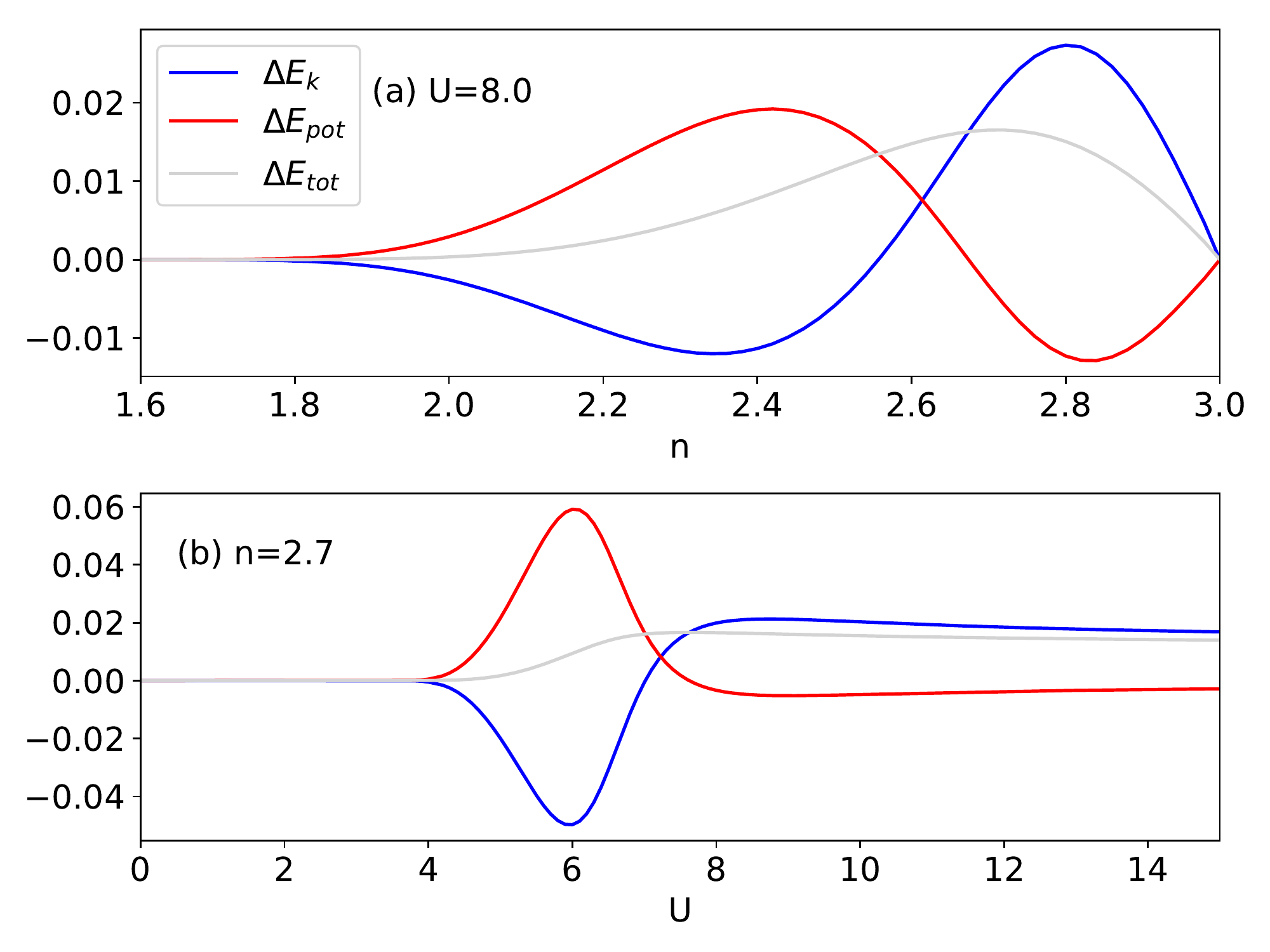}
\par\end{centering}
\caption{The kinetic energy $\Delta E_{\text{k}}$, the potential energy $\Delta E_{\text{pot}}$,
and the total energy gain $\Delta E_{\text{tot}}$ for the superconducting
paring state for (a) as a function of electron filling $n$ at $U=8t$
and $T=10^{-4}t$ and (b) as a function of Coulomb interaction $U$ and $J=U/4$ 
at $n=2.7$ and $T=10^{-4}t$.\label{fig:Energy}}
\end{figure}

\section{Total energy and weak to strong-coupling crossover}

We now discuss the energetic of the $s$-wave spin-triplet pairing
state. Figure \ref{fig:Energy} (a) and (b) shows the kinetic energy
gain $\Delta E_{\text{k}}=E_{\text{k}}^{N}-E_{\text{k}}^{sc}$, the
potential energy gain $\Delta E_{\text{pot}}=E_{\text{pot}}^{N}-E_{\text{pot}}^{sc}$,
and the total energy gain $\Delta E_{\text{tot}}=E_{\text{tot}}^{N}-E_{\text{tot}}^{sc}$
for forming the $s$-wave spin-triplet pairing state, where the superscript
$N$ and $sc$ corresponds to the energy in the normal state and the
superconducting state, respectively. The energetic in both Fig. (a)
and (b) shows a typical weak-coupling to strong-coupling crossover
behavior \citep{Sofo1992,Bulka1996,Bak1998}, where the energy gain
is dominated from the potential energy in the weak-coupling limit, 
and from the kinetic energy in the strong-coupling limit. Interestingly,
we find this crossover locates around the Hund's metal crossover where
the quasiparticle weight drops rapidly and the superconducting order
parameter shows a pronounce peak.

\section{Application to density matrix embedding theory \label{sec:dmet}}
In this section, we outline the equations for computing the susceptibility in {\color{black} the ``non-interacting bath'' DMET (NIB-DMET)} formalism. Since the {\color{black}NIB-DMET} self-consistent equations can be reproduced from the RISB Lagrangian by enforcing $\mathbf{R}=I$ and an additional constraint in Eq.~\eqref{qpcondition}, the formalism in Sec.~\ref{sec:QP_sus} can be directly applied to {\color{black}NIB-DMET} by removing the $r_s$ sector of the fluctuation basis in Eq.~\eqref{eq:X_zeta}, i.e., no fluctuation in $\mathbf{R}$. Hence, the {\color{black}NIB-DMET} fluctuation basis becomes:
\begin{align}
\mathbf{x}_{\mathbf{q}}=( & l_{\text{ch},\mathbf{q}},d_{\text{ch},\mathbf{q}},D_{\text{ch},\mathbf{q}},l_{\text{ch},\mathbf{q}}^{c},\zeta_{\text{ch},\mathbf{q}},...,l_{s,\mathbf{q}}, d_{s,\mathbf{q}},D_{s,\mathbf{q}},\nonumber \\
 &l_{s,\mathbf{q}}^{c} ,\zeta_{s,\mathbf{q}},...,l_{\text{P},\mathbf{q}},d_{\text{P},\mathbf{q}},D_{\text{P},\mathbf{q}},l_{\text{P},\mathbf{q}}^{c},\zeta_{\text{P},\mathbf{q}})\label{eq:X_zeta_dmet},
\end{align}
where, differ from RISB (Eq.~\eqref{eq:X_zeta}), the variables $r_s$ is absent. Following the same derivation in Sec.~\ref{sec:QP_sus}, the {\color{black}NI-DMET} susceptibility of an arbitrary {\color{black}operator} $\hat{\mathcal{O}}$ has the following form:
\begin{equation}
\chi^{}_{\mathcal{O}\mathcal{O}}(\mathbf{q}) =\chi^{(0)}_{\mathcal{O}\mathcal{O}}(\mathbf{q})+\sum_{\mu\nu}\chi_{\mu\mathcal{O}}(\mathbf{q})\mathcal{M}_{\mu\nu}^{-1}(\mathbf{q})\chi_{\nu\mathcal{O}}(\mathbf{q}),\label{chiqp_dmet}
\end{equation}
where the fluctuation matrix $\mathcal{M}$ is given in Appx.~\ref{sec:M_fluc} and we have to enforce $\mathbf{R}=I$ in each element. We have also introduced the following susceptibilities: 
\begin{align}
\chi^{(0)}_{\mathcal{O}\mathcal{O}}(\mathbf{q}) & =-\frac{T}{2N}\sum_{\omega_n\mathbf{k}}\text{Tr}\Big[\mathbf{G}_{\omega_n,\mathbf{k}+\mathbf{q}}\mathcal{O}\mathbf{G}_{\omega_n,\mathbf{k}}\mathcal{O}\Big],\label{eq:chi0qp_dmet}
\end{align}
\begin{align}
\chi_{\mu\mathcal{O}}(\mathbf{q}) & =\frac{T}{2N}\sum_{\omega_n,\mathbf{k}}\partial_{x_{\mu,\mathbf{q}}}\text{Tr}\left.\big[\mathbf{G}_{\omega_n,\mathbf{k}+\mathbf{q},\mathbf{k}}[\xi,\mathbf{x}]\mathcal{O}\big]\right|_{(\xi=0,\mathbf{x}(\xi=0))},\label{eq:chiXqp_dmet}
\end{align}
where $\mathcal{O}$ is the single-particle matrix representation of a generic {\color{black}operator}.
The Green's function has the following form:
\begin{equation}
\big[\mathbf{G}_{\omega_{n},\mathbf{k}_{1},\mathbf{k}_{2}}[\mathbf{x},\xi]\big]^{-1}=i\omega_{n}-\big[H_{\mathbf{k}_{1}\mathbf{k}_{2}}^{\text{qp}}\big]_{ab}+\xi_{\mathbf{k}_{1}-\mathbf{k}_{2}}\big[\mathcal{O}\big]_{ab},
\end{equation}
and  $\mathbf{G}_{\omega_n,\mathbf{k}}$ is the Green's function evaluated at $\xi=0$. We also introduce the quasiparticle Hamiltonian (low-level mean-field Hamiltonian):
\begin{align}
\big[H_{\mathbf{k}_{1}\mathbf{k}_{2}}^{\text{qp}}\big]_{ab} & \equiv \big[\tilde{\boldsymbol{\epsilon}}_{\mathbf{k}_1}\big]_{ab} \delta_{\mathbf{k}_{1},\mathbf{k}_{2}}+\big[\boldsymbol{\Lambda}_{\mathbf{k}_{1}-\mathbf{k}_{2}}\big]_{ab},
\end{align}
where $\boldsymbol{\Lambda}$ corresponds to the correlation potential in {\color{black}NIB-DMET}.

For the degenerate model considered in this work, the susceptibility can be written as:
\begin{align}
\chi^{}_{\mathcal{O}_s\mathcal{O}_s}(\mathbf{q}) =\chi^{(0)}_{\mathcal{O}_s\mathcal{O}_s}(\mathbf{q}) + \chi_{l_{s}\mathcal{O}_s}(\mathbf{q})\mathcal{M}_{l_{s}l_{s}}^{-1}(\mathbf{q})\chi_{l_{s}\mathcal{O}_s}(\mathbf{q}),\label{eq:chiqp_deg_dmet}
\end{align}
where 
\begin{align}
\chi_{l_{s}\mathcal{O}_s}(\mathbf{q}) & =-\frac{T}{2N}\sum_{\omega_n\mathbf{k}}\text{Tr}\Big[\mathbf{G}_{\omega_n,\mathbf{k}+\mathbf{q}}\mathbf{h}_{s}\mathbf{G}_{\omega_n,\mathbf{k}}\mathbf{h}_{s}\Big]=\chi^{(0)}_{\mathcal{O}_s\mathcal{O}_s}(\mathbf{q})\label{eq:chiqp_l_dmet}.
\end{align}
$\mathcal{M}_{l_{s}l_{s}}^{-1}(\mathbf{q})$ denotes the $\mu=l_{s}$
and $\nu=l_{s}$ component of $\mathcal{M}_{\mu\nu}^{-1}(\mathbf{q})$. The $\mathbf{G}_{\omega_n,\mathbf{k}}=[i\omega_n-\tilde{\epsilon}_{\mathbf{k}}-\boldsymbol{\Lambda}]^{-1}$ is the saddle-point Green's function.
%Comparing Eq.~\eqref{eq:chiqp_deg_dmet} to Eq.~\eqref{eq:chiqp_deg}, we see that Eq.~\eqref{eq:chiqp_deg_dmet} contains only the fluctuations in ${l_s}$, corresponding to the fluctuations of the correlation potential $\boldsymbol{\Lambda}$.

Finally, we comment on the advantages and the disadvantages between RISB and {\color{black}NIB-DMET}. One advantage of RISB with respect to {\color{black}NIB-DMET} is the presence of the renormalization matrix $\mathbf{R}$. It allows the description of the Mott transition within the single-site approach~\cite{BrinkmanRice1970}, while in the standard {\color{black}NIB-DMET}, one has to use at least a two-site cluster to capture the Mott transition~\cite{DMET_2012}. On the other hand, the additional determination of $\mathbf{R}$ in RISB may require more self-consistency iterations with respect to {\color{black}NI-DMET}, leading to more diagonalization of the embedding Hamiltonian $\hat{H}_{\text{emb}}$. Nevertheless, the performance and the accuracy of the two methods are similar~\citep{RISB_DMET_Lee_2019}. {\color{black} Note that our approach does not apply to the ``interacting bath'' construction of DMET (IB-DMET), which produces more accurate results than the NIB-DMET~\citep{DMET_2012,Kawano2020,DET_2014}. The extension of our approach to IB-DMET will be an interesting future topic.}
%However, as pointed out in Ref. 12, the gauge-invariance of the RISB Lagrangian greatly mitigates this problem, as it generates a manifold of physical equivalent solutions, reducing significantly the complexity of the optimization problem.

%\bibliographystyle{apsrev}
\bibliography{ref}

\end{document}